\documentclass[nojss,article]{jss}\usepackage[]{graphicx}\usepackage[]{color}
\makeatletter
\def\maxwidth{ %
  \ifdim\Gin@nat@width>\linewidth
    \linewidth
  \else
    \Gin@nat@width
  \fi
}
\makeatother

\usepackage[utf8]{inputenc}   

\author{Sebastian Meyer\\University of Zurich \And
        Leonhard Held\\University of Zurich \And
        Michael Höhle\\Stockholm University}
\title{\vspace{4cm}Spatio-Temporal Analysis of Epidemic Phenomena
  Using the \proglang{R} Package \pkg{surveillance}}

\Plainauthor{Sebastian Meyer, Leonhard Held, Michael Höhle}
\Plaintitle{Spatio-Temporal Analysis of Epidemic Phenomena Using the R Package surveillance}
\Shorttitle{\pkg{surveillance}: Spatio-Temporal Analysis of Epidemic Phenomena}

\Abstract{
The availability of geocoded health data and the inherent temporal
structure of communicable diseases have led to
an increased interest in statistical models and software for 
spatio-temporal data with epidemic features.
The open source \proglang{R} package \pkg{surveillance} can handle
various levels of aggregation at which infective events have been recorded:
individual-level time-stamped geo-referenced data (case reports) in either
continuous space or discrete space,
as well as counts aggregated by period and region. 
For each of these data types, the \pkg{surveillance} package implements tools for
visualization, likelihoood inference and simulation from recently developed
statistical regression frameworks capturing endemic and epidemic dynamics.
Altogether, this paper is a guide to the spatio-temporal modeling of
epidemic phenomena, exemplified by analyses of
public health surveillance data on measles and invasive meningococcal disease.
}


\Address{
  Sebastian Meyer\\
  Epidemiology, Biostatistics and Prevention Institute\\
  University of Zurich\\
  Hirschengraben 84\\
  CH-8001 Zurich, Switzerland\\
  E-mail: \email{sebastian.meyer@uzh.ch}\\
  URL: \url{http://www.ebpi.uzh.ch/en/aboutus/departments/biostatistics.html}

  \medskip
  Leonhard Held\\
  Epidemiology, Biostatistics and Prevention Institute\\
  University of Zurich\\
  E-mail: \email{leonhard.held@uzh.ch}

  \medskip
  Michael Höhle\\
  Department of Mathematics\\
  Stockholm University\\
  E-mail: \email{hoehle@math.su.se}\\
  URL: \url{http://www.math.su.se/~hoehle}
}

\DefineVerbatimEnvironment{Sinput}{Verbatim}{fontshape=sl, fontsize=\small}
\DefineVerbatimEnvironment{Soutput}{Verbatim}{fontsize=\small}

\usepackage{fixltx2e}         
\usepackage{lmodern}          
\usepackage[english]{babel}   

\usepackage{amsmath,amsfonts} 
\usepackage{mathtools}        
\usepackage{bm}               
\usepackage{bbm}              

\usepackage{booktabs}         
\usepackage{multirow}         
\usepackage{subcaption}       
\newcommand{\subfloat}[2][need a sub-caption]{\subcaptionbox{#1}{#2}} 

\usepackage{thumbpdf}         

\newcommand{\abs}[1]{\lvert#1\rvert}
\newcommand{\norm}[1]{\lVert#1\rVert}

\newcommand{\dif}{\,\mathrm{d}}

\newcommand{\ind}{\mathbbm{1}}
\DeclareMathOperator{\Po}{Po}

\DeclareMathOperator{\N}{N}

\newcommand{\class}[1]{\code{#1}}  

\shortcites{keeling.etal2001,mossong.etal2008}
\IfFileExists{upquote.sty}{\usepackage{upquote}}{}
\begin{document}




\section{Introduction} \label{sec:intro}

Epidemic data are realizations of
spatio-temporal processes with autoregressive or ``self-exciting'' behavior.
Examples of epidemic phenomena beyond infectious diseases include
earth quakes \citep{ogata1999},
crimes \citep{johnson2010,mohler.etal2011},
invasive species \citep{balderama.etal2012},
and forest fires \citep{vrbik.etal2012}.
Epidemic data are special with regard to at least three aspects,
which hinder the application of classical statistical approaches:
the data are rarely a result of planned experiments,
the observations (cases, events) are not independent, and
often the process is only partially observable.

Since 2005, 
the open source \proglang{R} \citep{R:base} package
\pkg{surveillance} provides a growing computational framework for methodological
developments and practical tools for the \emph{monitoring} and
\emph{modeling} of epidemic phenomena --
traditionally in the context of infectious diseases.
Monitoring is concerned with prospective aberration detection for which several
algorithms have been implemented as described by \citet{hoehle2007} and recently
updated and reviewed by \citet{salmon.etal2014}.
The other major purpose of the \pkg{surveillance} package and the focus of this
paper is the regression-oriented modeling of spatio-temporal epidemic data.
This enables the user to
a) assess the role of environmental factors, socio-demographic characteristics,
or control measures in shaping endemic and epidemic dynamics,
b) analyze the spatio-temporal interaction of events,
and c) simulate the epidemic spread from estimated models.

The implemented statistical modeling frameworks have
already been successfully applied to a broad range of surveillance data, e.g.,
human influenza \citep{paul.etal2008,paul.held2011,geilhufe.etal2012},
meningococcal disease \citep{paul.etal2008,paul.held2011,meyer.etal2011},
measles \citep{herzog.etal2011},
psychiatric hospital admissions \citep{meyer.etal2015},
rabies in foxes \citep{hoehle.etal2009},
coxiellosis in cows \citep{schroedle.etal2012},
and the classical swine fever virus \citep{hoehle2009}.
Although these applications all originate from public or animal health surveillance, we
stress that our methods also apply to the other epidemic phenomena
described above.


To the best of our knowledge, no other software can estimate regression models
for spatio-temporal epidemic data. There are, however, some related
\proglang{R}~packages that we like to mention here, since they also deal with
epidemic phenomena.
For instance, the \emph{\proglang{R}-epi project}\footnote{%
  \url{https://sites.google.com/site/therepiproject/}
} lists the package \pkg{EpiEstim} \citep{cori.etal2013}, which can estimate
the average number of secondary cases caused by an infected
individual, the so-called reproduction number, from a time series of disease
incidence. Similar functionality is provided by the package~\pkg{R0}
\citep{obadia.etal2012}. Other packages are designed to estimate transmission
characteristics from phylogenetic trees (\pkg{TreePar},
\citealp{stadler.bonhoeffer2013}), or to reconstruct transmission trees from
sequence data (\pkg{outbreaker}, \citealp{jombart.etal2014a}).
The package \pkg{amei} \citep{merl.etal2010} is targeted towards finding optimal
intervention strategies, e.g., the proportion of the population to be vaccinated
to prevent further disease spread, using purely temporal epidemic models.
The recently published package \pkg{tscount} \citep{liboschik.etal2015} is
dedicated to the analysis of count time series with serial correlation such as
the number of stock market transitions per minute or the weekly number of reported
infections of a particular disease. The \pkg{tscount} package can fit a
univariate version of the areal count time-series model presented in
Section~\ref{sec:hhh4}.
For a purely spatial analysis of disease occurrence, see, e.g., the recent paper
by \citet{brown2015} introducing the package \pkg{diseasemapping}.
One of the few packages fitting spatio-temporal epidemic models
is \pkg{etasFLP} \citep{adelfio.chiodi2015}.
The Epidemic-Type Aftershock-Sequences (ETAS) model for earthquakes
\citep{ogata1999} is closely related to the endemic-epidemic point process model
described in Section~\ref{sec:twinstim}, but incorporates seismological laws
rather than covariates.
The long-standing package \pkg{splancs} \citep{R:splancs} offers diagnostic
tools to investigate space-time clustering in a point pattern, i.e., to check if
the process at hand shows self-exciting epidemic behavior. Statistical tests for
space-time interaction are discussed in \citet{meyer.etal2015}, who propose a
test based on the regression framework of Section~\ref{sec:twinstim}.
An important recent development for spatio-temporal tasks in~\proglang{R} are
the basic data classes and utility functions provided by the dedicated package
\pkg{spacetime} \citep{pebesma2012}, which builds upon the quasi standards
\pkg{sp} \citep{Bivand.etal2013} for spatial data and \pkg{xts} \citep{R:xts}
for time-indexed data, respectively.
For a more general overview of \proglang{R}~packages for spatio-temporal data,
see the CRAN Task View ``Handling and Analyzing Spatio-Temporal Data''
\citep{CTV:SpatioTemporal}.
A non-\proglang{R} option is the \emph{Spatiotemporal Epidemiological Modeler} (STEM)
tool\footnote{\url{https://www.eclipse.org/stem/}}.
It has a graphical user interface 
and can simulate the evolution of disease incidence in a population.
The ability to estimate model
parameters from surveillance data, however, is limited to simple non-spatial
models.
WinBUGS has been used for Bayesian inference of 
specialized spatio-temporal epidemic models \citep{malesios.etal2014}.

The remainder of this paper is organized as follows:
Section~\ref{sec:models} gives a brief overview of the three 
statistical models for spatio-temporal epidemic data
implemented in \pkg{surveillance}.
Each of the subsequent model-specific Sections~\ref{sec:twinstim} to \ref{sec:hhh4}
first describes the associated methodology and then illustrates the model
implementation
-- including data handling, visualization, inference, and simulation --
by applications to infectious disease surveillance data.
Section~\ref{sec:conclusion} concludes the paper.


\section{Spatio-temporal endemic-epidemic modeling} \label{sec:models}

Epidemic models traditionally describe the spread of a communicable
disease in a population.  Often, a compartmental view of the
population is taken, placing individuals into one of the three states
(S)usceptible, (I)nfectious, or (R)emoved. Modeling the transitions
between these states in a closed population using deterministic
differential equations dates back to the work of
\citet{kermack.mckendrick1927}. Considering a stochastic version of
the simplest homogeneous SIR model in a closed population of size~$N$,
the hazard rate for a susceptible individual~$i\in S(t)$ to become infectious at
time~$t$ -- the so-called force of infection -- is
\begin{equation} \label{eqn:simpleSIR}
  \lambda_i(t) = \sum_{j \in I(t)} \beta \:.
\end{equation}
Here, $S(t), I(t) \subseteq \{1,\dotsc,N\}$ denote the index sets of
currently susceptible and infectious individuals, respectively, and the parameter
$\beta>0$ is called the transmission rate. The stochastic SIR model is
complemented by a distributional assumption about how long individuals
are infective, where typical choices are the exponential or the gamma
distribution. The set of recovered individuals at time~$t$ is found as
$R(t) = \{1,\dotsc,N\} \setminus ( S(t) \cup I(t) )$.
The above homogeneous SIR model has since been extended
in a multitude of ways, e.g., by additional states (addressing
population heterogeneities arising from age groups, spatial location
or vaccination) or population demographics. Overviews of SIR
modeling approaches can be found in \citet{Anderson.May1991},
\citet{Daley.Gani1999}, and \citet{Keeling.Rohani2008}.
The estimation
of SIR model parameters from actual observed data is, however, often only
treated marginally in such descriptions. In contrast, a number of
more statistically flavored epidemic models have emerged recently.
This includes, e.g., the
TSIR model~\citep{finkenstaedt.grenfell2000},
two-component time-series models~\citep{held.etal2005,held.etal2006a},
and point process models~\citep{lawson.leimich2000,diggle2006}.
An overview of temporal and spatio-temporal epidemic models and their
relation to the underlying metapopulation SIR models can be found
in \citet{Hoehle2016}.

At the heart of any statistical analysis is the subject-matter
scientific problem, which a data-driven analysis seeks to address. Due
to the generality and complexity of such problems we adopt here a
technocratic view and let the available data guide what a ``useful''
epidemic model is. The \pkg{surveillance} package offers
regression-oriented modeling frameworks for three different types of
spatio-temporal data distinguished by the spatial and temporal resolution
(Table~\ref{tab:models}).
First, if an entire region is continuously monitored for infective events,
which are time-stamped, geo-referenced, and
potentially enriched with further event-specific data,
then a (marked) spatio-temporal point pattern arises.
Such continuous space-time epidemic data can be viewed as a realization of a
self-exciting spatio-temporal point process (Section~\ref{sec:twinstim}).
The second data type we consider comprises the event history of a discrete set
of units followed over time -- e.g., farms during
livestock epidemics -- while registering when they become susceptible, 
infected, and potentially removed (neither at risk nor infectious).
These data fit into the framework of a spatial SIR model represented as a
multivariate temporal point process (Section~\ref{sec:twinSIR}).
Our third data type 
is often encountered as a result of privacy protection or
reporting regimes, and is an aggregated version of the individual event data
mentioned first: event counts by region and period.
Such areal count time series can be fitted with the multivariate negative
binomial time-series model presented in Section~\ref{sec:hhh4}.

The three aforementioned model
classes are all inspired by the Poisson branching process with
immigration approach
proposed by \citet{held.etal2005}.  Its main characteristic is the additive
decomposition of disease risk into \emph{endemic} and \emph{epidemic}
features, similar to the \emph{background} and \emph{triggered} components in the ETAS model
for earthquake occurrence.
The endemic component describes the risk of
new events by external factors independent of the history of the
epidemic process. In the context of infectious diseases, such factors may include
seasonality, population density, socio-demographic
variables, and vaccination coverage -- all potentially varying in time
and/or space.  Explicit dependence between events is then
introduced through an epidemic component driven by the observed past.

Each of the following three model-specific sections starts with a brief theoretical
introduction to the respective spatio-temporal endemic-epidemic model,
before we describe the implementation using the example data
mentioned in Table~\ref{tab:models}.


\begin{table}[thb]
\centering
\footnotesize
\begin{tabular}{l|c|c|c}
\toprule
& \bfseries\code{twinstim} (Section~\ref{sec:twinstim}) &
\bfseries\code{twinSIR} (Section~\ref{sec:twinSIR}) &
\bfseries\code{hhh4} (Section~\ref{sec:hhh4}) \\
\midrule
\textbf{Data class} & \class{epidataCS} & \class{epidata} & \class{sts}\\[1ex]
\multirow{2}{*}{\textbf{Resolution}} & individual events in & individual SI[R][S] event
& event counts aggregated\\
& continuous space-time & history of a fixed population & by region and time period\\[1ex]
\multirow{2}{*}{\textbf{Example}} & cases of meningococcal &
measles outbreak among & weekly counts of measles by\\
& disease, Germany, 2002--8 & children in Hagelloch, 1861 &
district, Weser-Ems, 2001--2\\[1ex]
\multirow{2}{*}{\textbf{Model}} & (marked) spatio-temporal & multivariate temporal &
multivariate time series\\
& point process & point process & (Poisson or NegBin)\\[1ex]
\textbf{Reference} & \citet{meyer.etal2011} & \citet{hoehle2009} & \citet{held.paul2012} \\
\bottomrule
\end{tabular}
\caption{Spatio-temporal endemic-epidemic models and corresponding data classes
  implemented in the \proglang{R}~package \pkg{surveillance}.}
\label{tab:models}
\end{table}


\section{Spatio-temporal point pattern of infective events} \label{sec:twinstim}

The endemic-epidemic spatio-temporal point process model ``\code{twinstim}'' is
designed for point-referenced, individual-level surveillance data.
As an illustrative example, we use case reports of invasive meningococcal
disease (IMD) caused by the two most common bacterial finetypes of meningococci
in Germany, 2002--2008, as previously analyzed by \citet{meyer.etal2011}
and \citet{meyer.held2013}.
We start by describing the general model class in Section~\ref{sec:twinstim:methods}.
Section~\ref{sec:twinstim:data} introduces the example data and the associated class
\class{epidataCS},
Section~\ref{sec:twinstim:fit} presents the core functionality of
fitting and analyzing such data using \code{twinstim}, and
Section~\ref{sec:twinstim:simulation} shows how to simulate realizations from a
fitted model.

\subsection[Model class]{Model class: \code{twinstim}} \label{sec:twinstim:methods}

Infective events occur at specific points in continuous space and
time, which gives rise to a spatio-temporal point pattern
$\{(\bm{s}_i,t_i): i = 1,\dotsc,n\}$
from a region~$\bm{W}$ observed during a period~$(0,T]$.
The locations~$\bm{s}_i$ and time points~$t_i$ of the $n$~events can be regarded
as a realization of a self-exciting spatio-temporal point process,
which can be characterized by its
conditional intensity function (CIF, also termed intensity process)
$\lambda(\bm{s},t)$.
It represents the instantaneous event rate at location~$\bm{s}$ at time point~$t$
given all past events, and is often more verbosely denoted by~$\lambda^*$
or by explicit conditioning on the ``history''~$\mathcal{H}_t$ of the process.
\citet[Chapter~7]{Daley.Vere-Jones2003} provide a rigorous mathematical
definition of this concept, which is key to likelihood analysis and simulation
of ``evolutionary'' point processes.

\citet{meyer.etal2011} formulated the model class ``\code{twinstim}'' --
a \emph{two}-component \emph{s}patio-\emph{t}emporal \emph{i}ntensity \emph{m}odel --
by a superposition of an endemic and an epidemic component:
\begin{equation} \label{eqn:twinstim}
  \lambda(\bm{s},t) = \nu_{[\bm{s}][t]} +
  \sum_{j \in I(\bm{s},t)} \eta_j \, f(\norm{\bm{s}-\bm{s}_j}) \, g(t-t_j) \:.
\end{equation}

This model constitutes a branching process with immigration,
where part of the event rate is due to the first, endemic component, which
reflects sporadic events caused by unobserved sources of infection.
This background rate of new events
is modelled by a piecewise constant log-linear predictor
$\nu_{[\bm{s}][t]}$ incorporating regional and/or time-varying characteristics.
Here, the space-time index $[\bm{s}][t]$ refers to the region covering $\bm{s}$
during the period containing $t$ and thus spans a whole spatio-temporal grid on
which the involved covariates are measured, e.g., district $\times$ month.
We will later see that the endemic component therefore simply equals an
inhomogeneous Poisson process for the event counts by cell of that grid.

The second, observation-driven epidemic component adds ``infection pressure''
from the set
\begin{equation*}
  I(\bm{s},t) = \big\{ j : t_j < t \:\wedge\: t-t_j \le \tau_j
  \:\wedge\: \norm{\bm{s}-\bm{s}_j} \le \delta_j \big\}
\end{equation*}
of past events and hence makes the process ``self-exciting''.
During its infectious period of length~$\tau_j$
and within its spatial interaction radius~$\delta_j$,
the model assumes each event~$j$ to trigger further events, which are
called offspring, secondary cases, or aftershocks, depending on the application.
The triggering rate (or force of infection) is proportional
to a log-linear predictor~$\eta_j$ associated with event-specific
characteristics (``marks'') $\bm{m}_j$, which are usually
attached to the point pattern of events.
The decay of infection pressure with increasing spatial and temporal
distance from the infective event is modelled by parametric interaction
functions~$f$ and~$g$, respectively \citep[Section~4]{lawson.leimich2000}.
A simple assumption for the time course of infectivity is $g(t) = 1$.
Alternatives include exponential decay, a step function, 
or empirically derived functions such as Omori's law for aftershock intervals
\citep{utsu.etal1995}.
With regard to spatial interaction, the statistician's standard choice is a
Gaussian kernel $f(x) = \exp\left\{-x^2/(2 \sigma^2)\right\}$.
However, in modeling the spread of human infectious diseases on larger scales,
a heavy-tailed power-law kernel $f(x) = (x+\sigma)^{-d}$ was found to
perform better \citep{meyer.held2013}.
The (possibly infinite) upper bounds~$\tau_j$ and~$\delta_j$ provide a way of
modeling event-specific interaction ranges. However, since these need to be
pre-specified, a common assumption is $\tau_j \equiv \tau$ and
$\delta_j \equiv \delta$, where the infectious period~$\tau$ and the spatial
interaction radius~$\delta$ are determined by subject-matter considerations.

\subsubsection{Model-based effective reproduction numbers}

Similar to the simple SIR model
\citep[see, e.g.,][Section 2.1]{Keeling.Rohani2008}, the above point process
model~\eqref{eqn:twinstim} features a reproduction number derived from its
branching process interpretation. 
As soon as an event occurs (individual becomes infected), it triggers offspring
(secondary cases) around its origin $(\bm{s_j}, t_j)$
according to an inhomogeneous Poisson process with rate
$\eta_j \, f(\norm{\bm{s}-\bm{s_j}}) \, g(t-t_j)$.
Since this triggering process is independent of the event's parentage and of other
events, the expected number $\mu_j$ of events triggered by event $j$
can be obtained by integrating the triggering rate over
the observed interaction domain: 
\begin{gather} \label{eqn:R0:twinstim}
  \mu_j = \eta_j \cdot
  \left[ \int_0^{\min(T-t_j,\tau_j)} g(t) \,dt \right] \cdot
  \left[ \int_{\bm{R}_j} f(\norm{\bm{s}}) \,d\bm{s} \right] \:,
\shortintertext{where}
  \label{eqn:twinstim:IR}
  \bm{R}_j = (b(\bm{s}_j,\delta_j) \cap \bm{W}) - \bm{s}_j
\end{gather}
is event $j$'s influence region centered at $\bm{s}_j$, and
$b(\bm{s}_j, \delta_j)$ denotes the disc centered at $\bm{s}_j$ with radius $\delta_j$.
Note that the above model-based reproduction number $\mu_j$ is event-specific
since it depends on event marks through $\eta_j$, on the ranges
of interaction $\delta_j$ and $\tau_j$, as well as on the event location
$\bm{s}_j$ and time point $t_j$.

Equation \ref{eqn:R0:twinstim} can also be motivated by looking at
a spatio-temporal version of the simple SIR model~\eqref{eqn:simpleSIR}
wrapped into the \class{twinstim} class~\eqref{eqn:twinstim}. This means:
no endemic component,
homogeneous force of infection ($\eta_j \equiv \beta$),
homogeneous mixing in space ($f(x) = 1$, $\delta_j \equiv \infty$),
and exponential decay of infectivity ($g(t) = e^{-\alpha t}$, $\tau_j
\equiv \infty$).
Then, for $T \rightarrow \infty$,
\begin{equation*}
  \mu = \beta \cdot
  \left[ \int_0^\infty e^{-\alpha t} \,dt \right] \cdot
  \left[ \int_{\bm{W}-\bm{s_j}} 1 \,d\bm{s} \right] =
  \beta \cdot \abs{\bm{W}} / \alpha \:,
\end{equation*}
which corresponds to the basic reproduction number
known from the simple SIR model by interpreting $\abs{\bm{W}}$ as the population
size, $\beta$ as the transmission rate and $\alpha$ as the removal rate.
Like in classic epidemic models, the process is sub-critical if $\mu < 1$ holds,
which means that its eventual extinction is almost sure.

However, it is crucial to understand that in a full model with an
endemic component, new infections may always occur via ``immigration''.
Hence, reproduction numbers in \class{twinstim} are adjusted for
infections occurring independently of previous infections.
This also means that a misspecified endemic component may distort model-based
reproduction numbers \citep{meyer.etal2015}.
Furthermore, under-reporting and implemented control measures imply that the
estimates are to be thought of as \emph{effective} reproduction numbers.

\subsubsection{Likelihood inference}

The log-likelihood of the point process model~\eqref{eqn:twinstim} is a function
of all parameters in the log-linear predictors $\nu_{[\bm{s}][t]}$ and $\eta_j$ and in
the interaction functions $f$ and $g$. It has the form
\begin{equation} \label{eqn:twinstim:loglik}
  \left[ \sum_{i=1}^{n} \log\lambda(\bm{s}_i,t_i) \right] -
  \int_0^T \int_{\bm{W}} \lambda(\bm{s},t) \dif\bm{s} \dif t \:.
\end{equation}
To estimate the model parameters, we maximize the above log-likelihood
numerically using the quasi-Newton algorithm available through the \proglang{R}
function \code{nlminb}. We thereby make use of the
analytical score function and an approximation of the expected Fisher information 
worked out by \citet[Web Appendices A and B]{meyer.etal2011}.

The space-time integral in the log-likelihood poses no difficulties for
the endemic component of $\lambda(\bm{s},t)$ since it is piecewise constant.
However, integration of the epidemic component has a clear computational
bottleneck: two-dimensional integrals
$\int_{\bm{R}_i} f(\norm{\bm{s}}) \dif\bm{s}$
over the influence regions~$\bm{R}_i$ of Equation~\ref{eqn:twinstim:IR},
which are computationally represented by polygons (as is~$\bm{W}$).
Similar integrals appear in the score function, where $f(\norm{\bm{s}})$ is
replaced by partial derivatives with respect to kernel parameters,
e.g., $\partial f(\norm{\bm{s}})/\partial \log\sigma$ for the Gaussian kernel
with standard deviation estimated on the log-scale.
Calculation of these integrals is trivial for (piecewise) constant~$f$, but otherwise requires
numerical integration. For this purpose, the \proglang{R}~package
\pkg{polyCub} \citep{R:polyCub} offers cubature methods for polygonal
domains as described in \citet[Section~2]{meyer.held2013:suppB}.
For Gaussian~$f$, we apply the two-dimensional midpoint rule with a
$\sigma$-adaptive bandwidth, combined with an analytical formula via the $\chi^2$
distribution if the $6\sigma$-circle around $\bm{s}_i$ is contained in
$\bm{R}_i$ \citep{meyer.etal2011}. The integrals
in the score function are
approximated by product Gauss cubature \citep{sommariva.vianello2007}.
For the recently implemented power-law kernels \citep{meyer.held2013},
we apply a particularly appealing method which takes analytical
advantage of the assumed isotropy of spatial interaction
in such a way that numerical integration remains in only one dimension
\citep[Section 2.4]{meyer.held2013:suppB}.
As a general means to reduce the computational burden during numerical
log-likelihood maximization, we \pkg{memoise} \citep{R:memoise}
the cubature function, which avoids redundant re-evaluations of the integral
with identical parameters of~$f$.

\subsubsection[Special case: Endemic-only model]{Special case: Endemic-only \code{twinstim}}

As mentioned above, a \code{twinstim} model \emph{without} an epidemic component
can actually be represented as a Poisson regression model for aggregated counts.
This provides a nice link to 
ecological regression approaches in general \citep{Waller.Gotway2004} 
and to the count data model \code{hhh4} illustrated in Section~\ref{sec:hhh4}.
To see this, recall that the endemic component $\nu_{[\bm{s}][t]}$ of a
\code{twinstim}~\eqref{eqn:twinstim}
is piecewise constant on the spatio-temporal grid with cells $([\bm{s}],[t])$.
Hence the log-likelihood~\eqref{eqn:twinstim:loglik} of an endemic-only
\code{twinstim} simplifies to a sum over all these cells,
\begin{equation*}
\sum_{[\bm{s}],[t]} \left\{
  Y_{[\bm{s}][t]} \log\nu_{[\bm{s}][t]} -
  \abs{[\bm{s}]} \, \abs{[t]} \, \nu_{[\bm{s}][t]}
\right\} \:,
\end{equation*}
where $Y_{[\bm{s}][t]}$ is the aggregated number of events observed in cell
$([\bm{s}],[t])$, and $\abs{[\bm{s}]}$ and $\abs{[t]}$ denote cell area and length,
respectively.
Except for an additive constant, the above log-likelihood is equivalently
obtained from the Poisson model 
$Y_{[\bm{s}][t]} \sim \Po( \abs{[\bm{s}]} \, \abs{[t]} \, \nu_{[\bm{s}][t]})$.
This relation offers a means of code validation using the established
\code{glm} function to fit an endemic-only \code{twinstim} model,
see the examples in \code{help("glm_epidataCS")}.

\subsubsection[Extension: Event types]{Extension: \code{twinstim} with event types}

To model the example data on invasive meningococcal disease in the remainder of this section,
we actually need to use an extended version $\lambda(\bm{s},t,k)$ of
Equation~\ref{eqn:twinstim}, which accounts for different event types~$k$ with
own transmission dynamics.
This introduces a further dimension in the point process, and the second
log-likelihood component in Equation~\ref{eqn:twinstim:loglik}
accordingly splits into a sum over all event types.
We refer to \citet[Sections~2.4 and~3]{meyer.etal2011} for the technical details
of this type-specific \code{twinstim} class.
The basic idea is that the meningococcal finetypes share the same endemic pattern
(e.g., seasonality), while infections of different finetypes are not associated
via transmission. This means that the force of infection is restricted to
previously infected individuals with the same bacterial finetype~$k$,
i.e., the epidemic sum in Equation~\ref{eqn:twinstim} is over the set
$I(\bm{s},t,k) = I(\bm{s},t) \cap \{j: k_j = k\}$.
The implementation has limited support for type-dependent interaction
functions $f_{k_j}$ and $g_{k_j}$ (not further considered here).

\subsection[Data structure]{Data structure: \class{epidataCS}} \label{sec:twinstim:data}

The first step toward fitting a \code{twinstim} is to turn the relevant
data into an object of the dedicated class \class{epidataCS}.\footnote{
  The suffix ``CS'' indicates that the data-generating point process is indexed
  in continuous space.
}
The primary ingredients of this class are a spatio-temporal point pattern
(\code{events}) and its underlying observation region (\code{W}).
An additional spatio-temporal grid (\code{stgrid}) holds (time-varying)
areal-level covariates for the endemic regression part.
We exemplify this data class by the \class{epidataCS} object for the
636 cases of invasive meningococcal disease in Germany
originally analyzed by \citet{meyer.etal2011}.
It is already contained in the \pkg{surveillance} package as
\code{data("imdepi")} and has been constructed as follows:
\begin{Schunk}
\begin{Sinput}
R> imdepi <- as.epidataCS(events = events, W = stateD, stgrid = stgrid,
+    qmatrix = diag(2), nCircle2Poly = 16)
\end{Sinput}
\end{Schunk}
The function \code{as.epidataCS} checks the consistency of the
three data ingredients described in detail below.
It also pre-computes auxiliary variables for model fitting, e.g.,
the individual influence regions~\eqref{eqn:twinstim:IR},
which are intersections of the observation region with discs
approximated by polygons with \code{nCircle2Poly = 16} edges.
The intersections are computed using 
functionality of the package \pkg{polyclip} \citep{R:polyclip}.
For multitype epidemics as in our example, the additional indicator matrix
\code{qmatrix} specifies transmissibility across event types.
An identity matrix corresponds to an independent spread of the event
types, i.e., cases of one type can not produce cases of another type.

\subsubsection{Data ingredients}

The core \code{events} data must be provided in the form of a
\class{SpatialPointsDataFrame} as defined by the package \pkg{sp}
\citep{Bivand.etal2013}:
\begin{Schunk}
\begin{Sinput}
R> summary(events)
\end{Sinput}
\end{Schunk}
\begin{Schunk}
\begin{Soutput}
Object of class SpatialPointsDataFrame
Coordinates:
   min  max
x 4039 4665
y 2710 3525
Is projected: TRUE 
proj4string :
[+init=epsg:3035 +units=km +proj=laea +lat_0=52 +lon_0=10 +x_0=4321000 +y_0=3210000
+ellps=GRS80 +no_defs]
Number of points: 636
Data attributes:
     time          tile    type       eps.t       eps.s        sex          agegrp   
Min.   :   0  05354  : 34  B:336  Min.   :30  Min.   :200  female:292  [0,3)   :194  
1st Qu.: 539  05370  : 27  C:300  1st Qu.:30  1st Qu.:200  male  :339  [3,19)  :279  
Median :1155  11000  : 27         Median :30  Median :200  NA's  :  5  [19,Inf):162  
Mean   :1193  05358  : 13         Mean   :30  Mean   :200              NA's    :  1  
3rd Qu.:1808  05162  : 12         3rd Qu.:30  3rd Qu.:200                            
Max.   :2543  05382  : 12         Max.   :30  Max.   :200                            
              (Other):511                                                            
\end{Soutput}
\end{Schunk}
The associated event coordinates are residence postcode centroids, projected in
the \emph{European Terrestrial Reference System 1989} (in kilometer units)
to enable Euclidean geometry.
See the \code{spTransform}-methods in package \pkg{rgdal} \citep{R:rgdal}
for how to project latitude and longitude coordinates
into a planar coordinate reference system (CRS).
The data frame associated with these spatial coordinates ($\bm{s}_i$) contains a
number of required variables and additional event marks
(in the notation of Section~\ref{sec:twinstim:methods}:
$\{(t_i,[\bm{s}_i],k_i,\tau_i,\delta_i,\bm{m}_i): i = 1,\dotsc,n\}$).
For the IMD data, the event
\code{time} is measured in days since the beginning of the observation period
2002--2008 and is subject to a tie-breaking procedure (described later).
The \code{tile} column refers to the region of the spatio-temporal grid where
the event occurred and here contains the official key of the administrative
district of the patient's residence.
There are two \code{type}s of events labeled as \code{"B"} and \code{"C"}, which refer
to the serogroups of the two meningococcal finetypes \emph{B:P1.7-2,4:F1-5} and
\emph{C:P1.5,2:F3-3} contained in the data.
The \code{eps.t} and \code{eps.s} columns specify upper limits
for temporal and spatial interaction, respectively.
Here, the infectious period is assumed to last a maximum of 30 days and
spatial interaction is limited to a 200 km radius for all cases.
The latter has numerical advantages for a Gaussian interaction function $f$ with a
relatively small standard deviation. For a power-law kernel, however, this
restriction will be dropped to enable occasional long-range transmission.
The last two data attributes displayed in the above \code{event} summary are
covariates from the case reports: the gender and age group of the patient.

For the observation region \code{W}, we use a polygon representation
of Germany's boundary.
Since the observation region defines the integration domain
in the point process log-likelihood~\eqref{eqn:twinstim:loglik},
the more detailed the polygons of \code{W} are
the longer it will take to fit a \code{twinstim}.
It is thus advisable to sacrifice some shape details for speed by reducing
the polygon complexity, e.g., by applying one of the simplification methods
available at \href{www.MapShaper.org}{MapShaper.org}
\citep{harrower.bloch2006}.
Alternative tools in \proglang{R} are
\pkg{spatstat}'s \code{simplify.owin} procedure \citep{R:spatstat} and the function
\code{thinnedSpatialPoly} in package \pkg{maptools} \citep{R:maptools}, which
implements the \citet{douglas.peucker1973} reduction method.
The \pkg{surveillance} package already contains a simplified representation of
Germany's boundaries:
\begin{Schunk}
\begin{Sinput}
R> load(system.file("shapes", "districtsD.RData", package = "surveillance"))
\end{Sinput}
\end{Schunk}
This file contains both the \class{SpatialPolygonsDataFrame} \code{districtsD}
of Germany's 413 administrative districts as at
January 1, 2009,
as well as their union \code{stateD}.
These boundaries are projected in the same CRS as the \code{events} data.

The \code{stgrid} input specific to the endemic model component
is a simple data frame with (time-dependent) areal-level covariates,
e.g., socio-economic or ecological characteristics.
For our IMD example, we have:
\begin{Schunk}
\begin{Soutput}
      start stop  tile   area popdensity
1         0   31 01001   56.4     1557.1
2         0   31 01002  118.7     1996.6
3         0   31 01003  214.2      987.6
...     ...  ...   ...    ...        ...
34690  2526 2557 16075 1148.5       79.2
34691  2526 2557 16076  843.5      133.6
34692  2526 2557 16077  569.1      181.5
\end{Soutput}
\end{Schunk}
Numeric (\code{start},\code{stop}] columns index the time periods and
the factor variable \code{tile} identifies the regions of the grid.
Note that the given time intervals (here: months) also define the resolution of
possible time trends and seasonality of the piecewise constant endemic intensity.
We choose monthly intervals to reduce package size and computational cost
compared to the weekly resolution originally used by \citet{meyer.etal2011}
and \citet{meyer.held2013}.
The above \code{stgrid} data frame thus consists of 7 (years) times 12 (months)
blocks of 413 (districts) rows each.
The \code{area} column gives the area of the respective \code{tile} in square
kilometers (compatible with the CRS used for \code{events} and \code{W}).
A geographic representation of the regions in \code{stgrid} is not required for
model estimation, and is thus not part of the \class{epidataCS} class.
In our example, the areal-level data only consists of the population density
\code{popdensity}, whereas \citet{meyer.etal2011} additionally incorporated
(lagged) weekly influenza counts by district as a time-dependent covariate.

\subsubsection{Data handling and visualization}

The generated \class{epidataCS} object \code{imdepi} is a simple list of the
checked ingredients
\code{events}, \code{stgrid}, \code{W} and \code{qmatrix}.
Several methods for data handling and visualization are available for
such objects as listed in Table~\ref{tab:methods:epidataCS}
and briefly presented in the remainder of this section.

Printing an \class{epidataCS} object presents some
metadata and the first 6
events by default:
\begin{Schunk}
\begin{Sinput}
R> imdepi
\end{Sinput}
\begin{Soutput}
Observation period: 0 - 2557 
Observation window (bounding box): [4031, 4672] x [2684, 3550] 
Spatio-temporal grid (not shown): 84 time blocks x 413 tiles 
Types of events: "B" "C"
Overall number of events: 636 

   coordinates   time  tile type eps.t eps.s    sex   agegrp BLOCK start popdensity
1 (4110, 3200)  0.212 05554    B    30   200   male   [3,19)     1     0        261
2 (4120, 3080)  0.712 05382    C    30   200   male   [3,19)     1     0        519
3 (4410, 2920)  5.591 09574    B    30   200 female [19,Inf)     1     0        209
4 (4200, 2880)  7.117 08212    B    30   200 female   [3,19)     1     0       1666
5 (4130, 3220) 22.060 05554    C    30   200   male   [3,19)     1     0        261
6 (4090, 3180) 24.954 05170    C    30   200   male   [3,19)     1     0        455
[....]
\end{Soutput}
\end{Schunk}
During conversion to \class{epidataCS}, the last three columns \code{BLOCK}
(time interval index), \code{start} and \code{popdensity} have been merged from
the checked \code{stgrid} to the \code{events} data frame.
The event marks including time and location can be extracted in a standard
data frame by \code{marks(imdepi)}, 
and this is summarized by \code{summary(imdepi)}.

\begin{table}[ht]
\centering
{\small
\begin{tabular}{lllll}
  \toprule
Display & Subset & Extract & Modify & Convert \\ 
  \midrule
\code{print} & \code{[} & \code{nobs} & \code{update} & \code{as.epidata} \\ 
  \code{summary} & \code{head} & \code{marks} & \code{untie} & \code{\textit{epidataCS2sts}} \\ 
  \code{plot} & \code{tail} &  &  &  \\ 
  \code{animate} & \code{subset} &  &  &  \\ 
  \code{as.stepfun} &  &  &  &  \\ 
   \bottomrule
\end{tabular}
}
\caption{Generic and \textit{non-generic} functions applicable to \class{epidataCS} objects.} 
\label{tab:methods:epidataCS}
\end{table}

A simple plot of the number of infectives as a function of time
(Figure~\ref{fig:imdepi_stepfun})
can be obtained by the step function converter:
\begin{Schunk}
\begin{Sinput}
R> plot(as.stepfun(imdepi), xlim = summary(imdepi)$timeRange, xaxs = "i",
+    xlab = "Time [days]", ylab = "Current number of infectives", main = "")
\end{Sinput}
\begin{figure}[hb]

{\centering \includegraphics[width=0.8\linewidth]{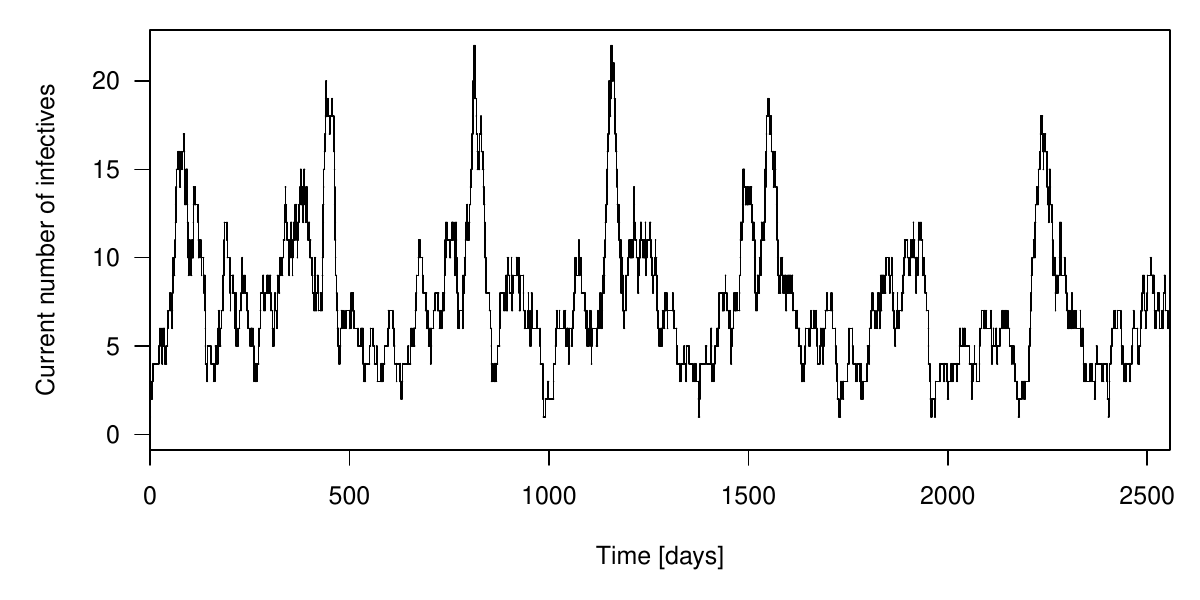} 

}

\caption{Time course of the number of infectives assuming infectious periods of 30 days.}\label{fig:imdepi_stepfun}
\end{figure}
\end{Schunk}

The \code{plot}-method for \class{epidataCS} offers aggregation of the events over time or space:
\begin{Schunk}
\begin{Sinput}
R> plot(imdepi, "time", col = c("indianred", "darkblue"), ylim = c(0, 20))
R> plot(imdepi, "space", lwd = 2,
+    points.args = list(pch = c(1, 19), col = c("indianred", "darkblue")))
R> layout.scalebar(imdepi$W, scale = 100, labels = c("0", "100 km"), plot = TRUE)
\end{Sinput}
\begin{figure}

{\centering \subfloat[Temporal pattern.\label{fig:imdepi_plot1}]{\includegraphics[width=0.49\linewidth]{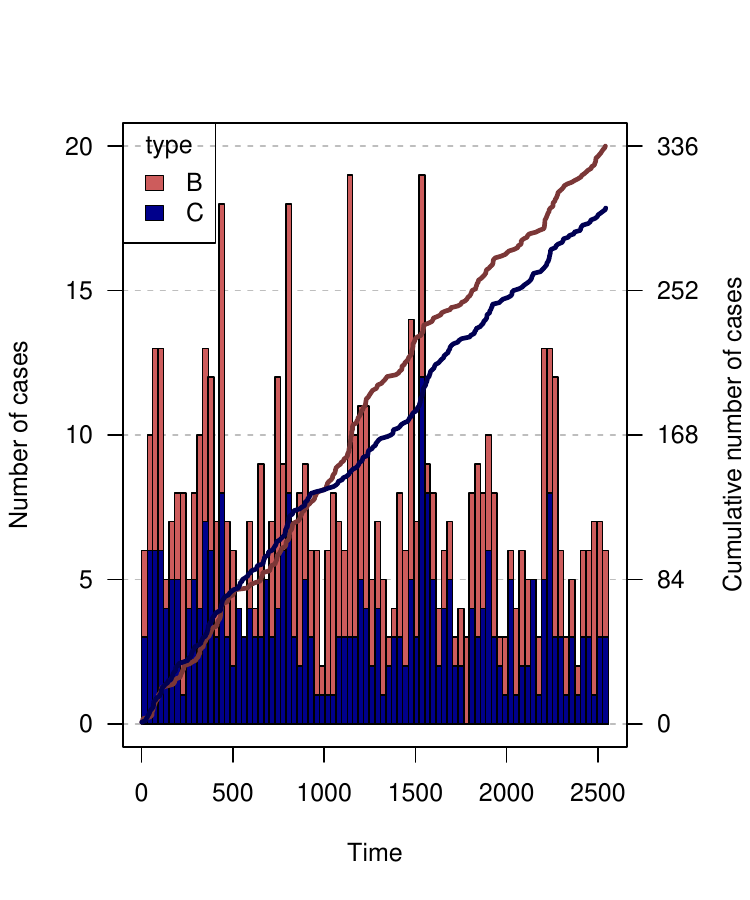} }\subfloat[Spatial pattern.\label{fig:imdepi_plot2}]{\includegraphics[width=0.49\linewidth]{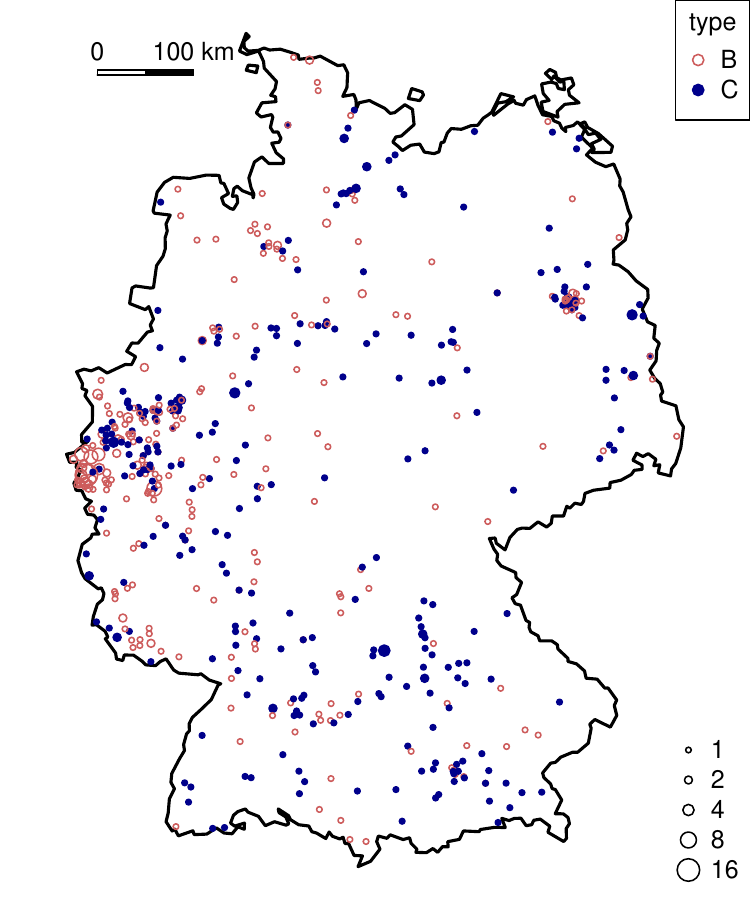} }

}

\caption{Occurrence of the two finetypes viewed in the temporal and spatial dimensions.}\label{fig:imdepi_plot}
\end{figure}
\end{Schunk}
The time-series plot (Figure~\ref{fig:imdepi_plot1}) shows the monthly aggregated number of cases by finetype in
a stacked histogram as well as each type's cumulative number over time.
The spatial plot (Figure~\ref{fig:imdepi_plot2}) shows the observation window \code{W} with the locations of all
cases (by type), where the areas of the points are proportional to the number of
cases at the respective location.
Additional shading by the population is possible and exemplified in
\code{help("plot.epidataCS")}.

The above static plots do not capture the space-time dynamics of epidemic
spread. An animation may provide additional insight and can be produced by the
corresponding \code{animate}-method. For instance, to look at the first year of
the B-type in a weekly sequence of snapshots in a web browser (using facilities of the
\pkg{animation} package of \citealp{R:animation}):
\begin{Schunk}
\begin{Sinput}
R> animation::saveHTML(
+    animate(subset(imdepi, type == "B"), interval = c(0, 365), time.spacing = 7),
+    nmax = Inf, interval = 0.2, loop = FALSE,
+    title = "Animation of the first year of type B events")
\end{Sinput}
\end{Schunk}
Selecting events from \class{epidataCS} as for the animation above is enabled
by the \code{[}- and \code{subset}-methods, which return a new
\class{epidataCS} object containing only the selected \code{events}.

A limited data sampling resolution may lead to tied event times or locations,
which are in conflict with a continuous spatio-temporal point process model.
For instance, a temporal residual analysis
would suggest model deficiencies \citep[Figure 4]{meyer.etal2011},
and a power-law kernel for spatial interaction may diverge if
there are events with zero distance to potential source events
\citep{meyer.held2013}. The function \code{untie} breaks ties by
random shifts. This has already been applied to the event \emph{times} in
the provided \code{imdepi} data by subtracting a U(0,1)-distributed random
number from the original dates.
The event \emph{coordinates} in the IMD data are subject to interval censoring
at the level of Germany's postcode regions. A possible replacement for the given
centroids would thus be a random location within the corresponding postcode area.
Lacking a suitable shapefile,
\citet{meyer.held2013} shifted all locations by a random
vector with length up to half the observed minimum spatial separation:
\begin{Schunk}
\begin{Sinput}
R> eventDists <- dist(coordinates(imdepi$events))
R> (minsep <- min(eventDists[eventDists > 0]))
\end{Sinput}
\begin{Soutput}
[1] 1.17
\end{Soutput}
\begin{Sinput}
R> set.seed(321)
R> imdepi_untied <- untie(imdepi, amount = list(s = minsep / 2))
\end{Sinput}
\end{Schunk}
Note that random tie-breaking requires sensitivity analyses as
discussed by \citet{meyer.held2013}, but skipped here for the sake of brevity.

The \code{update}-method is useful to change the values of the maximum
interaction ranges \code{eps.t} and \code{eps.s}, since it takes care of the
necessary updates of the hidden auxiliary variables in an \class{epidataCS}
object. For an unbounded interaction radius:
\begin{Schunk}
\begin{Sinput}
R> imdepi_untied_infeps <- update(imdepi_untied, eps.s = Inf)
\end{Sinput}
\end{Schunk}

Last but not least, \class{epidataCS} can be converted to the other
classes \class{epidata} (Section~\ref{sec:twinSIR})
and \class{sts} (Section~\ref{sec:hhh4}) by aggregation.
The method \code{as.epidata.epidataCS} aggregates events by region (\code{tile}),
and the function \code{epidataCS2sts} yields counts by region and time interval.
The data could then, e.g., be analyzed by the
multivariate time-series model presented in Section~\ref{sec:hhh4}. We can
also use visualization tools of the \class{sts} class, e.g., to produce
Figure~\ref{fig:imdsts_plot}:
\begin{Schunk}
\begin{Sinput}
R> imdsts <- epidataCS2sts(imdepi, freq = 12, start = c(2002, 1), tiles = districtsD)
R> plot(imdsts, type = observed ~ time)
R> plot(imdsts, type = observed ~ unit, population = districtsD$POPULATION / 100000)
\end{Sinput}
\begin{figure}[bh]

{\centering \subfloat[Time series of monthly counts.\label{fig:imdsts_plot1}]{\includegraphics[width=0.49\linewidth]{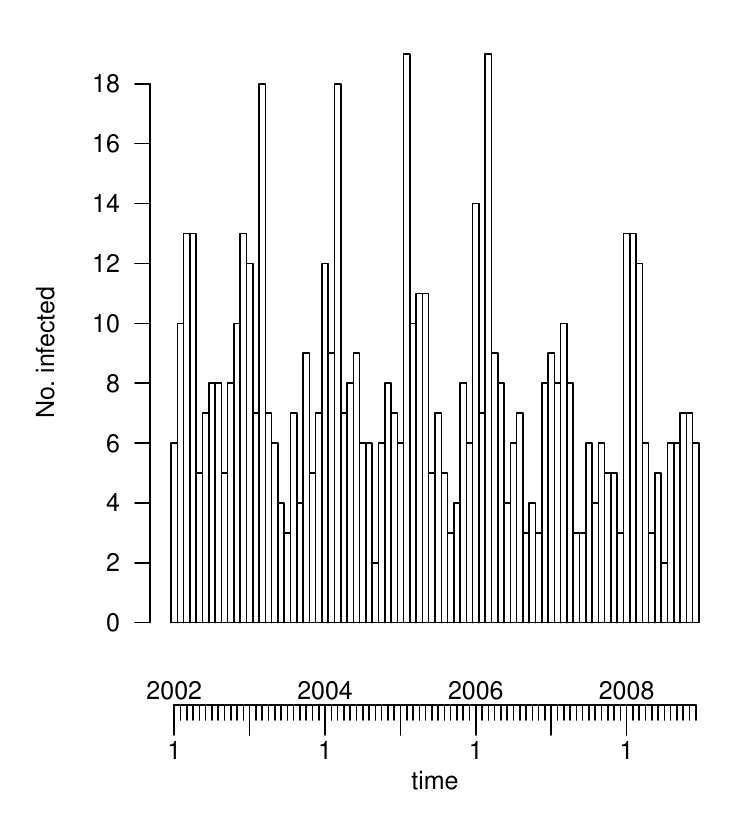} }\subfloat[Disease incidence (per 100\,000 inhabitants).\label{fig:imdsts_plot2}]{\includegraphics[width=0.49\linewidth]{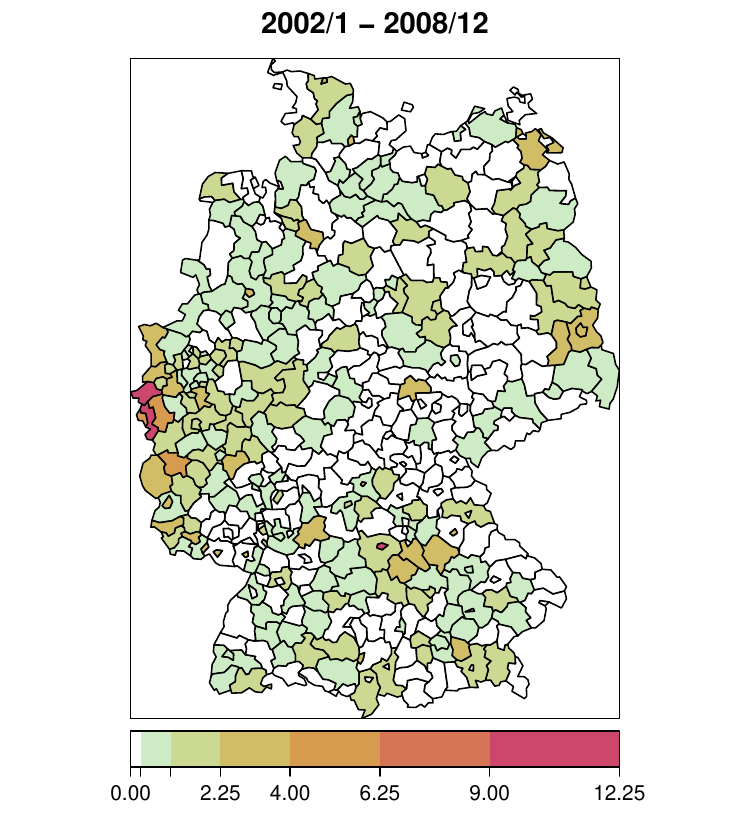} }

}

\caption{IMD cases (joint types) aggregated as an \class{sts} object by month and district.}\label{fig:imdsts_plot}
\end{figure}
\end{Schunk}

\subsection{Modeling and inference} \label{sec:twinstim:fit}

Having prepared the data as an object of class \class{epidataCS},
the function \code{twinstim} can be used to perform likelihood inference
for conditional intensity models of the form~\eqref{eqn:twinstim}.
The main arguments for \code{twinstim}
are the formulae of the \code{endemic} and \code{epidemic} linear predictors
($\nu_{[\bm{s}][t]} = \exp$(\code{endemic}) and $\eta_j = \exp$(\code{epidemic})), and
the spatial and temporal interaction functions \code{siaf} ($f$) and \code{tiaf} ($g$), respectively.
Both formulae are parsed internally using the standard \code{model.frame}
toolbox from package \pkg{stats} and thus can handle factor variables and
interaction terms.
While the \code{endemic} linear predictor incorporates time-dependent and/or
areal-level covariates from \code{stgrid}, 
the \code{epidemic} formula may use both \code{stgrid} variables and event marks
to be associated with the force of infection. 
For the interaction functions, several alternatives are predefined as listed in
Table~\ref{tab:iafs}. They are applicable out-of-the-box and
illustrated as part of the following modeling exercise for the IMD data.
Own interaction functions can also be used provided their implementation obeys a
certain structure, see \code{help("siaf")} and \code{help("tiaf")}, respectively.

\begin{table}[ht]
\centering
{\small
\begin{tabular}{ll}
  \toprule
Spatial (\code{siaf.*}) & Temporal (\code{tiaf.*}) \\ 
  \midrule
\code{constant} & \code{constant} \\ 
  \code{gaussian} & \code{exponential} \\ 
  \code{powerlaw} & \code{step} \\ 
  \code{powerlawL} & \code{} \\ 
  \code{step} & \code{} \\ 
  \code{student} & \code{} \\ 
   \bottomrule
\end{tabular}
}
\caption{Predefined spatial and temporal interaction functions.} 
\label{tab:iafs}
\end{table}

\subsubsection{Basic example}

To illustrate statistical inference with \code{twinstim},
we will estimate several models for the simplified and ``untied'' IMD data
presented in Section~\ref{sec:twinstim:data}.
In the endemic component, we include the district-specific population density as a multiplicative
offset, a (centered) time trend, and 
a sinusoidal wave of frequency $2\pi/365$ to capture seasonality,
where the \code{start} variable from \code{stgrid} measures time:
\begin{Schunk}
\begin{Sinput}
R> (endemic <- addSeason2formula(~offset(log(popdensity)) + I(start / 365 - 3.5),
+    period = 365, timevar = "start"))
\end{Sinput}
\begin{Soutput}
~offset(log(popdensity)) + I(start/365 - 3.5) + sin(2 * pi * 
    start/365) + cos(2 * pi * start/365)
\end{Soutput}
\end{Schunk}
See \citet[Section~2.2]{held.paul2012} for how such sine/cosine terms
reflect seasonality.
Because of the aforementioned integrations in the log-likelihood~\eqref{eqn:twinstim:loglik}, it
is advisable to first fit an endemic-only model to obtain reasonable
start values for more complex epidemic models:
\begin{Schunk}
\begin{Sinput}
R> imdfit_endemic <- twinstim(endemic = endemic, epidemic = ~0,
+    data = imdepi_untied, subset = !is.na(agegrp))
\end{Sinput}
\end{Schunk}
We exclude the single case with unknown age group from this analysis since we
will later estimate an effect of the age group on the force of infection.

\begin{table}[ht]
\centering
{\small
\begin{tabular}{llll}
  \toprule
Display & Extract & Modify & Other \\ 
  \midrule
\code{print} & \code{nobs} & \code{update} & \code{simulate} \\ 
  \code{summary} & \code{vcov} & \code{add1} & \code{\textit{epitest}} \\ 
  \code{xtable} & \code{coeflist} & \code{drop1} &  \\ 
  \code{plot} & \code{logLik} & \code{\textit{stepComponent}} &  \\ 
  \code{intensityplot} & \code{extractAIC} &  &  \\ 
  \code{\textit{iafplot}} & \code{profile} &  &  \\ 
  \code{\textit{checkResidualProcess}} & \code{residuals} &  &  \\ 
   & \code{terms} &  &  \\ 
   & \code{R0} &  &  \\ 
   \bottomrule
\end{tabular}
}
\caption{Generic and \textit{non-generic} functions applicable to \class{twinstim} objects. Note that there is no need for specific \code{coef}, \code{confint}, \code{AIC} or \code{BIC} methods, since the respective default methods from package \pkg{stats} apply outright.} 
\label{tab:methods:twinstim}
\end{table}

Many of the standard functions to access model fits in \proglang{R} are also
implemented for \class{twinstim} fits (see Table~\ref{tab:methods:twinstim}).
For example, we can produce the usual model summary:
\begin{Schunk}
\begin{Sinput}
R> summary(imdfit_endemic)
\end{Sinput}
\begin{Soutput}
Call:
twinstim(endemic = endemic, epidemic = ~0, data = imdepi_untied, 
    subset = !is.na(agegrp))

Coefficients of the endemic component:
                          Estimate Std. Error z value Pr(>|z|)    
h.(Intercept)             -20.3683     0.0419 -486.24  < 2e-16 ***
h.I(start/365 - 3.5)       -0.0444     0.0200   -2.22    0.027 *  
h.sin(2 * pi * start/365)   0.2733     0.0576    4.75  2.0e-06 ***
h.cos(2 * pi * start/365)   0.3509     0.0581    6.04  1.5e-09 ***
---
Signif. codes:  0 '***' 0.001 '**' 0.01 '*' 0.05 '.' 0.1 ' ' 1

No epidemic component.

AIC:  19166
Log-likelihood: -9579
\end{Soutput}
\end{Schunk}

Because of the aforementioned equivalence of the endemic component with a
Poisson regression model, the coefficients can be interpreted as
log rate ratios in the usual way. For instance, the endemic rate is estimated to
decrease by \code{1 - exp(coef(imdfit_endemic)[2])} $=$
4.3\% per year.
Coefficient correlations can be retrieved by the argument
\code{correlation = TRUE} in the \code{summary} call just like for
\code{summary.glm}, but may also be extracted via the standard
\code{cov2cor(vcov(imdfit_endemic))}.

We now update the endemic model to take additional spatio-temporal dependence between
events into account. Infectivity shall depend on the meningococcal finetype and
the age group of the patient, and is assumed to be constant over time (default),
$g(t)=\ind_{(0,30]}(t)$, with a Gaussian distance-decay
$f(x) = \exp\left\{-x^2/(2 \sigma^2)\right\}$.
This model was originally selected by \citet{meyer.etal2011} and can be fitted as follows:
\begin{Schunk}
\begin{Sinput}
R> imdfit_Gaussian <- update(imdfit_endemic, epidemic = ~type + agegrp,
+    siaf = siaf.gaussian(), start = c("e.(Intercept)" = -12.5, "e.siaf.1" = 2.75),
+    control.siaf = list(F = list(adapt = 0.25), Deriv = list(nGQ = 13)),
+    cores = 2 * (.Platform$OS.type == "unix"), model = TRUE)
\end{Sinput}
\end{Schunk}
To reduce the runtime of this example, we specified convenient \code{start} values
for some parameters (others start at 0) and set
\code{control.siaf} with a rather low number of nodes for the cubature
of $f(\norm{\bm{s}})$ in the log-likelihood (via the midpoint rule) and
$\frac{\partial f(\norm{\bm{s}})}{\partial \log\sigma}$
in the score function (via product Gauss cubature).
On Unix-alikes, these numerical integrations can be performed in parallel 
using the ``multicore'' functions
\code{mclapply} \textit{et al.}\ from the base package \pkg{parallel},
here with \code{cores = 2} processes.
For later generation of an \code{intensityplot}, the \code{model}
environment is retained.

\begin{table}[ht]
\centering
{\small
\begin{tabular}{lrrr}
  \toprule
 & RR & 95\% CI & p-value \\ 
  \midrule
\code{h.I(start/365 - 3.5)} & 0.955 & 0.91--1.00 & 0.039 \\ 
  \code{h.sin(2 * pi * start/365)} & 1.243 & 1.09--1.41 & 0.0008 \\ 
  \code{h.cos(2 * pi * start/365)} & 1.375 & 1.21--1.56 & $<$0.0001 \\ 
  \code{e.typeC} & 0.402 & 0.24--0.68 & 0.0007 \\ 
  \code{e.agegrp[3,19)} & 2.000 & 1.06--3.78 & 0.033 \\ 
  \code{e.agegrp[19,Inf)} & 0.776 & 0.32--1.91 & 0.58 \\ 
   \bottomrule
\end{tabular}
}
\caption{Estimated rate ratios (RR) and associated Wald confidence intervals (CI) for endemic (\code{h.}) and epidemic (\code{e.}) terms. This table was generated by \code{xtable(imdfit\_Gaussian)}.} 
\label{tab:imdfit_Gaussian}
\end{table}

Table~\ref{tab:imdfit_Gaussian} shows the output of \code{twinstim}'s
\code{xtable} method \citep{R:xtable},
which provides rate ratios for the endemic and epidemic effects. 
The alternative \code{toLatex} method simply translates the
\code{summary} table of coefficients to \LaTeX\ without \code{exp}-transformation.
On the subject-matter level, we can conclude from 
Table~\ref{tab:imdfit_Gaussian} that the meningococcal finetype of
serogroup~C is less than half as infectious as the B-type, and that patients in
the age group 3 to 18 years are estimated to cause twice as many secondary infections as infants
aged 0 to 2 years.

\subsubsection{Model-based effective reproduction numbers}

The event-specific reproduction numbers~\eqref{eqn:R0:twinstim} can be extracted
from fitted \class{twinstim} objects via the \code{R0} method.
For the above IMD model, we obtain the following mean numbers of
secondary infections by finetype:
\begin{Schunk}
\begin{Sinput}
R> R0_events <- R0(imdfit_Gaussian)
R> tapply(R0_events, marks(imdepi_untied)[names(R0_events), "type"], mean)
\end{Sinput}
\begin{Soutput}
     B      C 
0.2161 0.0958 
\end{Soutput}
\end{Schunk}
Confidence intervals 
can be obtained via Monte Carlo simulation,
where Equation~\ref{eqn:R0:twinstim} is repeatedly evaluated
with parameters sampled from the asymptotic multivariate normal distribution of the
maximum likelihood estimate. For this purpose, the \code{R0}-method
takes an argument \code{newcoef}, which is exemplified in \code{help("R0")}.

\subsubsection{Interaction functions}

Figure~\ref{fig:imdfit_siafs} shows several estimated spatial interaction
functions, which can be plotted by, e.g.,
\code{plot(imdfit_Gaussian, which = "siaf")}.
\citet{meyer.held2013} found that a power-law decay of spatial interaction is
more appropriate than a Gaussian kernel to describe the spread of human infectious
diseases.
The power-law kernel concentrates on short-range interaction, but also exhibits a
heavier tail reflecting occasional transmission over large distances.
To use the power-law kernel $f(x) = (x+\sigma)^{-d}$, we switch to the
prepared \class{epidataCS} object with \code{eps.s = Inf} and update the
previous Gaussian model as follows:
\begin{Schunk}
\begin{Sinput}
R> imdfit_powerlaw <- update(imdfit_Gaussian, data = imdepi_untied_infeps,
+    siaf = siaf.powerlaw(), control.siaf = NULL,
+    start = c("e.(Intercept)" = -6.2, "e.siaf.1" = 1.5, "e.siaf.2" = 0.9))
\end{Sinput}
\end{Schunk}

Table~\ref{tab:iafs} also lists the step function kernel as an alternative, which is
particularly useful for two reasons. First, it is a more flexible approach since it
estimates interaction between the given knots without assuming an overall functional form.
Second, the spatial integrals in the log-likelihood can be computed
analytically for the step function kernel, which therefore offers a quick estimate
of spatial interaction. We update the Gaussian model to use four
steps at log-equidistant knots up to an interaction range of 100 km:
\begin{Schunk}
\begin{Sinput}
R> imdfit_step4 <- update(imdfit_Gaussian, data = imdepi_untied_infeps,
+    siaf = siaf.step(exp(1:4 * log(100) / 5), maxRange = 100), control.siaf = NULL,
+    start = c("e.(Intercept)" = -10, setNames(-2:-5, paste0("e.siaf.", 1:4))))
\end{Sinput}
\end{Schunk}
Figure~\ref{fig:imdfit_siafs} suggests that the estimated step function is in
line with the power law.

For the temporal interaction function $g(t)$, model updates and plots are
similarly possible, e.g.,
\code{update(imdfit_Gaussian, tiaf = tiaf.exponential())}. However, the events
in the IMD data are too rare to infer the time-course of infectivity with
confidence. 


\begin{Schunk}
\begin{figure}[hbt]

{\centering \includegraphics[width=0.8\linewidth]{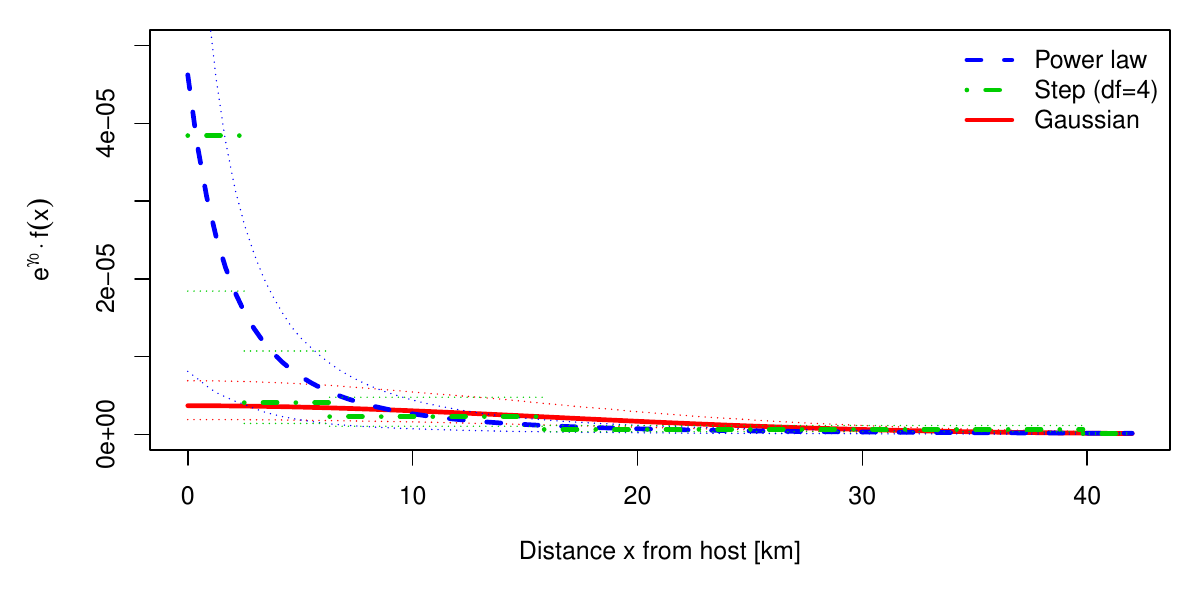} 

}

\caption{Various estimates of spatial interaction (scaled by the epidemic intercept $\gamma_0$). The standard deviation of the Gaussian kernel is estimated to be $\hat\sigma=$ 16.00 (95\% CI: 13.65--18.75), and the estimated power-law parameters are $\hat\sigma=$ 4.64 (95\% CI: 1.82--11.84) and $\hat{d}=$ 2.49 (95\% CI: 1.81--3.42).}\label{fig:imdfit_siafs}
\end{figure}
\end{Schunk}

\subsubsection{Model selection}

\begin{Schunk}
\begin{Sinput}
R> AIC(imdfit_endemic, imdfit_Gaussian, imdfit_powerlaw, imdfit_step4)
\end{Sinput}
\begin{Soutput}
                df   AIC
imdfit_endemic   4 19166
imdfit_Gaussian  9 18967
imdfit_powerlaw 10 18940
imdfit_step4    12 18933
\end{Soutput}
\end{Schunk}

Akaike's Information Criterion (AIC)
suggests superiority of the power-law vs.\ the Gaussian model and the
endemic-only model. The more flexible step function yields the best AIC value
but its shape strongly depends on the chosen knots and is not guaranteed to be
monotonically decreasing.
The function \code{stepComponent} -- a wrapper around the \code{step} function
from \pkg{stats} -- can be used to perform AIC-based stepwise
selection within a given model component.

\subsubsection{Model diagnostics}

Two other plots are implemented for \class{twinstim} objects.
Figure~\ref{fig:imdfit_powerlaw_intensityplot_time} shows
an \code{intensityplot} of
the fitted ``ground'' intensity
$\sum_{k=1}^2 \int_{\bm{W}} \hat\lambda(\bm{s},t,k) \dif \bm{s}$
aggregated over both event types:
\begin{Schunk}
\begin{Sinput}
R> intensityplot(imdfit_powerlaw, which = "total", aggregate = "time", types = 1:2)
\end{Sinput}
\end{Schunk}
\begin{Schunk}
\begin{figure}[bth]

{\centering \includegraphics[width=0.8\linewidth]{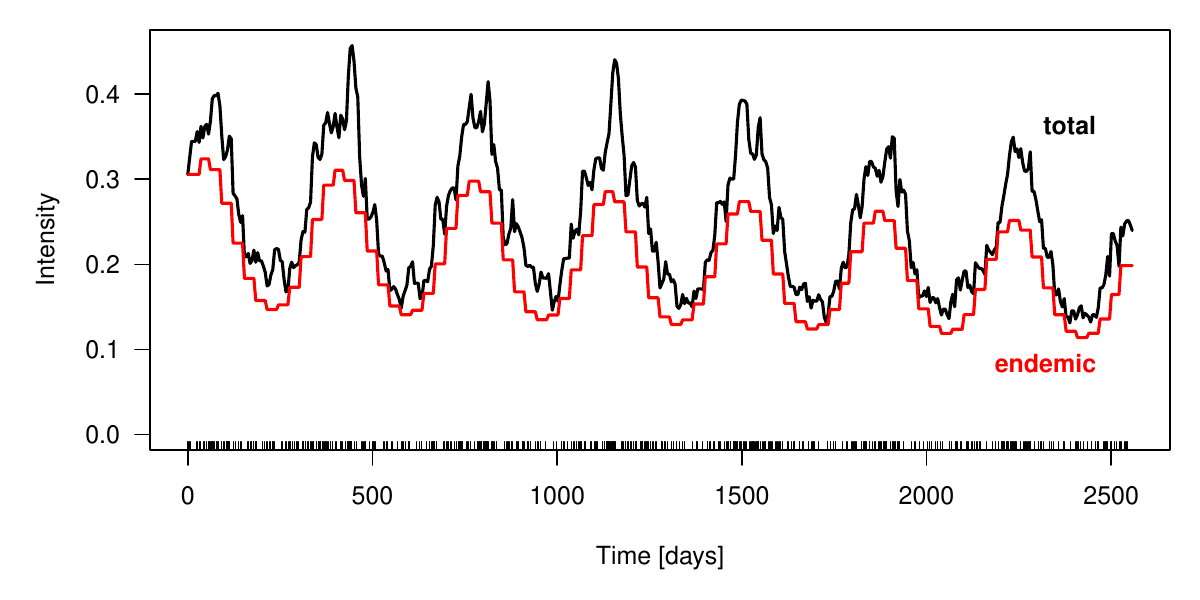} 

}

\caption{Fitted ``ground'' intensity process aggregated over space and both types.}\label{fig:imdfit_powerlaw_intensityplot_time}
\end{figure}
\end{Schunk}

The estimated endemic intensity component has also been added to the plot.
It exhibits strong seasonality and a slow negative trend. The proportion
of the endemic intensity is rather constant along time since
no major outbreaks occurred.
This proportion can be visualized separately by specifying
\code{which = "endemic proportion"} in the above call.

Spatial \code{intensityplot}s can be produced via
\code{aggregate = "space"} and require a geographic representation of \code{stgrid}.
Figure~\ref{fig:imdfit_powerlaw_intensityplot_space} shows the
accummulated epidemic proportion by event type.
It is naturally high in regions with a
large number of cases and even more so if the population density is low.
The function \code{epitest} offers a model-based global test for epidemicity,
while \code{knox} and \code{stKtest} implement related classical approaches
\citep{meyer.etal2015}.

\begin{Schunk}
\begin{figure}

{\centering \subfloat[Type B.\label{fig:imdfit_powerlaw_intensityplot_space1}]{\includegraphics[width=0.49\linewidth]{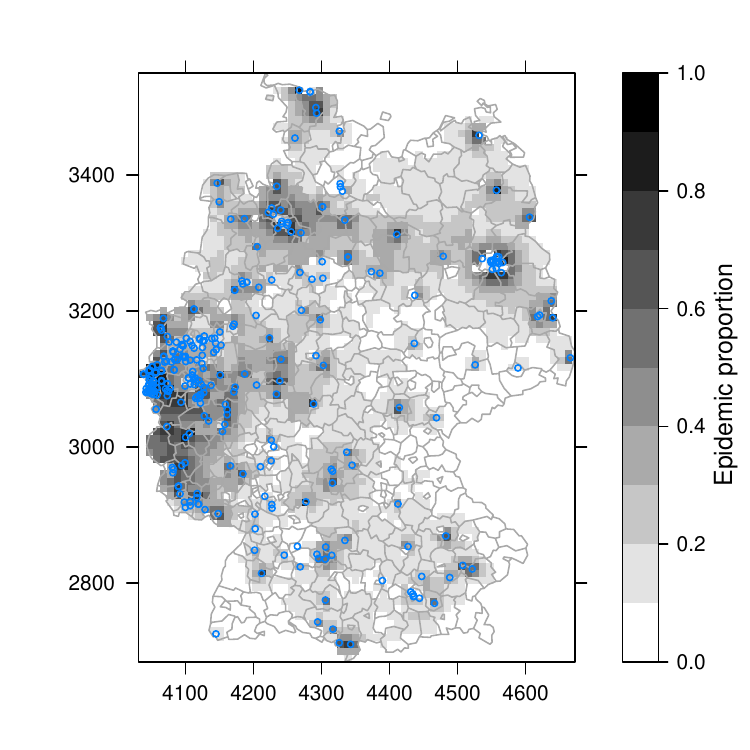} }\subfloat[Type C.\label{fig:imdfit_powerlaw_intensityplot_space2}]{\includegraphics[width=0.49\linewidth]{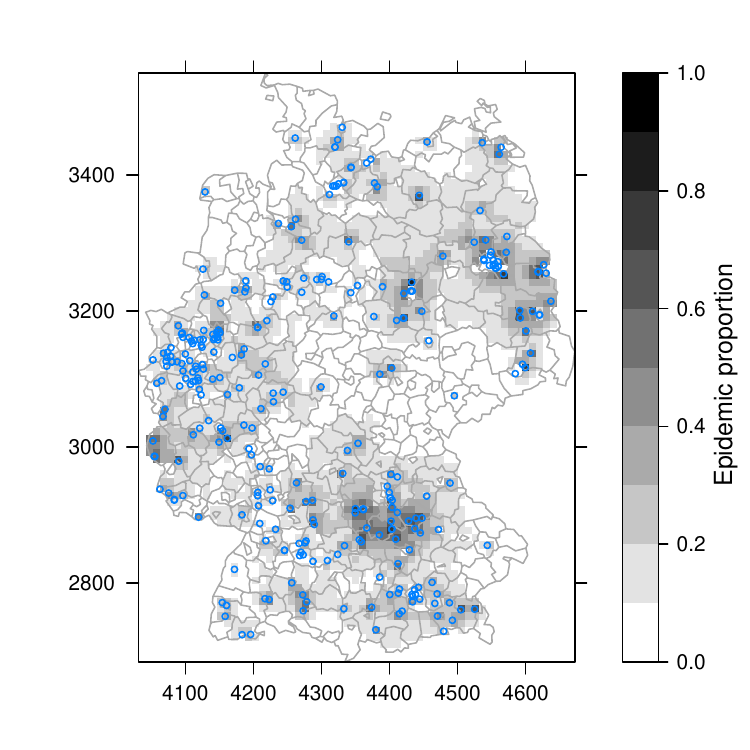} }

}

\caption{Epidemic proportion of the fitted intensity process accumulated over time by type.}\label{fig:imdfit_powerlaw_intensityplot_space}
\end{figure}
\end{Schunk}

Another diagnostic tool is the function \code{checkResidualProcess}, which transforms the temporal ``residual
process'' in such a way that it exhibits a uniform distribution and lacks serial
correlation if the fitted model describes the true CIF well
\citep[see][Section~3.3]{ogata1988}. 
These properties can be checked graphically as in
Figure~\ref{fig:imdfit_checkResidualProcess} produced by:
\begin{Schunk}
\begin{Sinput}
R> checkResidualProcess(imdfit_powerlaw)
\end{Sinput}
\begin{figure}[!h]

{\centering \includegraphics[width=0.9\linewidth]{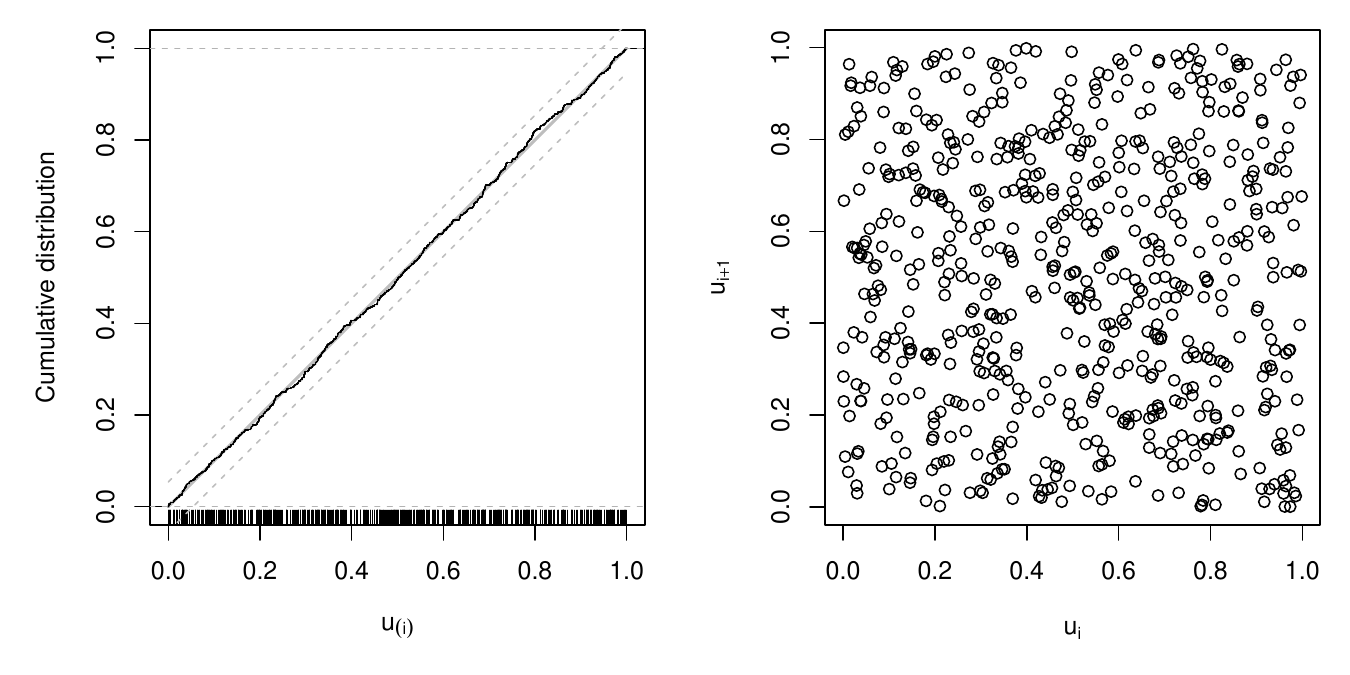} 

}

\caption{The left plot shows the \code{ecdf} of the transformed residuals with a 95\% confidence band obtained by inverting the corresponding Kolmogorov-Smirnov test (no evidence for deviation from uniformity). The right-hand plot suggests absence of serial correlation.}\label{fig:imdfit_checkResidualProcess}
\end{figure}
\end{Schunk}

\subsection{Simulation} \label{sec:twinstim:simulation}

To identify regions with unexpected IMD dynamics,
\citet{meyer.etal2011} compared the observed numbers of cases by district to the
respective 2.5\% and 97.5\% quantiles of 100 simulations from the selected model.
Furthermore, simulations allow us to investigate the stochastic volatility of the
endemic-epidemic process, to obtain probabilistic forecasts, and 
to perform parametric bootstrap of the spatio-temporal point pattern.

The simulation algorithm we apply is described in \citet[Section 4]{meyer.etal2011}.
It requires a geographic representation of the \code{stgrid},
as well as functionality for sampling locations from the spatial kernel
$f_2(\bm{s}) := f(\norm{\bm{s}})$. This is implemented for all predefined
spatial interaction functions listed in Table~\ref{tab:iafs}. 
Event marks are by default sampled from their respective empirical
distribution in the original data.
The following code runs 30 simulations over the last two years
based on the estimated power-law model:
\begin{Schunk}
\begin{Sinput}
R> imdsims <- simulate(imdfit_powerlaw, nsim = 30, seed = 1, t0 = 1826, T = 2555,
+    data = imdepi_untied_infeps, tiles = districtsD)
\end{Sinput}
\end{Schunk}
Figure~\ref{fig:imdsims_plot} shows the cumulative number of cases from the
simulations appended to the first five years of data.
Extracting a single simulation (e.g., \code{imdsims[[1]]}) yields
an object of the class \class{simEpidataCS}, which extends
\class{epidataCS}. It carries additional components from the
generating model to enable an \code{R0}-method and
\code{intensityplot}s for simulated data.
A special feature of such simulations is that the source of each event is
actually known:
\begin{Schunk}
\begin{Sinput}
R> table(imdsims[[1]]$events$source > 0, exclude = NULL)
\end{Sinput}
\begin{Soutput}
FALSE  TRUE  <NA> 
  112    25     8
\end{Soutput}
\end{Schunk}
The stored \code{source} value is 0 for endemic events, \code{NA} for events of
the prehistory but still infective at \code{t0},
and otherwise corresponds to the row index of the infective source.
Averaged over all 30 simulations, the proportion of events triggered by previous
events is 0.218.

\begin{Schunk}
\begin{figure}[!h]

{\centering \includegraphics[width=0.8\linewidth]{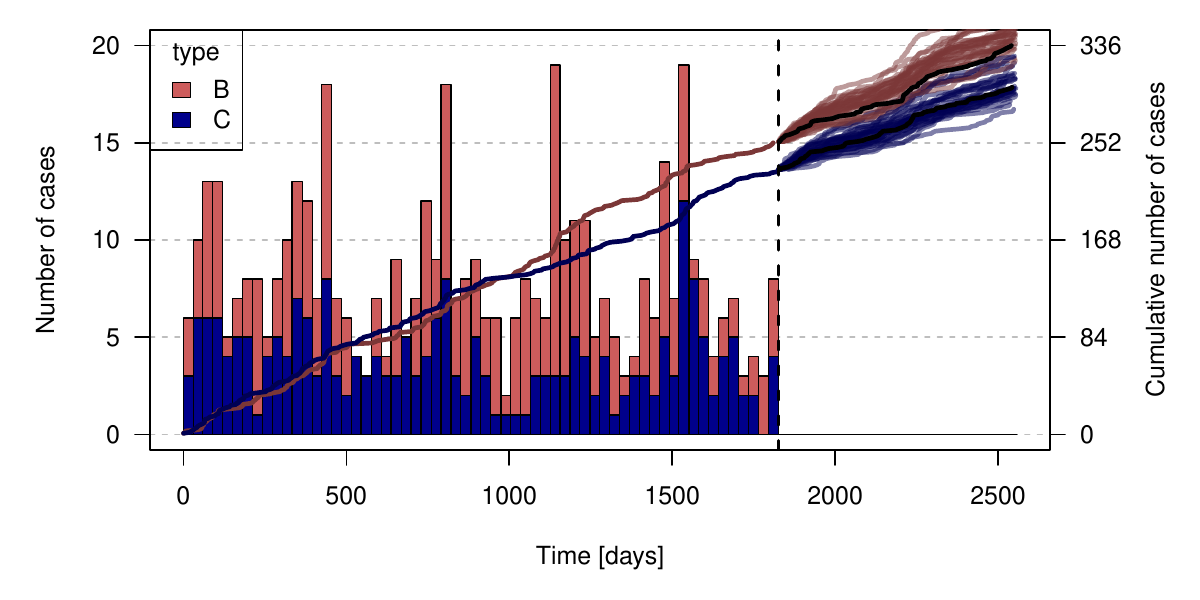} 

}

\caption{Simulation-based forecast of the cumulative number of cases by finetype in the last two years. The black lines correspond to the observed numbers.}\label{fig:imdsims_plot}
\end{figure}
\end{Schunk}

\section{SIR event history of a fixed population} \label{sec:twinSIR}

The endemic-epidemic multivariate point process model ``\code{twinSIR}'' is designed for
individual-level surveillance data of a fixed population of which
the complete SIR event history is assumed to be known.
As an illustrative example, we use a
particularly well-documented measles outbreak among children of the isolated
German village Hagelloch in the year 1861,
which has previously been analyzed by, e.g., \citet{neal.roberts2004}.
Other potential applications include farm-level data as well as epidemics across
networks.
We start by describing the general model class in Section~\ref{sec:twinSIR:methods}.
Section~\ref{sec:twinSIR:data} introduces the example data and the associated class \class{epidata}, and
Section~\ref{sec:twinSIR:fit} presents the core functionality of
fitting and analyzing such data using \code{twinSIR}.
Due to the many similarities with the \code{twinstim} framework covered in
Section~\ref{sec:twinstim}, we condense the \code{twinSIR} treatment accordingly.

\subsection[Model class]{Model class: \code{twinSIR}} \label{sec:twinSIR:methods}

The previously described point process model \code{twinstim} (Section~\ref{sec:twinstim})
is indexed in a continuous spatial domain, i.e., the set of
possible event locations 
consists of the whole observation region and is thus infinite.
However, if infections can only occur at a known discrete set of sites, such as for
livestock diseases among farms, 
the conditional intensity function
formally becomes $\lambda_i(t)$. It characterizes the
instantaneous rate of infection of individual $i$ at time $t$,
given the sets $S(t)$ and $I(t)$ of susceptible and infectious individuals,
respectively (just before time $t$).
In a similar regression view as in Section~\ref{sec:twinstim},
\citet{hoehle2009} proposed the endemic-epidemic
multivariate temporal point process ``\code{twinSIR}'':
\begin{equation} \label{eqn:twinSIR}
  \lambda_i(t) = \lambda_0(t) \, \nu_i(t) +
  \sum_{j \in I(t)} \left\{ f(d_{ij}) +
  \bm{w}_{ij}^\top \bm{\alpha}^{(w)} \right\} \:, 
\end{equation}
if $i \in S(t)$, i.e., if individual $i$ is currently susceptible,
and $\lambda_i(t) = 0$ otherwise.  The rate decomposes into two components. The
endemic component consists of a Cox proportional hazards formulation
containing a semi-parametric baseline hazard $\lambda_0(t)$ and a
log-linear predictor $\nu_i(t)=\exp\left( \bm{z}_i(t)^\top \bm{\beta}
\right)$ of covariates modeling infection from external
sources. Furthermore, an additive epidemic component captures
transmission from the set $I(t)$ of currently infectious individuals.
The force of infection of individual $i$ depends on the distance $d_{ij}$ to each
infective source $j \in I(t)$ through a distance kernel
\begin{equation} \label{eqn:twinSIR:f}
f(u) = \sum_{m=1}^M \alpha_m^{(f)} B_m(u) \: \geq 0 \:,
\end{equation}
which is represented by a linear combination of non-negative basis
functions $B_m$ with the $\alpha_m^{(f)}$'s being
the respective coefficients. 
For instance, $f$ could be modelled by a B-spline
\citep[Section~8.1]{Fahrmeir.etal2013},
and $d_{ij}$ could refer to the Euclidean distance $\norm{\bm{s}_i - \bm{s}_j}$
between the individuals' locations $\bm{s}_i$ and $\bm{s}_j$, or to the geodesic
distance between the nodes $i$ and $j$ in a network.
The distance-based force of infection
is modified additively by a linear predictor of covariates $\bm{w}_{ij}$ describing
the interaction of individuals $i$ and~$j$ further.
Hence, the whole epidemic component of Equation~\ref{eqn:twinSIR} can be written
as a single linear predictor $\bm{x}_i(t)^\top \bm{\alpha}$
by interchanging the summation order to
\begin{equation} \label{eqn:twinSIR:x}
  \sum_{m=1}^M \alpha^{(f)}_m \sum_{j \in I(t)} B_m(d_{ij}) +
  \sum_{k=1}^K \alpha^{(w)}_k \sum_{j \in I(t)} w_{ijk}
  = \bm{x}_i(t)^\top \bm{\alpha} \:,
\end{equation}
such that $\bm{x}_i(t)$ comprises all epidemic terms summed over $j\in
I(t)$. Note that the use of additive covariates $\bm{w}_{ij}$ on top
of the distance kernel in \eqref{eqn:twinSIR} is
different from \code{twinstim}'s multiplicative approach in
\eqref{eqn:twinstim}. One advantage of the additive approach is that
the subsequent linear decomposition of the distance kernel allows one
to gather all parts of the epidemic component in a single linear predictor.
Hence, the above model represents a CIF extension of what in
the context of survival analysis is known as an
additive-multiplicative hazard
model~\citep{martinussen_scheike2002}. As a consequence, the
\code{twinSIR} model could in principle be fitted with the \pkg{timereg}
package~\citep{R:timereg}, which yields estimates for the cumulative
hazards. However, \citet{hoehle2009} chooses a more direct inferential
approach: To ensure that the CIF $\lambda_i(t)$ is non-negative, all
covariates are encoded such that the components of $\bm{w}_{ij}$ are
non-negative. Additionally, the parameter vector $\bm{\alpha}$ is
constrained to be non-negative. Subsequent parameter inference is then
based on the resulting constrained penalized likelihood which gives
directly interpretable estimates of $\bm{\alpha}$.

\subsection[Data structure]{Data structure: \class{epidata}} \label{sec:twinSIR:data}

New SIR-type event data typically arrive in the form of a simple data frame with
one row per individual and the time points of the sequential events of the individual as columns.
For the 1861 Hagelloch measles epidemic, such a data set of the 188 affected
children is contained in the \pkg{surveillance} package:
\begin{Schunk}
\begin{Sinput}
R> data("hagelloch")
R> head(hagelloch.df, n = 5)
\end{Sinput}
\begin{Soutput}
  PN    NAME FN HN AGE    SEX        PRO        ERU        CL DEAD IFTO SI
1  1 Mueller 41 61   7 female 1861-11-21 1861-11-25 1st class <NA>   45 10
2  2 Mueller 41 61   6 female 1861-11-23 1861-11-27 1st class <NA>   45 12
3  3 Mueller 41 61   4 female 1861-11-28 1861-12-02 preschool <NA>  172  9
4  4 Seibold 61 62  13   male 1861-11-27 1861-11-28 2nd class <NA>  180 10
5  5  Motzer 42 63   8 female 1861-11-22 1861-11-27 1st class <NA>   45 11
                C PR CA NI GE TD   TM x.loc y.loc tPRO tERU tDEAD   tR   tI
1 no complicatons  4  4  3  1 NA   NA   142   100 22.7 26.2    NA 29.2 21.7
2 no complicatons  4  4  3  1  3 40.3   142   100 24.2 28.8    NA 31.8 23.2
3 no complicatons  4  4  3  2  1 40.5   142   100 29.6 33.7    NA 36.7 28.6
4 no complicatons  1  1  1  1  3 40.7   165   102 28.1 29.0    NA 32.0 27.1
5 no complicatons  5  3  2  1 NA   NA   145   120 23.1 28.4    NA 31.4 22.1
\end{Soutput}
\end{Schunk}
The \code{help("hagelloch")} contains a description of all columns. Here
we concentrate on the event columns \code{PRO} (appearance of prodromes),
\code{ERU} (eruption), and \code{DEAD} (day of death if during the outbreak).
We take the day on which the index case developed first symptoms,
30 October 1861 (\code{min(hagelloch.df$PRO)}), as the start of the epidemic,
i.e., we condition on this case being initially infectious.
As for \code{twinstim}, the property of point processes that concurrent
events have zero probability requires special treatment. Ties are due to the
interval censoring of the data to a daily basis -- we broke these ties
by adding random jitter to the event times within the given days. The resulting
columns \code{tPRO}, \code{tERU}, and \code{tDEAD} are relative to the defined
start time. Following \citet{neal.roberts2004}, we assume that each child
becomes infectious (S~$\rightarrow$~I event at time \code{tI}) one day before the appearance
of prodromes, and is removed from the epidemic (I~$\rightarrow$~R event at time \code{tR})
three days after the appearance of rash or at the time of death, whichever comes
first.

For further processing of the data, we convert \code{hagelloch.df} to
the standardized \class{epidata} structure for \code{twinSIR}.
This is done by the converter function \code{as.epidata},
which also checks consistency and optionally
pre-calculates the epidemic terms $\bm{x}_i(t)$ of Equation~\ref{eqn:twinSIR:x}
to be incorporated in a \code{twinSIR} model.
The following call generates the \class{epidata} object \code{hagelloch}:
\begin{Schunk}
\begin{Sinput}
R> hagelloch <- as.epidata(hagelloch.df,
+    t0 = 0, tI.col = "tI", tR.col = "tR",
+    id.col = "PN", coords.cols = c("x.loc", "y.loc"),
+    f = list(household    = function(u) u == 0,
+             nothousehold = function(u) u > 0),
+    w = list(c1 = function (CL.i, CL.j) CL.i == "1st class" & CL.j == CL.i,
+             c2 = function (CL.i, CL.j) CL.i == "2nd class" & CL.j == CL.i),
+    keep.cols = c("SEX", "AGE", "CL"))
\end{Sinput}
\end{Schunk}
The coordinates (\code{x.loc}, \code{y.loc}) correspond to the location of the
household the child lives in and are measured in meters.
Note that \class{twinSIR} allows for tied locations of individuals, but assumes
the relevant spatial location to be fixed during the entire observation period.
By default, the Euclidean distance between the given coordinates will be used.
Alternatively, \code{as.epidata} also accepts a pre-computed distance matrix via
its argument \code{D} without requiring spatial coordinates.
The argument \code{f} lists distance-dependent basis functions $B_m$ for
which the epidemic terms $\sum_{j\in I(t)} B_m(d_{ij})$
shall be generated. Here, \code{household} ($x_{i,H}(t)$) and
\code{nothousehold} ($x_{i,\bar{H}}(t)$) count for each child the number of
currently infective children in its household and outside its household,
respectively.
Similar to \citet{neal.roberts2004}, we also calculate the covariate-based
epidemic terms \code{c1} ($x_{i,c1}(t)$) and \code{c2} ($x_{i,c2}(t)$)
counting the number of currently infective classmates.
Note from the corresponding definitions of $w_{ij1}$ and $w_{ij2}$ in \code{w}
that \code{c1} is always zero for children of the second class and \code{c2} is
always zero for children of the first class. For pre-school children, both
variables equal zero over the whole period.
By the last argument \code{keep.cols}, we choose to only keep the covariates
\code{SEX}, \code{AGE}, and school \code{CL}ass from \code{hagelloch.df}.

The first few rows of the generated \class{epidata} object are shown below:
\begin{Schunk}
\begin{Sinput}
R> head(hagelloch, n = 5)
\end{Sinput}
\begin{Soutput}
  BLOCK id start stop atRiskY event Revent x.loc y.loc    SEX AGE        CL
1     1  1     0 1.14       1     0      0   142   100 female   7 1st class
2     1  2     0 1.14       1     0      0   142   100 female   6 1st class
3     1  3     0 1.14       1     0      0   142   100 female   4 preschool
4     1  4     0 1.14       1     0      0   165   102   male  13 2nd class
5     1  5     0 1.14       1     0      0   145   120 female   8 1st class
  household nothousehold c1 c2
1         0            1  0  0
2         0            1  0  0
3         0            1  0  0
4         0            1  0  1
5         0            1  0  0
\end{Soutput}
\end{Schunk}

The \class{epidata} structure inherits from counting processes as implemented by
the \class{Surv} class of package \pkg{survival}~\citep{R:survival} and also
used in, e.g., the \pkg{timereg} package~\citep{scheike.zhang2011}.
Specifically, the observation period is splitted up into consecutive time
intervals (\code{start}; \code{stop}] of constant conditional intensities.
As the CIF $\lambda_i(t)$ of Equation~\eqref{eqn:twinSIR} only changes at time points, where
the set of infectious individuals $I(t)$ or some endemic covariate in $\nu_i(t)$
change, those occurrences define the break points of the time intervals.
Altogether, the \code{hagelloch} event history consists of
375 time \code{BLOCK}s of
188 rows, where each row describes the state of
individual \code{id} during the corresponding time interval.
The susceptibility status and the I- and R-events are
captured by the columns \code{atRiskY}, \code{event} and
\code{Revent}, respectively. The \code{atRiskY} column indicates if
the individual is at risk of becoming infected in the current interval.
The event columns indicate, which individual was infected or removed at the
\code{stop} time. Note that at most one entry in the \code{event} and
\code{Revent} columns is 1, all others are 0.

Apart from being the input format for \code{twinSIR} models,
the \class{epidata} class has several associated methods
(Table~\ref{tab:methods:epidata}), which are similar in
spirit to the methods described for \class{epidataCS}.

\begin{table}[ht]
\centering
{\small
\begin{tabular}{lll}
  \toprule
Display & Subset & Modify \\ 
  \midrule
\code{print} & \code{[} & \code{update} \\ 
  \code{summary} &  &  \\ 
  \code{plot} &  &  \\ 
  \code{animate} &  &  \\ 
  \code{\textit{stateplot}} &  &  \\ 
   \bottomrule
\end{tabular}
}
\caption{Generic and \textit{non-generic} functions applicable to \class{epidata} objects.} 
\label{tab:methods:epidata}
\end{table}

For example, Figure~\ref{fig:hagelloch_plot} illustrates the course of
the Hagelloch measles epidemic by counting processes for the
number of susceptible, infectious and removed children,
respectively.
Figure~\ref{fig:hagelloch_households} shows the locations of the households.
An \code{animate}d map can also be produced
to view the households' states over time
and a \code{stateplot} shows the changes for a selected unit.

\begin{Schunk}
\begin{Sinput}
R> plot(hagelloch, xlab = "Time [days]")
\end{Sinput}
\begin{figure}[!h]

{\centering \includegraphics[width=0.8\linewidth]{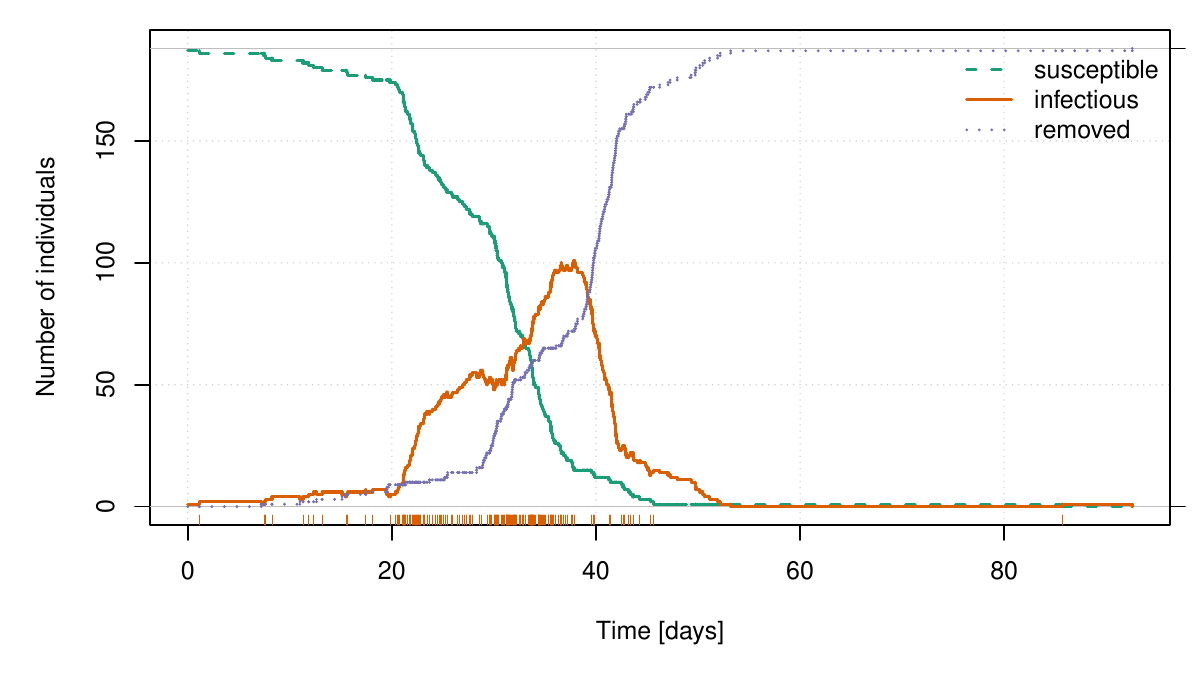} 

}

\caption{Evolution of the 1861 Hagelloch measles epidemic in terms of the numbers of susceptible, infectious, and recovered children. The bottom \code{rug} marks the infection times \code{tI}.}\label{fig:hagelloch_plot}
\end{figure}
\end{Schunk}

\begin{Schunk}
\begin{Sinput}
R> hagelloch_coords <- summary(hagelloch)$coordinates
R> plot(hagelloch_coords, xlab = "x [m]", ylab = "y [m]",
+    pch = 15, asp = 1, cex = sqrt(multiplicity(hagelloch_coords)))
R> legend(x = "topleft", pch = 15, legend = c(1, 4, 8), pt.cex = sqrt(c(1, 4, 8)),
+    title = "Household size")
\end{Sinput}
\begin{figure}[ht]

{\centering \includegraphics[width=0.8\linewidth]{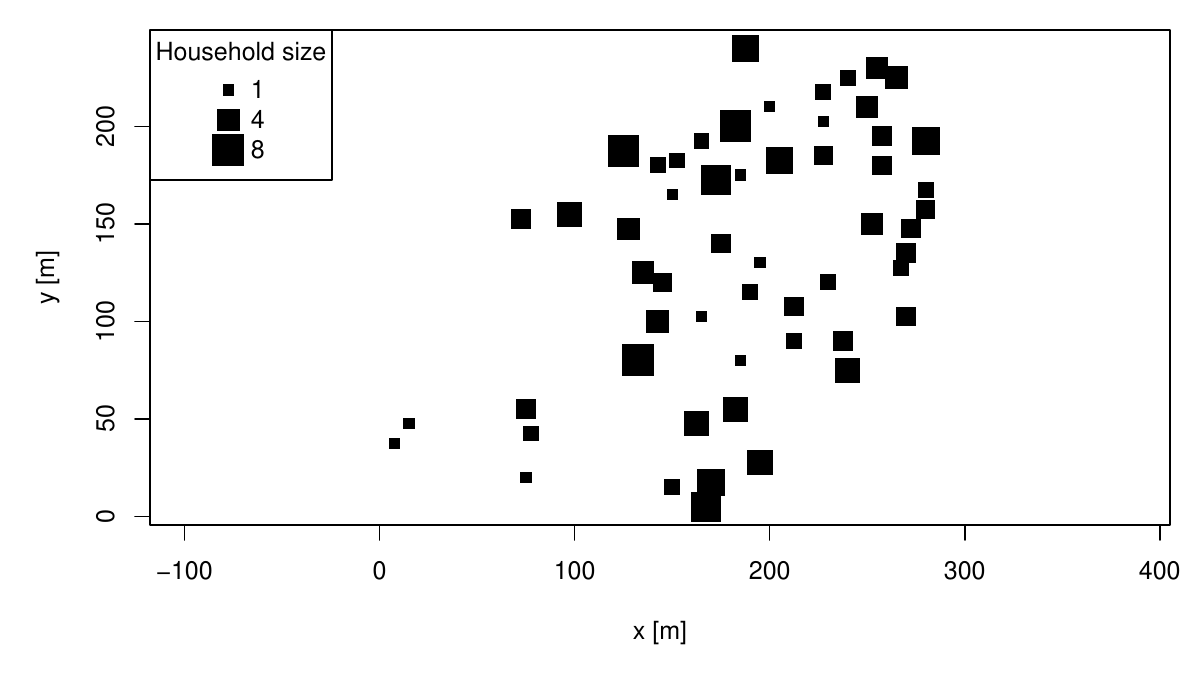} 

}

\caption{Spatial locations of the Hagelloch households. The size of each dot is proportional to the number of children in the household.}\label{fig:hagelloch_households}
\end{figure}
\end{Schunk}

\subsection{Modeling and inference} \label{sec:twinSIR:fit}

\subsubsection{Basic example}

To illustrate the flexibility of \code{twinSIR} we will analyze
the Hagelloch data using class room and household indicators similar to
\citet{neal.roberts2004}. We include an additional endemic background rate
$\exp(\beta_0)$, which allows for multiple outbreaks triggered
by external sources. Consequently, we do not
need to ignore the child that got infected about one month after the end of the
main epidemic (see the last event mark in Figure~\ref{fig:hagelloch_plot}),
as, e.g., done in a thorough network-based analysis of the Hagelloch data by
\citet{groendyke.etal2012}.
Altogether, the CIF for a child $i$ is modeled as
\begin{equation} \label{eqn:twinSIR:hagelloch}
  \lambda_i(t) = Y_i(t) \cdot \left[ \exp(\beta_0) +
    \alpha_H x_{i,H}(t) +
    \alpha_{c1} x_{i,c1}(t) + \alpha_{c2} x_{i,c2}(t) +
    \alpha_{\bar{H}} x_{i,\bar{H}}(t) \right] \:,
\end{equation}
where $Y_i(t) = \ind(i \in S(t))$ is the at-risk indicator.
By counting the number of infectious classmates separately for both school
classes as described in the previous section, we allow for class-specific
effects $\alpha_{c1}$ and $\alpha_{c2}$ on the force of infection.
The model is estimated by maximum likelihood \citep{hoehle2009}
using the following call:

\begin{Schunk}
\begin{Sinput}
R> hagellochFit <- twinSIR(~household + c1 + c2 + nothousehold, data = hagelloch)
R> summary(hagellochFit)
\end{Sinput}
\end{Schunk}
\begin{Schunk}
\begin{Soutput}
Call:
twinSIR(formula = ~household + c1 + c2 + nothousehold, data = hagelloch)

Coefficients:
                  Estimate Std. Error z value Pr(>|z|)    
household         0.026868   0.006113    4.39  1.1e-05 ***
c1                0.023892   0.005026    4.75  2.0e-06 ***
c2                0.002932   0.000755    3.88   0.0001 ***
nothousehold      0.000831   0.000142    5.87  4.3e-09 ***
cox(logbaseline) -7.362644   0.887989   -8.29  < 2e-16 ***
---
Signif. codes:  0 '***' 0.001 '**' 0.01 '*' 0.05 '.' 0.1 ' ' 1

Total number of infections:  187 

One-sided AIC: 1245	(simulated penalty weights)
Log-likelihood: -619
Number of log-likelihood evaluations: 119
\end{Soutput}
\end{Schunk}

The results show, e.g., a
0.0239 /
  0.0029 $=$
    8.15
times higher transmission between individuals in
the 1st class than in the 2nd class.
Furthermore, an infectious housemate adds
0.0269 /
  0.0008 $=$
    32.3
times as much infection pressure as infectious children outside the household.
The endemic background rate of infection in a population with no current measles
cases is estimated to be
$\exp(\hat{\beta}_0) = \exp(-7.36) = 0.000635$.
An associated Wald confidence interval (CI) based on the asymptotic normality of
the maximum likelihood estimator (MLE) can be obtained by
\code{exp}-transforming the \code{confint} for $\beta_0$:
\begin{Schunk}
\begin{Sinput}
R> exp(confint(hagellochFit, parm = "cox(logbaseline)"))
\end{Sinput}
\begin{Soutput}
                    2.5 %  97.5 %
cox(logbaseline) 0.000111 0.00362
\end{Soutput}
\end{Schunk}

Note that Wald confidence intervals for the epidemic parameters $\bm{\alpha}$
are to be treated carefully, because their construction does not
take the restricted parameter space into account.
For more adequate statistical inference,
the behavior of the log-likelihood near the MLE can be investigated using the
\code{profile}-method for \class{twinSIR} objects.
For instance, to evaluate the normalized profile log-likelihood of
$\alpha_{c1}$ and $\alpha_{c2}$ on an equidistant grid of 25 points within the
corresponding 95\% Wald CIs, we do:
\begin{Schunk}
\begin{Sinput}
R> prof <- profile(hagellochFit,
+    list(c(match("c1", names(coef(hagellochFit))), NA, NA, 25),
+         c(match("c2", names(coef(hagellochFit))), NA, NA, 25)))
\end{Sinput}
\end{Schunk}
The profiling result contains 95\% highest likelihood based CIs
for the parameters, as well as the Wald CIs for comparison:
\begin{Schunk}
\begin{Sinput}
R> prof$ci.hl
\end{Sinput}
\begin{Soutput}
   idx  hl.low   hl.up wald.low wald.up     mle
c1   2 0.01522 0.03497  0.01404 0.03374 0.02389
c2   3 0.00158 0.00454  0.00145 0.00441 0.00293
\end{Soutput}
\end{Schunk}
The entire functional form of the normalized profile log-likelihood on the
requested grid as stored in \code{prof$lp} can be visualized by:
\begin{Schunk}
\begin{Sinput}
R> plot(prof)
\end{Sinput}
\begin{figure}[ht]

{\centering \includegraphics[width=0.9\linewidth]{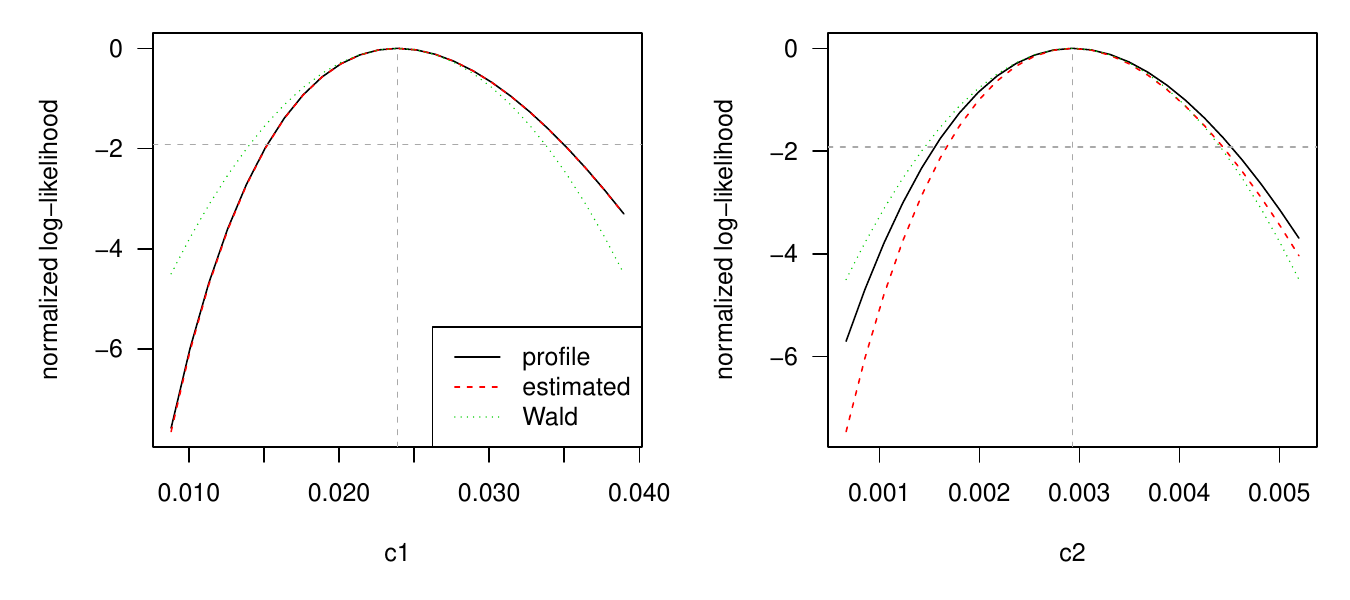} 

}

\caption{Normalized log-likelihood for $\alpha_{c1}$ and $\alpha_{c2}$ when fitting the \code{twinSIR} model formulated in Equation~\eqref{eqn:twinSIR:hagelloch} to the Hagelloch data.}\label{fig:hagellochFit_profile_plot}
\end{figure}
\end{Schunk}

\subsubsection{Model diagnostics}

\begin{table}[ht]
\centering
{\small
\begin{tabular}{lll}
  \toprule
Display & Extract & Other \\ 
  \midrule
\code{print} & \code{vcov} & \code{simulate} \\ 
  \code{summary} & \code{logLik} &  \\ 
  \code{plot} & \code{AIC} &  \\ 
  \code{intensityplot} & \code{extractAIC} &  \\ 
  \code{\textit{checkResidualProcess}} & \code{profile} &  \\ 
   & \code{residuals} &  \\ 
   \bottomrule
\end{tabular}
}
\caption{Generic and \textit{non-generic} functions for \class{twinSIR}. There are no specific \code{coef} or \code{confint} methods, since the respective default methods from package \pkg{stats} apply outright.} 
\label{tab:methods:twinSIR}
\end{table}

Table~\ref{tab:methods:twinSIR} lists all methods for the \class{twinSIR} class.
For example, to investigate how the CIF decomposes into endemic and epidemic
intensity over time, we produce Figure~\ref{fig:hagellochFit_plot1} by:
\begin{Schunk}
\begin{Sinput}
R> plot(hagellochFit, which = "epidemic proportion", xlab = "time [days]")
\end{Sinput}
\begin{figure}

{\centering \subfloat[Epidemic proportion.\label{fig:hagellochFit_plot1}]{\includegraphics[width=0.47\linewidth]{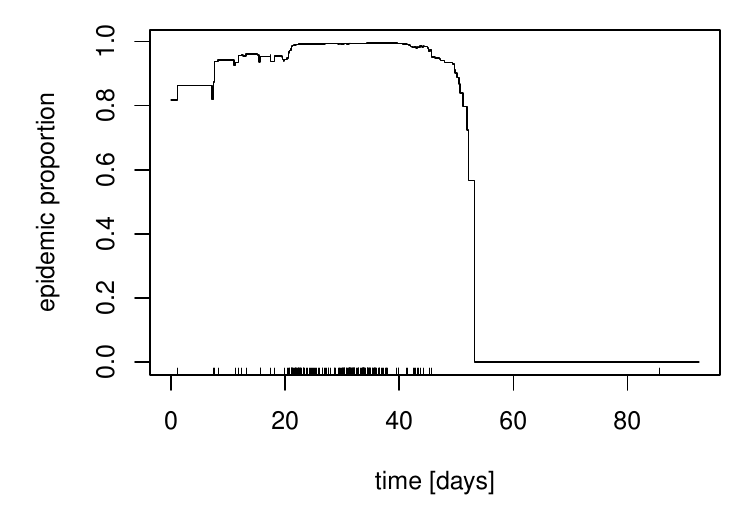} }\subfloat[Transformed residuals.\label{fig:hagellochFit_plot2}]{\includegraphics[width=0.47\linewidth]{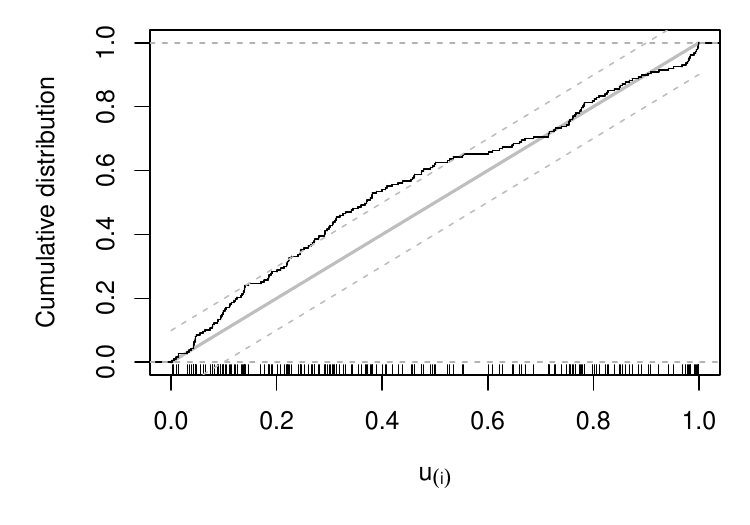} }

}

\caption{Diagnostic plots for the \code{twinSIR} model formulated in Equation~\ref{eqn:twinSIR:hagelloch}.}\label{fig:hagellochFit_plot}
\end{figure}
\end{Schunk}
Note that the last infection was necessarily caused by the endemic component
since there were no more infectious children in the observed population which
could have triggered the new case.
We can also inspect temporal Cox-Snell-like \code{residuals} of the fitted point
process using the function \code{checkResidualProcess} as for the
spatio-temporal point process models in
Section~\ref{sec:twinstim:fit}.
The resulting Figure~\ref{fig:hagellochFit_plot2} reveals some deficiencies of
the model in describing the waiting times between events, which might be related
to the assumption of fixed infection periods.

Finally, \class{twinSIR}'s \code{AIC}-method computes the one-sided
AIC~\citep{hughes.king2003} as described in \citet{hoehle2009}, which can be
used for model selection
under positivity constraints on $\bm{\alpha}$.
For instance, we may consider a more flexible model for local spread using a
step function for the distance kernel $f(u)$ in Equation \ref{eqn:twinSIR:f}.
An updated model with
$B_{1} = I_{(0;100)}(u)$, $B_{2} = I_{[100;200)}(u)$, $B_{3} = I_{[200;\infty)}(u)$
can be fitted as follows:
\begin{Schunk}
\begin{Sinput}
R> knots <- c(100, 200)
R> fstep <- list(
+    B1 = function(D) D > 0 & D < knots[1],
+    B2 = function(D) D >= knots[1] & D < knots[2],
+    B3 = function(D) D >= knots[2])
R> hagellochFit_fstep <- twinSIR(
+    ~household + c1 + c2 + B1 + B2 + B3,
+    data = update(hagelloch, f = fstep))
\end{Sinput}
\end{Schunk}

\begin{Schunk}
\begin{Sinput}
R> set.seed(1)
R> AIC(hagellochFit, hagellochFit_fstep)
\end{Sinput}
\begin{Soutput}
                   df  AIC
hagellochFit        5 1245
hagellochFit_fstep  7 1246
\end{Soutput}
\end{Schunk}
Hence the simpler model with just a \code{nothousehold} component instead
of the more flexible distance-based step function is preferred.
A random seed was set since the parameter penalty in the one-sided AIC is
determined by Monte Carlo simulation. The algorithm is described in
\citet[p.~79, Simulation 3]{Silvapulle.Sen2005} and involves quadratic
programming using package \pkg{quadprog} \citep{R:quadprog}.

\subsection{Simulation} \label{sec:twinsSIR:simulation}

Simulation from fitted \code{twinSIR} models is described in
detail in~\citet[Section~4]{hoehle2009}. The implementation is made
available by an appropriate \code{simulate}-method for class \class{twinSIR}.
Because both the algorithm and the call are similar to the invocation on
\class{twinstim} objects
(Section~\ref{sec:twinstim:simulation}), we skip the illustration here and
refer to \code{help("simulate.twinSIR")}.

\section{Areal time series of counts} \label{sec:hhh4}

In public health surveillance, routine reports of infections 
to public health authorities give rise to spatio-temporal data,
which are usually made available in the form of aggregated counts by region and period.
The Robert Koch Institute (RKI) in Germany, for example, maintains a database of
cases of notifiable diseases,
which can be queried via the
\emph{SurvStat@RKI}\footnote{\url{https://survstat.rki.de}} 
online service. 
As an illustrative example, we use weekly counts of measles
infections by district in the
Weser-Ems region of Lower Saxony, Germany, 2001--2002.
These spatio-temporal count data constitute the response
$Y_{it}$, $i=1,\dotsc,17$ (districts), $t=1,\dotsc,104$ (weeks), for our illustration of the
endemic-epidemic multivariate time-series model ``\code{hhh4}''.
We start by describing the general model class in Section~\ref{sec:hhh4:methods}.
Section~\ref{sec:hhh4:data} introduces the data and the associated \code{S4}-class
\class{sts} (``surveillance time series'').
In Section~\ref{sec:hhh4:fit}, a simple model for the measles data based on the
original analysis of \citet{held.etal2005} is introduced,
which is then sequentially improved by suitable model extensions.
The final Section~\ref{sec:hhh4:simulation} illustrates simulation from fitted
\class{hhh4} models.

\subsection[Model class]{Model class: \code{hhh4}} \label{sec:hhh4:methods}

An endemic-epidemic multivariate time-series model for infectious disease counts
$Y_{it}$ from units $i=1,\dotsc,I$ during periods $t=1,\dotsc,T$
was proposed by \citet{held.etal2005} and was later extended in
a series of papers
\citep{paul.etal2008,paul.held2011,held.paul2012,meyer.held2013}.
In its most general formulation, this so-called ``\code{hhh4}'' model assumes that, conditional on past
observations, $Y_{it}$ has a negative binomial distribution with mean
\begin{equation} \label{eqn:hhh4}
  \mu_{it} = e_{it} \, \nu_{it} + \lambda_{it} \, Y_{i,t-1} +
  \phi_{it} \sum_{j \ne i} w_{ji} \, Y_{j,t-1}
\end{equation}
and overdispersion parameter $\psi_i > 0$ such that the conditional
variance of $Y_{it}$ is $\mu_{it} (1+\psi_i \mu_{it})$.
Shared overdispersion parameters, e.g., $\psi_i\equiv\psi$, are supported
as well as replacing the negative binomial by a Poisson distribution,
which corresponds to the limit $\psi_i\equiv 0$.

Similar to the point process models of Sections~\ref{sec:twinstim}
and~\ref{sec:twinSIR}, the mean~\eqref{eqn:hhh4} decomposes additively into
endemic and epidemic components.
The endemic mean is usually modelled proportional to an offset of expected
counts~$e_{it}$. In spatial applications of the multivariate \code{hhh4} model
as in this paper, the ``unit''~$i$ refers to a geographical region and we
typically use (the fraction of) the population living in region~$i$ as the
endemic offset.
The observation-driven epidemic component splits up into
autoregressive effects, i.e., reproduction of the disease within region~$i$,
and neighbourhood effects, i.e., transmission from other regions~$j$.
Overall, Equation~\ref{eqn:hhh4} becomes a rich regression model by allowing for
log-linear predictors in all three components:
\begin{align} \label{eqn:hhh4:predictors}
  \log(\nu_{it}) &= \alpha_i^{(\nu)} + {\bm{\beta}^{(\nu)}}^\top \bm{z}^{(\nu)}_{it} \:, \\
  \log(\lambda_{it}) &= \alpha_i^{(\lambda)} + {\bm{\beta}^{(\lambda)}}^\top \bm{z}^{(\lambda)}_{it} \:, \\
  \log(\phi_{it}) &= \alpha_i^{(\phi)} + {\bm{\beta}^{(\phi)}}^\top \bm{z}^{(\phi)}_{it} \:.
\end{align}
The intercepts of these predictors can be assumed identical
across units, unit-specific, or random (and possibly correlated). 
The regression terms often involve sine-cosine effects of time
to reflect seasonally varying incidence, 
but may, e.g., also capture heterogeneous vaccination coverage
\citep{herzog.etal2011}.
Data on infections imported from outside the study region may enter the endemic
component \citep{geilhufe.etal2012}, which generally accounts for cases
not directly linked to other observed cases, e.g., due to edge effects.

For a single time series of counts $Y_t$, \code{hhh4} can be regarded as an
extension of \code{glm.nb} from package \pkg{MASS} \citep{R:MASS} to account for
autoregression. See the \code{vignette("hhh4")} for examples of modeling
univariate and bivariate count time series using \code{hhh4}.
With multiple regions, spatio-temporal dependence is adopted by the third
component in Equation~\ref{eqn:hhh4} with weights $w_{ji}$ reflecting the flow of
infections from region $j$ to region $i$. These transmission weights may be
informed by movement network data
\citep{paul.etal2008,schroedle.etal2012,geilhufe.etal2012},
but may also be estimated parametrically.
A suitable choice to reflect epidemiological coupling
between regions \citep[Chapter~7]{Keeling.Rohani2008}
is a power-law distance decay $w_{ji} = o_{ji}^{-d}$
defined in terms of the adjacency order~$o_{ji}$ in the
neighbourhood graph of the regions \citep{meyer.held2013}.
Note that we usually normalize the transmission weights such that
$\sum_i w_{ji} = 1$, i.e., the $Y_{j,t-1}$ cases are distributed
among the regions proportionally to the $j$'th row vector of the
weight matrix $(w_{ji})$.

Likelihood inference for the above multivariate time-series model
has been established by \citet{paul.held2011} with
extensions for parametric neighbourhood weights by \citet{meyer.held2013}.
Supplied with the analytical score function and Fisher information, the function
\code{hhh4} by default uses the quasi-Newton algorithm available through the
\proglang{R} function \code{nlminb} to maximize the log-likelihood.
Convergence is usually fast even for a large number of parameters.
If the model contains random effects, the penalized and marginal log-likelihoods
are maximized alternately until convergence.
Computation of the marginal Fisher information is accelerated using the
\pkg{Matrix} package \citep{R:Matrix}.

\subsection[Data structure]{Data structure: \class{sts}}  \label{sec:hhh4:data}

We briefly introduce the \code{S4}-class \class{sts} used for data input in
\code{hhh4} models. See \citet{hoehle.mazick2010} and \citet{salmon.etal2014}
for more detailed descriptions of this class, which is also used for the
prospective aberration detection facilities of the \pkg{surveillance} package.

The epidemic modeling of multivariate count time series essentially involves three 
data matrices: a $T \times I$ matrix of the observed counts, a corresponding
matrix with potentially time-varying population numbers (or fractions), and an
$I \times I$ neighbourhood matrix quantifying the coupling between the
$I$ units. In our example, the latter consists of the adjacency
orders~$o_{ji}$ between the districts.
A map of the districts in the form of a \code{SpatialPolygons} object (defined
by the \pkg{sp} package) can be used to 
derive the matrix of adjacency orders automatically using the functions
\code{poly2adjmat} and \code{nbOrder},
which wrap functionality of package \pkg{spdep} \citep{R:spdep}:
\begin{Schunk}
\begin{Sinput}
R> weserems_nbOrder <- nbOrder(poly2adjmat(map), maxlag = 10)
\end{Sinput}
\end{Schunk}
Given the aforementioned ingredients, the \class{sts} object
\code{data("measlesWeserEms")} included in \pkg{surveillance}
has been constructed as follows:
\begin{Schunk}
\begin{Sinput}
R> measlesWeserEms <- sts(observed = counts, start = c(2001, 1), frequency = 52,
+    neighbourhood = weserems_nbOrder, map = map, population = populationFrac)
\end{Sinput}
\end{Schunk}

Here, \code{start} and \code{frequency} have the same meaning as for classical
time-series objects of class \class{ts}, i.e., (year, sample number) of the
first observation and the number of observations per year.
Note that \code{data("measlesWeserEms")} constitutes a corrected version of
\code{data("measles.weser")} originally used by \citet{held.etal2005}.

We can visualize such \class{sts} data in four ways: 
individual time series, overall time series, map of accumulated counts by
district, or animated maps.
For instance, the two plots in Figure~\ref{fig:measlesWeserEms}
have been generated by the following code:
\begin{Schunk}
\begin{Sinput}
R> plot(measlesWeserEms, type = observed ~ time)
R> plot(measlesWeserEms, type = observed ~ unit,
+    population = measlesWeserEms@map$POPULATION / 100000,
+    labels = list(font = 2), colorkey = list(space = "right"),
+    sp.layout = layout.scalebar(measlesWeserEms@map, corner = c(0.05, 0.05),
+      scale = 50, labels = c("0", "50 km"), height = 0.03))
\end{Sinput}
\begin{figure}[!h]

{\centering \subfloat[Time series of weekly counts.\label{fig:measlesWeserEms1}]{\includegraphics[width=0.47\linewidth]{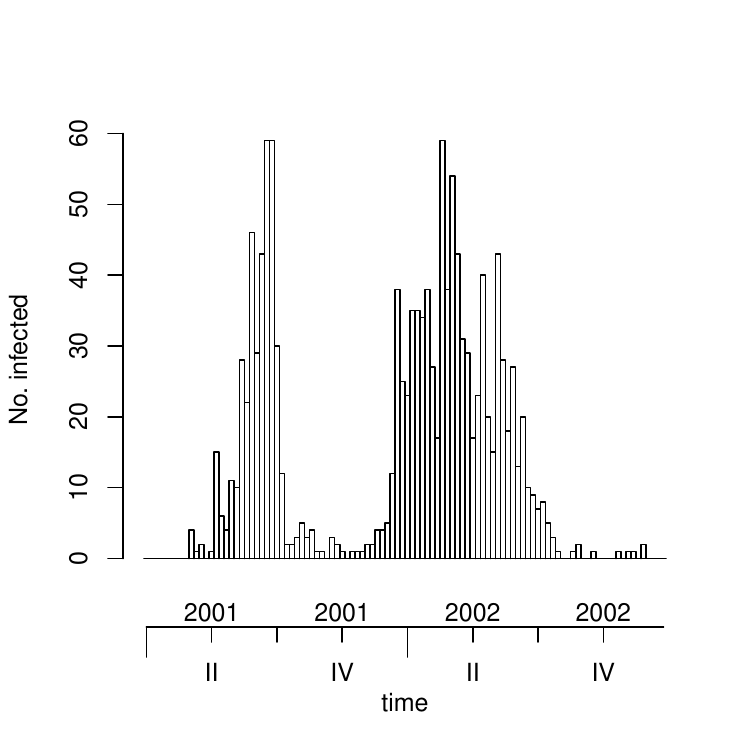} }\subfloat[Disease incidence (per 100\,000 inhabitants).\label{fig:measlesWeserEms2}]{\includegraphics[width=0.47\linewidth]{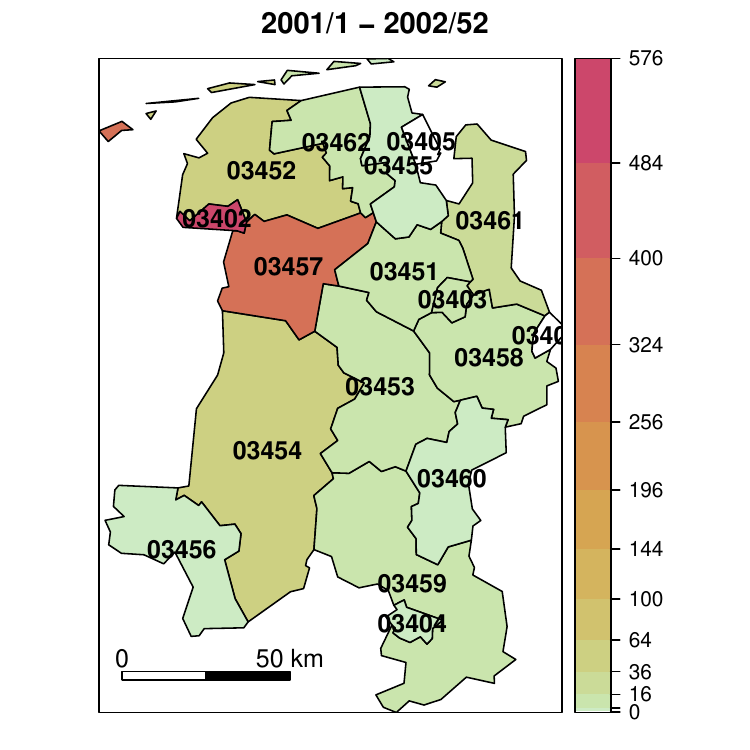} }

}

\caption{Measles infections in the Weser-Ems region, 2001--2002.}\label{fig:measlesWeserEms}
\end{figure}
\end{Schunk}
The overall time-series plot in Figure~\ref{fig:measlesWeserEms1} reveals strong
seasonality in the data with slightly different patterns in the two years.
The spatial plot in Figure~\ref{fig:measlesWeserEms2} is a tweaked \code{spplot}
(package \pkg{sp}) with colors from \pkg{colorspace} \citep{R:colorspace} using
$\sqrt{}$-equidistant cut points handled by package \pkg{scales}
\citep{R:scales}.
The default plot \code{type} is \code{observed ~ time | unit} and shows
the individual time series by district (Figure~\ref{fig:measlesWeserEms15}):
\begin{Schunk}
\begin{Sinput}
R> plot(measlesWeserEms, units = which(colSums(observed(measlesWeserEms)) > 0))
\end{Sinput}
\begin{figure}

{\centering \includegraphics[width=.95\linewidth]{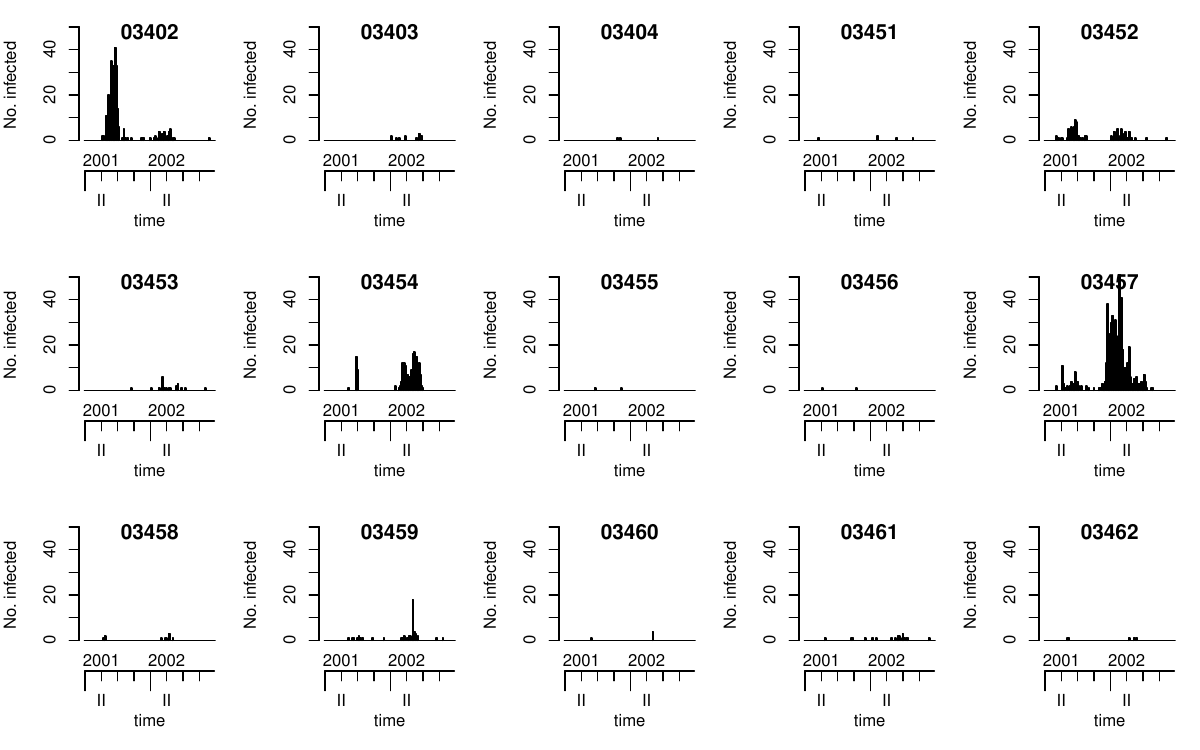} 

}

\caption{Count time series of the 15 affected districts.}\label{fig:measlesWeserEms15}
\end{figure}
\end{Schunk}
The plot excludes the districts
03401 (SK Delmenhorst) and 03405 (SK Wilhelmshaven)
without any reported cases. Obviously, the districts have been affected by
measles to a very heterogeneous extent during these two years.

An animation of the data can be easily produced as well. We recommend to use
converters of the \pkg{animation} package, e.g., to watch the series of plots in
a web browser.
The following code will generate weekly disease maps during the year 2001 with
the respective total number of cases shown in a legend and
-- if package \pkg{gridExtra} \citep{R:gridExtra} is available --
an evolving time-series plot at the bottom:
\begin{Schunk}
\begin{Sinput}
R> animation::saveHTML(
+    animate(measlesWeserEms, tps = 1:52, total.args = list()),
+    title = "Evolution of the measles epidemic in the Weser-Ems region, 2001",
+    ani.width = 500, ani.height = 600)
\end{Sinput}
\end{Schunk}

\subsection{Modeling and inference}  \label{sec:hhh4:fit}

For multivariate surveillance time series of counts such as the
\code{measlesWeserEms} data, the function \code{hhh4} fits models of the
form~\eqref{eqn:hhh4} via (penalized) maximum likelihood. 
We start by modeling the measles counts in the Weser-Ems region by a slightly
simplified version of the original negative binomial model by \citet{held.etal2005}.
Instead of district-specific intercepts $\alpha_i^{(\nu)}$ in the endemic component, we
first assume a common intercept $\alpha^{(\nu)}$ in order to not be forced to
exclude the two districts without any reported cases of measles.
After the estimation and illustration of this basic model, we will discuss the
following sequential extensions: 
covariates (district-specific vaccination coverage), 
estimated transmission weights,
and random effects to eventually account for
unobserved heterogeneity of the districts.

\subsubsection{Basic model}

Our initial model has the following mean structure:
\begin{align}
\mu_{it} &= e_i \, \nu_t + \lambda \, Y_{i,t-1} + \phi \sum_{j \ne i} w_{ji} Y_{j,t-1}\:,\label{eqn:hhh4:basic}\\
\log(\nu_t) &= \alpha^{(\nu)} + \beta_t t + \gamma \sin(\omega t) + \delta \cos(\omega t)\:. \label{eqn:hhh4:basic:end}
\end{align}
To account for temporal variation of disease incidence, the endemic log-linear
predictor $\nu_t$ incorporates an overall trend and a sinusoidal wave of frequency
$\omega=2\pi/52$. As a basic district-specific measure of disease incidence,
the population fraction $e_i$ is included as a multiplicative offset.
The epidemic parameters
$\lambda = \exp(\alpha^{(\lambda)})$ and $\phi = \exp(\alpha^{(\phi)})$ are
assumed homogeneous across districts and constant over time.
Furthermore, we define $w_{ji} = \ind(j \sim i) = \ind(o_{ji} = 1)$ for the time being, which means that the
epidemic can only arrive from directly adjacent districts. 
This \class{hhh4} model transforms into the following list of \code{control} arguments:
\begin{Schunk}
\begin{Sinput}
R> measlesModel_basic <- list(
+    end = list(f = addSeason2formula(~1 + t, period = measlesWeserEms@freq),
+               offset = population(measlesWeserEms)),
+    ar = list(f = ~1),
+    ne = list(f = ~1, weights = neighbourhood(measlesWeserEms) == 1),
+    family = "NegBin1")
\end{Sinput}
\end{Schunk}
The formulae of the three predictors $\log\nu_t$, $\log\lambda$ and $\log\phi$
are specified as element \code{f} of the \code{end}, \code{ar}, and \code{ne}
lists, respectively. For the endemic formula we use the convenient function
\code{addSeason2formula} to generate the sine-cosine terms, and we take the
multiplicative \code{offset} of population fractions $e_i$ from the
\code{measlesWeserEms} object.
The autoregressive part only consists of the intercept $\alpha^{(\lambda)}$,
whereas the neighbourhood component specifies the intercept $\alpha^{(\phi)}$
and also the matrix of transmission \code{weights} $(w_{ji})$ to use --
here a simple indicator of first-order adjacency.
The chosen \code{family} corresponds to a negative binomial model with a common
overdispersion parameter $\psi$ for all districts. Alternatives are
\code{"Poisson"}, \code{"NegBinM"} ($\psi_i$), or a factor determining
which groups of districts share a common overdispersion parameter. 
Together with the data, the complete list of control arguments is then fed into
the \code{hhh4} function to estimate the model,
a summary of which is printed below.

\begin{Schunk}
\begin{Sinput}
R> measlesFit_basic <- hhh4(stsObj = measlesWeserEms, control = measlesModel_basic)
R> summary(measlesFit_basic, idx2Exp = TRUE, amplitudeShift = TRUE, maxEV = TRUE)
\end{Sinput}
\begin{Soutput}
Call: 
hhh4(stsObj = measlesWeserEms, control = measlesModel_basic)

Coefficients:
                      Estimate  Std. Error
exp(ar.1)              0.64540   0.07927  
exp(ne.1)              0.01581   0.00420  
exp(end.1)             1.08025   0.27884  
exp(end.t)             1.00119   0.00426  
end.A(2 * pi * t/52)   1.16423   0.19212  
end.s(2 * pi * t/52)  -0.63436   0.13350  
overdisp               2.01384   0.28544  

Epidemic dominant eigenvalue:  0.72 

Log-likelihood:   -972 
AIC:              1957 
BIC:              1996 

Number of units:        17 
Number of time points:  103
\end{Soutput}
\end{Schunk}
The \code{idx2Exp} argument requests the estimates for $\lambda$, $\phi$,
$\alpha^{(\nu)}$ and $\exp(\beta_t)$ instead of their respective internal log-values.
For instance, \code{exp(end.t)} represents the seasonality-adjusted factor by which
the basic endemic incidence increases per week.
The \code{amplitudeShift} argument transforms the internal
coefficients $\gamma$ and $\delta$ of the sine-cosine terms to the amplitude $A$
and phase shift $\varphi$ of the corresponding sinusoidal wave
$A \sin(\omega t + \varphi)$ in $\log\nu_t$ \citep{paul.etal2008}. 
The multiplicative effect of seasonality on $\nu_t$ is shown in
Figure~\ref{fig:measlesFit_basic_endseason} produced by:
\begin{Schunk}
\begin{Sinput}
R> plot(measlesFit_basic, type = "season", components = "end", main = "")
\end{Sinput}
\begin{figure}[ht]

{\centering \includegraphics[width=.6\linewidth]{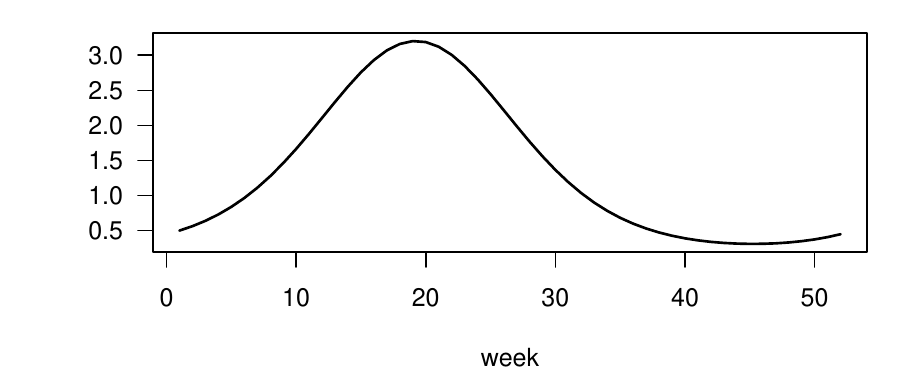} 

}

\caption{Estimated multiplicative effect of seasonality on the endemic mean.}\label{fig:measlesFit_basic_endseason}
\end{figure}
\end{Schunk}
The \code{overdisp} parameter and its 95\% confidence interval obtained by
\begin{Schunk}
\begin{Sinput}
R> confint(measlesFit_basic, parm = "overdisp")
\end{Sinput}
\begin{Soutput}
         2.5 % 97.5 %
overdisp  1.45   2.57
\end{Soutput}
\end{Schunk}
suggest that a negative binomial distribution with overdispersion is
more adequate than a Poisson model corresponding to
$\psi = 0$. We can underpin this finding by an AIC comparison, taking advantage
of the convenient \code{update} method for \class{hhh4} fits:
\begin{Schunk}
\begin{Sinput}
R> AIC(measlesFit_basic, update(measlesFit_basic, family = "Poisson"))
\end{Sinput}
\begin{Soutput}
                                             df  AIC
measlesFit_basic                              7 1957
update(measlesFit_basic, family = "Poisson")  6 2479
\end{Soutput}
\end{Schunk}

The epidemic potential of the process as determined by the parameters
$\lambda$ and $\phi$ is best investigated by a combined measure:
the dominant eigenvalue (\code{maxEV}) of the matrix $\bm{\Lambda}$ 
which has the entries
$(\Lambda)_{ii} = \lambda$ 
on the diagonal and
$(\Lambda)_{ij} = \phi w_{ji}$ 
for $j\ne i$ \citep{paul.etal2008}.
If the dominant eigenvalue is smaller than unity, it can be interpreted as the
epidemic proportion of disease incidence. In the above model, the
estimate is 72\%.
Another way of judging the relative importance of the three model
components is to plot the fitted mean components along with the observed counts.
Figure~\ref{fig:measlesFitted_basic} shows this for the six districts
with more than 20 cases:
\begin{Schunk}
\begin{Sinput}
R> districts2plot <- which(colSums(observed(measlesWeserEms)) > 20)
R> plot(measlesFit_basic, type = "fitted", units = districts2plot, hide0s = TRUE)
\end{Sinput}
\begin{figure}[tb]

{\centering \includegraphics[width=.95\linewidth]{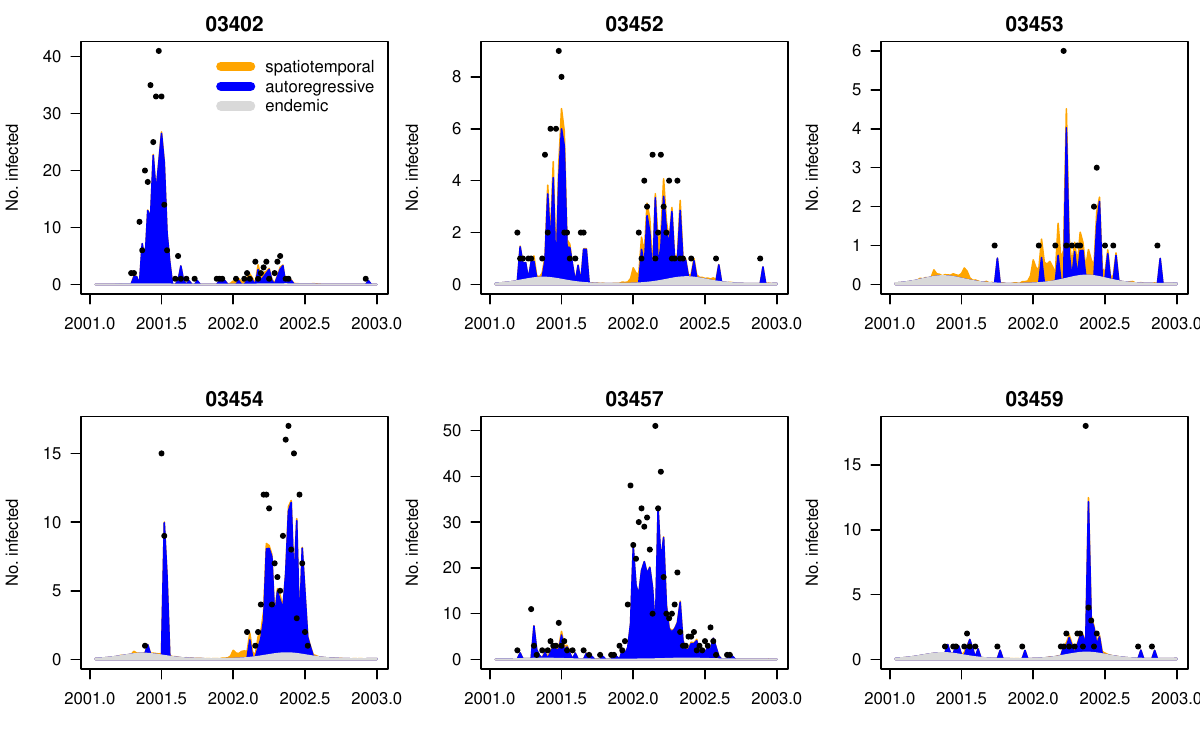} 

}

\caption{Fitted components in the initial model \code{measlesFit\_basic} for the six districts with more than 20 cases. Dots are only drawn for positive weekly counts.}\label{fig:measlesFitted_basic}
\end{figure}
\end{Schunk}
The largest portion of the fitted mean indeed results from the within-district
autoregressive component with very little contribution of cases from adjacent
districts and a rather small endemic incidence.

Other plot \code{type}s and methods for fitted \class{hhh4} models as listed in
Table~\ref{tab:methods:hhh4} will be applied in the course of the following
model extensions.

\begin{table}[ht]
\centering
{\small
\begin{tabular}{llll}
  \toprule
Display & Extract & Modify & Other \\ 
  \midrule
\code{print} & \code{nobs} & \code{update} & \code{predict} \\ 
  \code{summary} & \code{coef} &  & \code{simulate} \\ 
  \code{plot} & \code{fixef} &  & \code{pit} \\ 
   & \code{ranef} &  & \code{scores} \\ 
   & \code{vcov} &  & \code{calibrationTest} \\ 
   & \code{confint} &  & \code{all.equal} \\ 
   & \code{coeflist} &  & \code{\textit{oneStepAhead}} \\ 
   & \code{logLik} &  &  \\ 
   & \code{residuals} &  &  \\ 
   & \code{terms} &  &  \\ 
   & \code{formula} &  &  \\ 
   \bottomrule
\end{tabular}
}
\caption{Generic and \textit{non-generic} functions applicable to \class{hhh4} objects.} 
\label{tab:methods:hhh4}
\end{table}

\subsubsection{Covariates}

The \class{hhh4} model framework allows for covariate effects
on the endemic or epidemic contributions to disease
incidence. Covariates may vary over both regions and time and thus obey
the same $T \times I$ matrix structure as the observed counts.
For infectious disease models, the regional vaccination coverage is an important
example of such a covariate, since it reflects the (remaining)
susceptible population. 
In a thorough analysis of measles occurrence in the German federal
states, \citet{herzog.etal2011} found vaccination coverage to be
associated with outbreak size. We follow their approach of using the
district-specific proportion $1-v_i$ of unvaccinated children just starting
school as a proxy for the susceptible population. 
As $v_i$ we use the proportion of children vaccinated with at least one dose
among the ones presenting their vaccination card at school entry in district
$i$ in the year 2004.\footnote{%
  This is the first year with complete data for all 17
  districts, available from the public health department of Lower Saxony
  at \url{http://www.nlga.niedersachsen.de/portal/live.php?navigation_id=27093}.}
This time-constant covariate needs to be transformed to the common
matrix structure for incorporation in \code{hhh4}:
\begin{Schunk}
\begin{Sinput}
R> Sprop <- matrix(1 - measlesWeserEms@map@data$vacc1.2004,
+    nrow = nrow(measlesWeserEms), ncol = ncol(measlesWeserEms), byrow = TRUE)
R> summary(Sprop[1, ])
\end{Sinput}
\begin{Soutput}
   Min. 1st Qu.  Median    Mean 3rd Qu.    Max. 
 0.0306  0.0481  0.0581  0.0675  0.0830  0.1400 
\end{Soutput}
\end{Schunk}

There are several ways to account for the susceptible proportion in our model,
among which the simplest is to update the endemic population offset $e_i$ by
multiplication with $(1-v_i)$. 
\citet{herzog.etal2011} found that the susceptible proportion is best added as a
covariate in the autoregressive component in the form
\[ \lambda_i \, Y_{i,t-1} =
\exp\big(\alpha^{(\lambda)} + \beta_s \log(1-v_i)\big) \, Y_{i,t-1} = 
\exp\big(\alpha^{(\lambda)}\big) \, (1-v_i)^{\beta_s} \, Y_{i,t-1}
\]
according to the mass action principle \citep{Keeling.Rohani2008}.
A higher proportion of susceptibles in district $i$ is expected to boost the
generation of new infections, i.e., $\beta_s > 0$. Alternatively, this effect
could be assumed as an offset, i.e., $\beta_s \equiv 1$.
To choose between endemic and/or autoregressive effects, and multiplicative
offset vs.\ covariate modeling, we perform AIC-based model selection.
First, we set up a grid of all combinations of envisaged extensions for the
endemic and autoregressive components:
\begin{Schunk}
\begin{Sinput}
R> Soptions <- c("unchanged", "Soffset", "Scovar")
R> SmodelGrid <- expand.grid(end = Soptions, ar = Soptions)
R> row.names(SmodelGrid) <- do.call("paste", c(SmodelGrid, list(sep = "|")))
\end{Sinput}
\end{Schunk}
Then we update the initial model \code{measlesFit_basic} according to each row of
\code{SmodelGrid}:
\begin{Schunk}
\begin{Sinput}
R> measlesFits_vacc <- apply(X = SmodelGrid, MARGIN = 1, FUN = function (options) {
+    updatecomp <- function (comp, option) switch(option,
+      "unchanged" = list(),
+      "Soffset" = list(offset = comp$offset * Sprop),
+      "Scovar" = list(f = update(comp$f, ~. + log(Sprop))))
+    update(measlesFit_basic,
+      end = updatecomp(measlesFit_basic$control$end, options[1]),
+      ar = updatecomp(measlesFit_basic$control$ar, options[2]),
+      data = list(Sprop = Sprop))
+    })
\end{Sinput}
\end{Schunk}
The resulting object \code{measlesFits_vacc} is a list of
9 \class{hhh4} fits, which are named according to
the corresponding \code{Soptions} used for the endemic and autoregressive
component. 
We construct a call of the function \code{AIC} taking all list elements as
arguments:
\begin{Schunk}
\begin{Sinput}
R> aics_vacc <- do.call(AIC, lapply(names(measlesFits_vacc), as.name),
+    envir = as.environment(measlesFits_vacc))
R> aics_vacc[order(aics_vacc[, "AIC"]), ]
\end{Sinput}
\begin{Soutput}
                    df  AIC
`Scovar|unchanged`   8 1917
`Scovar|Scovar`      9 1919
`Soffset|unchanged`  7 1922
`Soffset|Scovar`     8 1924
`Scovar|Soffset`     8 1934
`Soffset|Soffset`    7 1937
unchanged|unchanged  7 1957
`unchanged|Scovar`   8 1959
`unchanged|Soffset`  7 1967
\end{Soutput}
\end{Schunk}

Hence, AIC increases if the susceptible proportion is only added to the autoregressive
component, but we see a remarkable improvement when adding it to the
endemic component. The best model is obtained by leaving the autoregressive
component unchanged ($\lambda$) and adding the term $\beta_s \log(1-v_i)$ to
the endemic predictor in Equation~\ref{eqn:hhh4:basic:end}.
\begin{Schunk}
\begin{Sinput}
R> measlesFit_vacc <- measlesFits_vacc[["Scovar|unchanged"]]
R> coef(measlesFit_vacc, se = TRUE)["end.log(Sprop)", ]
\end{Sinput}
\begin{Soutput}
  Estimate Std. Error 
     1.718      0.288 
\end{Soutput}
\end{Schunk}
The estimated exponent $\hat{\beta}_s$ is
both clearly positive and different from the offset assumption.
In other words, if a district's fraction of susceptibles is doubled,
the endemic measles incidence is estimated to multiply by
$2^{\hat{\beta}_s} =$
3.29 (95\% CI: 2.23--4.86).

\subsubsection{Spatial interaction}

Up to now, the model assumed that the epidemic can only arrive from directly
adjacent districts because $w_{ji} = \ind(j\sim i)$,
and that all districts have the same potential $\phi$ for importing cases from
neighbouring regions.
Given the ability of humans to travel further and preferrably to metropolitan
areas, both assumptions seem overly simplistic.
First, to reflect commuter-driven spread 
in our model, we scale the district's susceptibility according to its
population fraction by multiplying $\phi$ by $e_i^{\beta_{pop}}$:
\begin{Schunk}
\begin{Sinput}
R> measlesFit_nepop <- update(measlesFit_vacc,
+    ne = list(f = ~log(pop)), data = list(pop = population(measlesWeserEms)))
\end{Sinput}
\end{Schunk}
As in a similar analysis of influenza 
\citep{meyer.held2013}, we find strong evidence for such an agglomeration
effect: the estimated exponent is $\hat{\beta}_{pop} =$
2.85 (95\% CI: 1.83--3.87) and AIC decreases from 
1917 to 1887.
Models where attraction to a region scales with population size are
called ``gravity'' models \citep{xia.etal2004}.

To account for long-range transmission of cases, \citet{meyer.held2013} proposed
to estimate the weights $w_{ji}$ as a function of the adjacency order
$o_{ji}$ between the districts. For instance, a power-law model
assumes the form $w_{ji} = o_{ji}^{-d}$, for $j\ne i$ and $w_{jj}=0$,
where the decay parameter $d$ is to be estimated.
Normalization to $w_{ji} / \sum_k w_{jk}$ is recommended and applied by default
when supplying \code{W_powerlaw} as weights in the neighbourhood component:
\begin{Schunk}
\begin{Sinput}
R> measlesFit_powerlaw <- update(measlesFit_nepop,
+    ne = list(weights = W_powerlaw(maxlag = 5)))
\end{Sinput}
\end{Schunk}
The argument \code{maxlag} sets an upper bound for spatial interaction in
terms of adjacency order. Here we set no limit since
\code{max(neighbourhood(measlesWeserEms))}
is 5.
The resulting parameter estimate is
$\hat{d} =$~4.10 (95\% CI: 2.03--6.17),
which represents a strong decay of spatial interaction for higher-order neighbours.
As an alternative to the parametric power law, unconstrained weights up to
\code{maxlag} can be estimated by using \code{W_np} instead of \code{W_powerlaw}.
For instance, \code{W_np(maxlag = 2)} corresponds to a second-order model, i.e.,
\mbox{$w_{ji} = 1 \cdot \ind(o_{ji} = 1) + e^{\omega_2} \cdot \ind(o_{ji} = 2)$},
which is also row-normalized by default:
\begin{Schunk}
\begin{Sinput}
R> measlesFit_np2 <- update(measlesFit_nepop,
+    ne = list(weights = W_np(maxlag = 2)))
\end{Sinput}
\end{Schunk}

Figure~\ref{fig:measlesFit_neweights2} shows both the power law model
$o^{-\hat{d}}$ and the second-order model, where
$e^{\hat{\omega}_2} =$ 0.09 (95\% CI: 0.02--0.39). 
Alternatively, the plot \code{type = "neweights"} for \class{hhh4} fits
can produce a \code{stripplot} \citep{R:lattice} of $w_{ji}$ against $o_{ji}$ as shown in
Figure~\ref{fig:measlesFit_neweights1} for the power-law model:
\begin{Schunk}
\begin{Sinput}
R> library("lattice")
R> plot(measlesFit_powerlaw, type = "neweights", plotter = stripplot,
+    panel = function (...) {panel.stripplot(...); panel.average(...)},
+    jitter.data = TRUE, xlab = expression(o[ji]), ylab = expression(w[ji]))
\end{Sinput}
\begin{figure}

{\centering \subfloat[Normalized weights in the power-law model.\label{fig:measlesFit_neweights1}]{\includegraphics[width=0.47\linewidth]{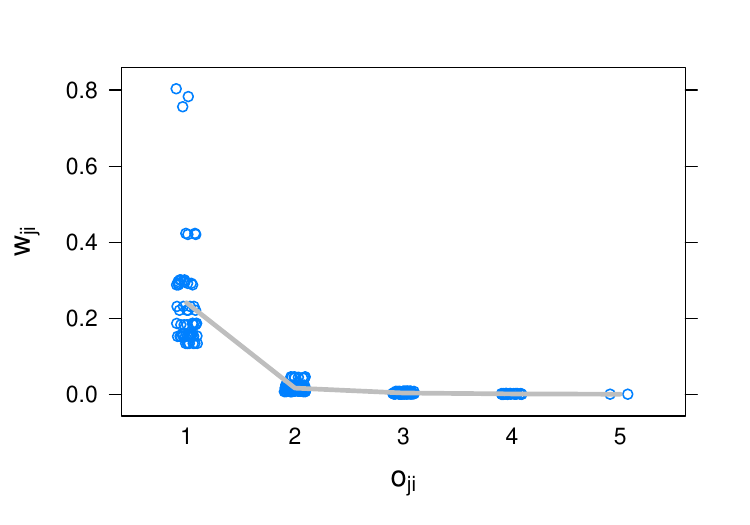} }\subfloat[Non-normalized weights with 95\% CIs.\label{fig:measlesFit_neweights2}]{\includegraphics[width=0.47\linewidth]{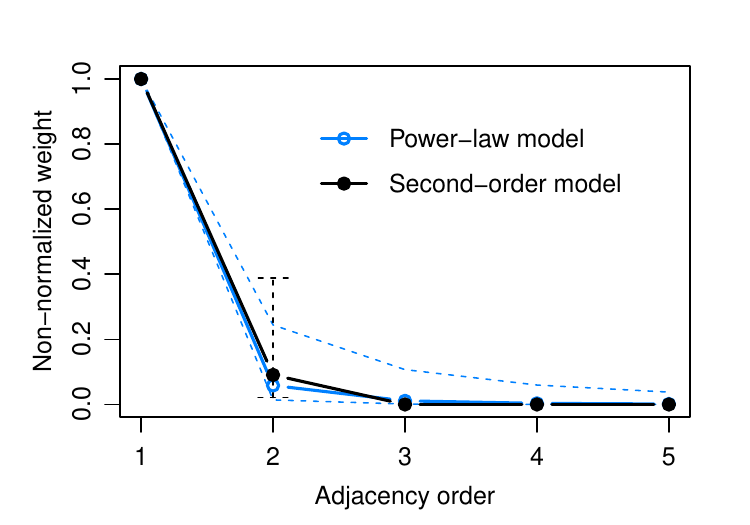} }

}

\caption{Estimated weights as a function of adjacency order.}\label{fig:measlesFit_neweights}
\end{figure}
\end{Schunk}
Note that only horizontal jitter is added in this case.
Because of normalization, the weight $w_{ji}$ for transmission from district $j$ to
district $i$ is determined not only by the districts' neighbourhood $o_{ji}$ but
also by the total amount of neighbourhood of district $j$ in the form of
$\sum_{k\ne j} o_{jk}^{-d}$, which causes some variation of the weights for a
specific order of adjacency.


An AIC comparison of the different models for the transmission weights yields:
\begin{Schunk}
\begin{Sinput}
R> AIC(measlesFit_nepop, measlesFit_powerlaw, measlesFit_np2)
\end{Sinput}
\begin{Soutput}
                    df  AIC
measlesFit_nepop     9 1887
measlesFit_powerlaw 10 1882
measlesFit_np2      10 1881
\end{Soutput}
\end{Schunk}
AIC improves when accounting for transmission between higher-order neighbours
by a power law or a second-order model.
In spite of the latter resulting in a slightly better fit, we will
use the power-law model as a basis for further model extensions since the
stand-alone second-order effect is not always identifiable in more complex
models and is scientifically implausible.

\subsubsection{Random effects}

\citet{paul.held2011} introduced random effects for \class{hhh4} models,
which are useful if the districts exhibit heterogeneous incidence levels not
explained by observed covariates, and especially if the number of districts is
large. For infectious disease surveillance data, a typical example of
unobserved heterogeneity is under-reporting \citep{bernard.etal2014}.
Our measles data even contain two districts
without any reported cases, while the district with the
smallest population (03402, SK Emden) had the second-largest number of cases
reported and the highest overall incidence 
(see Figures~\ref{fig:measlesWeserEms2} and~\ref{fig:measlesWeserEms15}).
Hence, allowing for district-specific intercepts in the endemic or epidemic
components is expected to improve the model fit.
For independent random effects
$\alpha_i^{(\nu)} \stackrel{iid}{\sim} \N(\alpha^{(\nu)}, \sigma_\nu^2)$,
$\alpha_i^{(\lambda)} \stackrel{iid}{\sim} \N(\alpha^{(\lambda)}, \sigma_\lambda^2)$, and
$\alpha_i^{(\phi)} \stackrel{iid}{\sim} \N(\alpha^{(\phi)}, \sigma_\phi^2)$
in all three components, we update the corresponding formulae as follows:
\begin{Schunk}
\begin{Sinput}
R> measlesFit_ri <- update(measlesFit_powerlaw,
+    end = list(f = update(formula(measlesFit_powerlaw)$end, ~. + ri() - 1)),
+    ar  = list(f = update(formula(measlesFit_powerlaw)$ar,  ~. + ri() - 1)),
+    ne  = list(f = update(formula(measlesFit_powerlaw)$ne,  ~. + ri() - 1)))
\end{Sinput}
\end{Schunk}
\begin{Schunk}
\begin{Sinput}
R> summary(measlesFit_ri, amplitudeShift = TRUE, maxEV = TRUE)
\end{Sinput}
\begin{Soutput}
Call: 
hhh4(stsObj = object$stsObj, control = control)

Random effects:
            Var   Corr   
ar.ri(iid)  1.076        
ne.ri(iid)  1.294 0      
end.ri(iid) 1.312 0    0 

Fixed effects:
                      Estimate  Std. Error
ar.ri(iid)            -1.61389   0.38197  
ne.log(pop)            3.42406   1.07722  
ne.ri(iid)             6.62429   2.81553  
end.t                  0.00578   0.00480  
end.A(2 * pi * t/52)   1.20359   0.20149  
end.s(2 * pi * t/52)  -0.47916   0.14205  
end.log(Sprop)         1.79350   0.69159  
end.ri(iid)            4.42260   1.94605  
neweights.d            3.60640   0.77602  
overdisp               0.97723   0.15132  

Epidemic dominant eigenvalue:  0.84 

Penalized log-likelihood:  -869 
Marginal log-likelihood:   -54.2 

Number of units:        17 
Number of time points:  103
\end{Soutput}
\end{Schunk}
The summary now contains an extra section with the
estimated variance components $\sigma_\lambda^2$, $\sigma_\phi^2$, and
$\sigma_\nu^2$ of the random effects.
We did not assume correlation between the three intercepts, but this is
possible by specifying \code{ri(corr = "all")} in the component formulae.
The implementation also supports a conditional autoregressive
formulation \citep{BesagYorkMollie1991} for spatially correlated intercepts by
using \code{ri(type = "car")}.
The estimated district-specific intercepts can be extracted by the
\code{ranef}-method:
\begin{Schunk}
\begin{Sinput}
R> head(ranef(measlesFit_ri, tomatrix = TRUE), n = 3)
\end{Sinput}
\begin{Soutput}
      ar.ri(iid) ne.ri(iid) end.ri(iid)
03401      0.000    -0.0567       -1.00
03402      1.223     0.0431        1.53
03403     -0.827     1.5588       -0.62
\end{Soutput}
\end{Schunk}
They can also be visualized in a map by the plot \code{type = "ri"}:
\begin{Schunk}
\begin{Sinput}
R> for (comp in c("ar", "ne", "end")) {
+    print(plot(measlesFit_ri, type = "ri", component = comp,
+      col.regions = rev(cm.colors(100)), labels = list(cex = 0.6),
+      at = seq(-1.6, 1.6, length.out = 15)))
+  }
\end{Sinput}
\begin{figure}[htb]

{\centering \subfloat[Autoregressive $\alpha_i^{(\lambda)}$\label{fig:measlesFit_ri_map1}]{\includegraphics[width=0.31\linewidth]{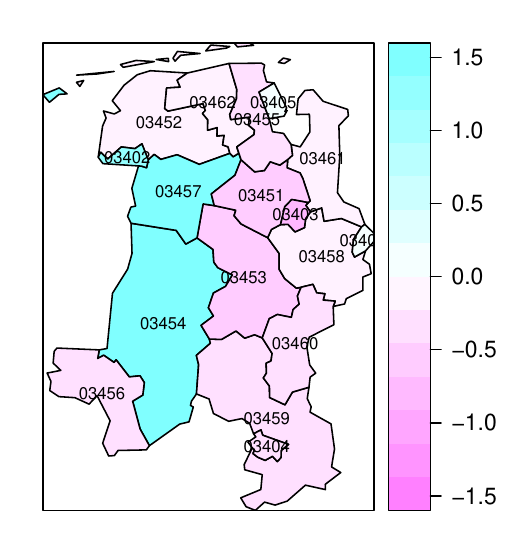} }\subfloat[Spatio-temporal $\alpha_i^{(\phi)}$\label{fig:measlesFit_ri_map2}]{\includegraphics[width=0.31\linewidth]{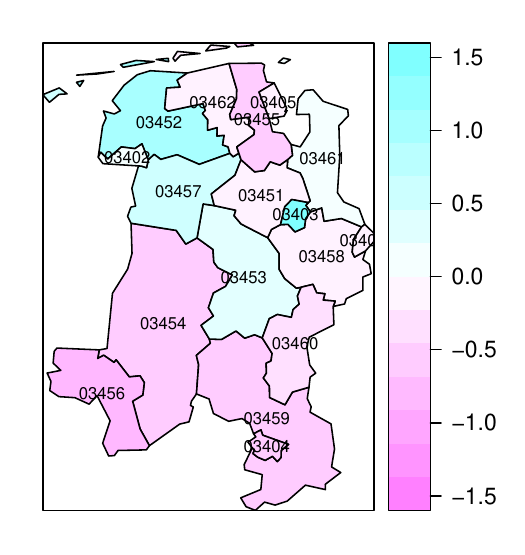} }\subfloat[Endemic $\alpha_i^{(\nu)}$\label{fig:measlesFit_ri_map3}]{\includegraphics[width=0.31\linewidth]{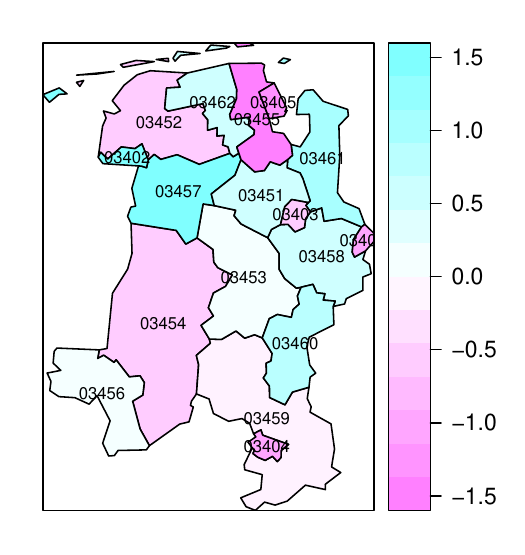} }

}

\caption{Maps of the estimated random intercepts.}\label{fig:measlesFit_ri_map}
\end{figure}
\end{Schunk}

For the autoregressive component in Figure~\ref{fig:measlesFit_ri_map1}, we see
a pronounced heterogeneity between the three western districts in blue and the
remaining districts. These three districts have been affected by large local
outbreaks and are also the ones with the highest overall numbers of cases.
In contrast, the city of Oldenburg (03403) is estimated with a relatively low
autoregressive factor $\lambda_i = \exp(\alpha^{(\lambda)} + \alpha_i^{(\lambda)}) =
0.087$,
but it seems to import more cases from other districts than explained by its
population (Figure~\ref{fig:measlesFit_ri_map2}).
In Figure~\ref{fig:measlesFit_ri_map3}, the two districts without any reported
measles cases (03401 and 03405) appear in dark pink, which means that they
exhibit a relatively low endemic incidence after adjusting for the
population and susceptible proportion.
Such districts could be suspected of a larger amount of under-reporting.

Note that the extra flexiblility of the random effects model comes at a price. 
First, the estimation runtime increases considerably from
0.1 seconds for the
previous power-law model \code{measlesFit_powerlaw} to 
4 seconds with additional
random effects. Furthermore, we no longer obtain AIC values in the model
summary, since random effects invalidate simple AIC-based model comparisons
\citep{greven.kneib2010}.
Of course we can plot the fitted values 
and visually compare their quality
with the initial fit shown in Figure~\ref{fig:measlesFitted_basic}:
\begin{Schunk}
\begin{Sinput}
R> plot(measlesFit_ri, type = "fitted", units = districts2plot, hide0s = TRUE)
\end{Sinput}
\begin{figure}[htb]

{\centering \includegraphics[width=.95\linewidth]{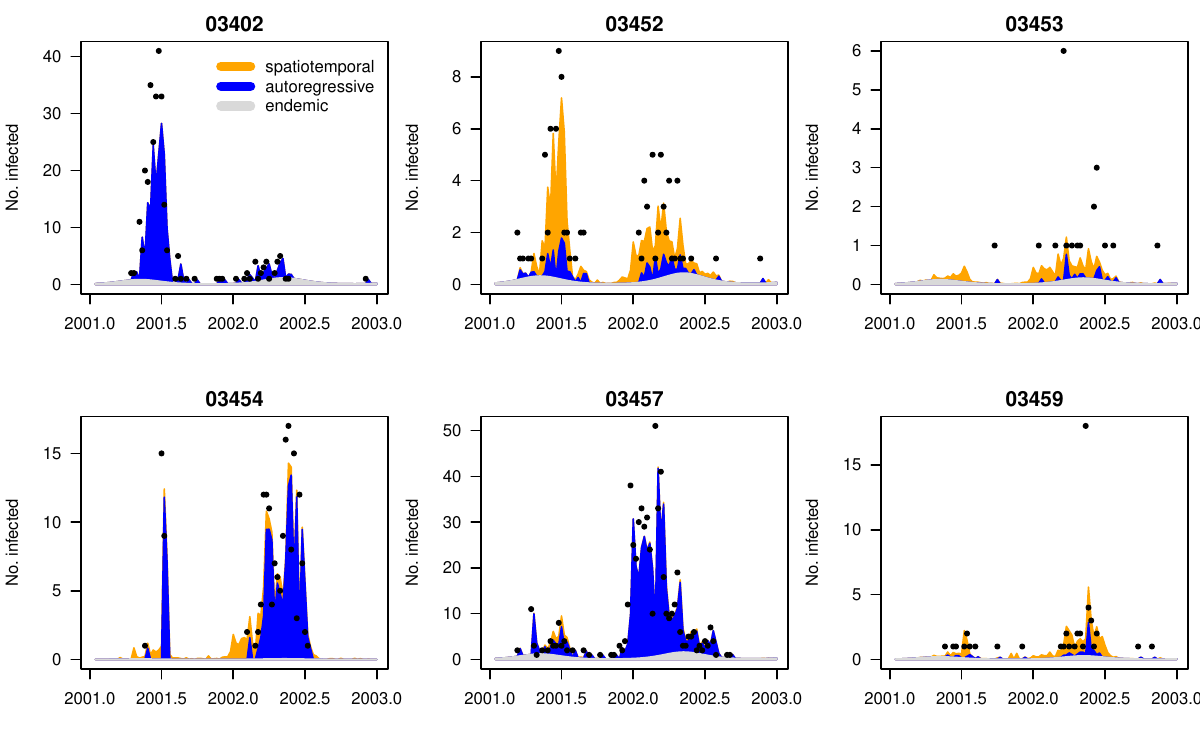} 

}

\caption{Fitted components in the random effects model \code{measlesFit\_ri} for the six districts with more than 20 cases. Compare to Figure~\ref{fig:measlesFitted_basic}.}\label{fig:measlesFitted_ri}
\end{figure}
\end{Schunk}
For some of these districts, a great amount of cases is now explained via transmission from
neighbouring regions while others are mainly influenced by the local
autoregression. Note that the estimated decomposition of the mean by district
can also be seen from the related plot \code{type = "maps"} (not shown).
However, for quantitative comparisons of model performance we have to resort to
more sophisticated techniques presented in the next section.

\subsubsection{Predictive model assessment}

\citet{paul.held2011} suggest to evaluate one-step-ahead forecasts from 
competing models by proper scoring rules for count data \citep{czado-etal-2009}.
These scores measure the discrepancy between the predictive distribution $P$
from a fitted model and the later observed value $y$.
A well-known example is the squared error score (``ses'') $(y-\mu_P)^2$,
which is usually averaged over a suitable set of forecasts to obtain the mean
squared error. More elaborate scoring rules such as the logarithmic score
(``logs'') or the ranked probability score (``rps'') take into account the whole
predictive distribution to assess calibration and sharpness simultaneously --
see the recent review by \citet{gneiting.katzfuss2014}.
The so-called Dawid-Sebastiani score (``dss'') is another option.
Lower scores correspond to better predictions. 

In the \class{hhh4} framework, predictive model assessment is made available by
the functions \code{oneStepAhead}, \code{scores}, \code{pit}, and
\code{calibrationTest}.
We will use the second quarter of 2002 as the test period, and compare
the basic model, the power-law model, and the random effects model. 
First, we use the \code{"final"} fits on the complete time series to
compute the predictions, which then simply correspond to the fitted values
during the test period:
\begin{Schunk}
\begin{Sinput}
R> tp <- c(65, 77)
R> models2compare <- paste0("measlesFit_", c("basic", "powerlaw", "ri"))
R> measlesPreds1 <- lapply(mget(models2compare), oneStepAhead,
+    tp = tp, type = "final")
\end{Sinput}
\end{Schunk}

Note that in this case, the log-score for a model's prediction in
district $i$ in week $t$ equals the associated negative log-likelihood
contribution. Comparing the mean scores from different models is thus
essentially a goodness-of-fit assessment:

\begin{Schunk}
\begin{Sinput}
R> SCORES <- c("logs", "rps", "dss", "ses")
R> measlesScores1 <- lapply(measlesPreds1, scores, which = SCORES, individual = TRUE)
R> t(sapply(measlesScores1, colMeans, dims = 2))
\end{Sinput}
\begin{Soutput}
                    logs   rps   dss  ses
measlesFit_basic    1.09 0.736 1.291 5.29
measlesFit_powerlaw 1.10 0.731 2.222 5.39
measlesFit_ri       1.01 0.638 0.966 4.82
\end{Soutput}
\end{Schunk}

All scoring rules claim that the random effects model gives the best fit during
the second quarter of 2002. 
Now we turn to true one-week-ahead predictions of \code{type = "rolling"},
which means that we always refit the model up to week $t$ to get
predictions for week $t+1$:
\begin{Schunk}
\begin{Sinput}
R> measlesPreds2 <- lapply(mget(models2compare), oneStepAhead,
+    tp = tp, type = "rolling", which.start = "final",
+    cores = 2 * (.Platform$OS.type == "unix"))
R> measlesScores2 <- lapply(measlesPreds2, scores, which = SCORES, individual = TRUE)
R> t(sapply(measlesScores2, colMeans, dims = 2))
\end{Sinput}
\begin{Soutput}
                    logs   rps  dss  ses
measlesFit_basic    1.10 0.748 1.34 5.40
measlesFit_powerlaw 1.14 0.765 2.93 5.87
measlesFit_ri       1.11 0.763 2.35 7.08
\end{Soutput}
\end{Schunk}

Thus, the most parsimonious initial model \code{measlesFit_basic}
gives the best one-week-ahead predictions in terms of overall mean scores.
Statistical significance of the differences in mean scores can be investigated
by a \code{permutationTest} for paired data or a paired t-test:
\begin{Schunk}
\begin{Sinput}
R> set.seed(321)
R> sapply(SCORES, function (score) permutationTest(
+    measlesScores2$measlesFit_ri[, , score],
+    measlesScores2$measlesFit_basic[, , score]))
\end{Sinput}
\begin{Soutput}
            logs    rps    dss   ses  
diffObs     0.00782 0.0154 1.01  1.68 
pVal.permut 0.867   0.72   0.518 0.19 
pVal.t      0.854   0.717  0.374 0.171
\end{Soutput}
\end{Schunk}

Hence, there is no clear evidence for a difference between the basic and the
random effects model with regard to predictive performance during the test
period. 
Whether predictions of a particular model are well calibrated can be formally
investigated by \code{calibrationTest}s for count data as recently 
proposed by \citep{wei.held2013}. For example:
\begin{Schunk}
\begin{Sinput}
R> calibrationTest(measlesPreds2[["measlesFit_ri"]], which = "rps")
\end{Sinput}
\begin{Soutput}
Calibration Test for Count Data (based on RPS)

data:  measlesPreds2[["measlesFit_ri"]]
z = 0.80671, n = 221, p-value = 0.4198
\end{Soutput}
\end{Schunk}
Thus, there is no evidence of miscalibrated predictions from the random effects
model. \citet{czado-etal-2009} describe an alternative informal approach to
assess calibration: probability integral transform
(PIT) histograms for count data (Figure~\ref{fig:measlesPreds2_pit}). 
\begin{Schunk}
\begin{Sinput}
R> for (m in models2compare)
+    pit(measlesPreds2[[m]], plot = list(ylim = c(0, 1.25), main = m))
\end{Sinput}
\begin{figure}[hbt]

{\centering \includegraphics[width=0.95\linewidth]{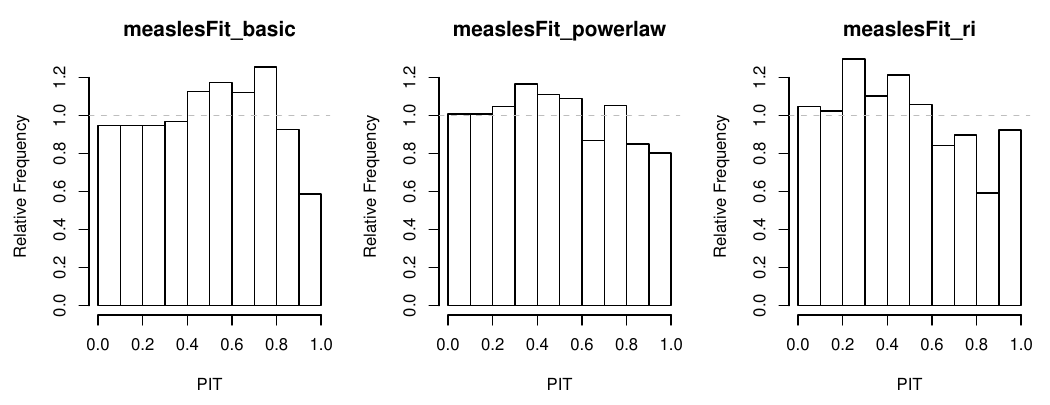} 

}

\caption{PIT histograms of competing models to check calibration of the one-week-ahead predictions during the second quarter of 2002.}\label{fig:measlesPreds2_pit}
\end{figure}
\end{Schunk}

Under the hypothesis of calibration, i.e., 
$y_{it} \sim P_{it}$ for all predictive distributions $P_{it}$ in the test period, the PIT histogram
is uniform. Underdispersed predictions lead to U-shaped histograms,
and bias causes skewness.
In this aggregate view of the predictions
over all districts and weeks of the test period, predictive performance is
comparable between the models, and there is no evidence of badly dispersed
predictions. However, the right-hand decay in all histograms suggests that all
models tend to predict higher counts than observed. This is most likely related
to the seasonal shift between the years 2001 and 2002. In 2001, the peak of the
epidemic was in the second quarter, while it already occured in the first
quarter in 2002 (cp.\ Figure~\ref{fig:measlesWeserEms1}).

\subsubsection{Further modeling options}

In the previous sections we extended our model for measles in the Weser-Ems
region with respect to spatial variation of the counts and their interaction.
Temporal variation was only accounted for in the endemic component,
which included a long-term trend and a sinusoidal wave on the
log-scale. \citet{held.paul2012} suggest to also allow seasonal
variation of the epidemic force by adding a superposition of $S$ harmonic waves
of fundamental frequency~$\omega$,
$\sum_{s=1}^S \left\{ \gamma_s \sin(s\,\omega t) + \delta_s \cos(s\,\omega t) \right\}$,
to the log-linear predictors of the autoregressive and/or neighbourhood component --
just like for $\log\nu_t$ in Equation~\ref{eqn:hhh4:basic:end} with $S=1$.
However, given only two years of measles surveillance and the apparent shift of
seasonality with regard to the start of the outbreak in 2002 compared to 2001,
more complex seasonal models are likely to overfit the data. 
Concerning the coding in \proglang{R}, sine-cosine terms can be added to the
epidemic components without difficulties by again using the convenient function
\code{addSeason2formula}.
Updating a previous model for different numbers of harmonics is even simpler,
since the \code{update}-method has a corresponding argument \code{S}.
The plots of \code{type = "season"} and \code{type = "maxEV"} for \class{hhh4} fits
can visualize the estimated component seasonality.

All of our models for the measles surveillance data incorporated an
epidemic effect of the counts from the local district and its neighbours.
Without further notice, we thereby assumed a lag equal to the observation
interval of one week. However, the generation time of measles is around 10 days
\citep{Anderson.May1991}, which is why some studies, e.g.,
\citet{finkenstaedt.etal2002} or \citet{herzog.etal2011}, aggregate their weekly measles 
surveillance data into biweekly intervals. 
\citet{fine.clarkson1982} used weekly counts in their analysis and report that
biweekly aggregation would have little effect on the results.
We can also perform such a sensitivity analysis by running the whole code of the
current section based on \code{aggregate(measlesWeserEms, nfreq = 26)}.
Doing so, the parameter estimates of the various models retain their order of
magnitude and conclusions remain the same. However, with the number of time
points halved, the complex random effects model would not always be identifiable
when calculating one-week-ahead predictions during
the test period.

We have shown several options to account for the spatio-temporal dynamics of
infectious disease spread. However, for directly transmitted human diseases, 
the social phenomenon of ``like seeks like'' results in contact patterns
between subgroups of a population, which extend the pure distance decay
of interaction.
Especially for school children, social contacts are known to be highly
assortative with respect to age \citep{mossong.etal2008}.
A useful epidemic model should therefore be additionally stratified by age group
and take the inherent contact structure into account.
How this extension can be incorporated in the spatio-temporal endemic-epidemic
modeling framework \class{hhh4} is the focus of current research. 

\subsection{Simulation}  \label{sec:hhh4:simulation}

Simulation from fitted \class{hhh4} models is enabled by an associated
\code{simulate}-method. Compared to the point process models of
Sections~\ref{sec:twinstim} and \ref{sec:twinSIR}, simulation is less complex
since it essentially consists of sequential calls of \code{rnbinom} (or
\code{rpois}). At each time point $t$,
the mean $\mu_{it}$ is determined by plugging in the parameter estimates and the
counts $Y_{i,t-1}$ simulated at the previous time point. In addition to a model
fit, we thus need to specify an initial vector of counts \code{y.start}.
As an example, we simulate 100 realizations of the evolution of measles during
the year 2002 based on the fitted random effects model and the counts of the
last week of the year 2001 in the 17 districts:
\begin{Schunk}
\begin{Sinput}
R> (y.start <- observed(measlesWeserEms)[52, ])
\end{Sinput}
\begin{Soutput}
03401 03402 03403 03404 03405 03451 03452 03453 03454 03455 03456 03457 03458 03459 
    0     0     0     0     0     0     0     0     0     0     0    25     0     0 
03460 03461 03462 
    0     0     0 
\end{Soutput}
\begin{Sinput}
R> measlesSim <- simulate(measlesFit_ri,
+    nsim = 100, seed = 1, subset = 53:104, y.start = y.start)
\end{Sinput}
\end{Schunk}
The simulated counts are returned as a
$52\times 17\times 100$ array instead of a list of 100 \class{sts} objects.
We can, e.g., look at the final size distribution of the simulations:
\begin{Schunk}
\begin{Sinput}
R> summary(colSums(measlesSim, dims = 2))
\end{Sinput}
\begin{Soutput}
   Min. 1st Qu.  Median    Mean 3rd Qu.    Max. 
    223     326     424     550     582    3970 
\end{Soutput}
\end{Schunk}
A few large outbreaks have been simulated, but the mean size
is below the observed number of \code{sum(observed(measlesWeserEms)[53:104, ])}
$= 779$ cases in the year 2002.
Using the \code{plot}-method associated with such \code{hhh4} simulations,
Figure~\ref{fig:measlesSim_plot_time} shows the weekly number of observed cases
compared to the long-term forecast:
\begin{Schunk}
\begin{Sinput}
R> plot(measlesSim, "time", ylim = c(0, 100))
\end{Sinput}
\begin{figure}[htb]

{\centering \includegraphics[width=0.8\linewidth]{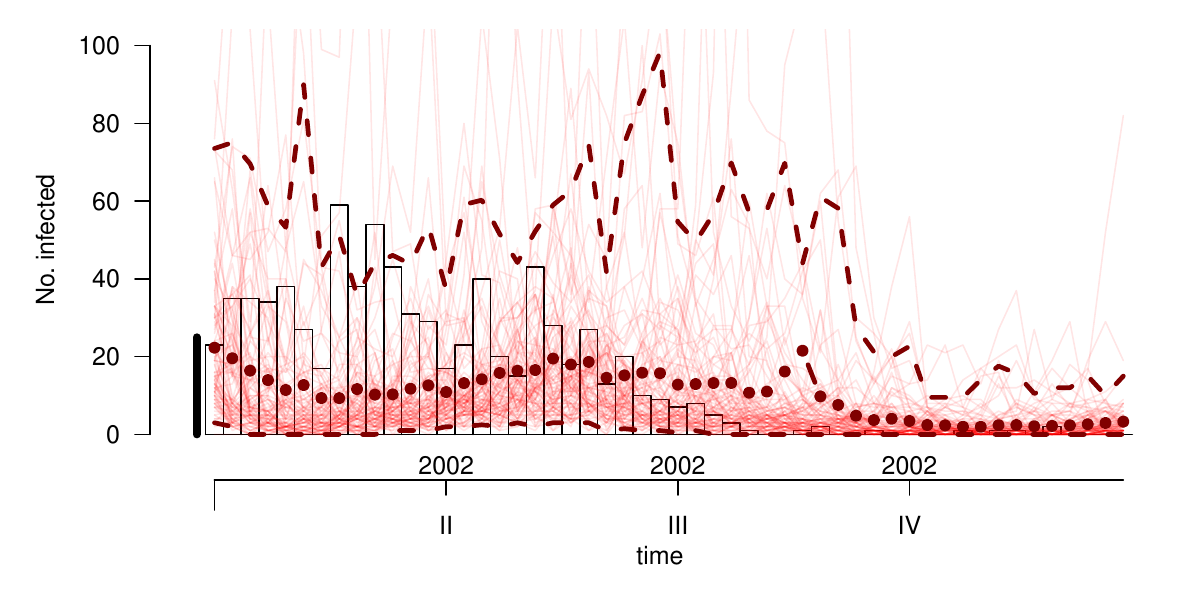} 

}

\caption{Simulation-based long-term forecast starting from the last week in 2001 (vertical bar on the left), showing the counts aggregated over all districts. The weekly mean of the simulations is represented by dots and the dashed lines correspond to the pointwise 2.5\% and 97.5\% quantiles. The actually observed counts are shown in the background.}\label{fig:measlesSim_plot_time}
\end{figure}
\end{Schunk}
We refer to \code{help("simulate.hhh4")} 
for further examples.

\section{Conclusion} \label{sec:conclusion}

In the present work we have introduced the \proglang{R} package \pkg{surveillance}
as a comprehensive statistical framework for the analysis of spatio-temporal
surveillance data covering individual-level event data as well as
aggregated count data time series. The package offers a
multitude of methods for visualization, likelihood inference and
simulation of endemic-epidemic models.
Additional functionality beyond the illustrations in
Sections~\ref{sec:twinstim} to \ref{sec:hhh4}
can be found via \code{help(package = "surveillance")}.
By the open-source implementation of recently developed statistical
methodology in a readily available \proglang{R} package,
we support reproducibility of research and
hope to serve an increased need in analyzing
spatio-temporal epidemic data using statistical models.

%
%
%

\section*{Acknowledgements}

The implementation of the \code{hhh4} model is mainly due to Michaela Paul, to
whom we are thankful for all methodological advances and code contributions in
the past years.
We also acknowledge all other code contributors in the long history
of the \pkg{surveillance} package (in alphabetical order):
Thais Correa, Mathias Hofmann, Christian Lang, Juliane Manitz, Andrea Riebler, Daniel Sabanés Bové, Maëlle Salmon, Dirk Schumacher, Stefan Steiner, Mikko Virtanen, Wei Wei, Valentin Wimmer.
Many have also helped us by investigating the package and giving feedback:
Doris Altmann, Johannes Dreesman, Johannes Elias, Marc Geilhufe, Kurt Hornik, Mayeul Kauffmann, Marcos Prates, Brian D. Ripley, Barry Rowlingson, Christopher W. Ryan, Klaus Stark, Yann Le Strat, André Michael Toschke, Wei Wei, George Wood, Achim Zeileis, Bing Zhang.
We appreciate the helpful comments from two anonymous reviewers of
an earlier version of this manuscript.

Financial support by the Munich Center of Health Sciences (2007--2010)
and the Swiss National Science Foundation (2007--2015)
is gratefully acknowledged.

\section*{Involved \proglang{R} packages and versions}

This paper is based on \pkg{surveillance}~1.10-0 \citep{R:surveillance}
in R version 3.2.2 (2015-08-14)
using \pkg{knitr} \citep{Xie2015} for dynamic report generation.
The implementations of the three presented endemic-epidemic modeling frameworks
rely on several other \proglang{R} packages.
In the following we list all packages involved as first-order dependencies of
\pkg{surveillance} with the versions used in this paper:
\pkg{sp}~1.2-1 \citep{R:sp}, \pkg{xtable}~1.8-0 \citep{R:xtable}, \pkg{polyCub}~0.5-2 \citep{R:polyCub}, \pkg{MASS}~7.3-44 \citep{R:MASS}, \pkg{Matrix}~1.2-2 \citep{R:Matrix}, \pkg{spatstat}~1.43-0 \citep{R:spatstat}, \pkg{lattice}~0.20-33 \citep{R:lattice}, \pkg{colorspace}~1.2-6 \citep{R:colorspace}, \pkg{scales}~0.3.0 \citep{R:scales}, \pkg{quadprog}~1.5-5 \citep{R:quadprog}, \pkg{memoise}~0.2.1 \citep{R:memoise}, \pkg{polyclip}~1.3-2 \citep{R:polyclip}, \pkg{maptools}~0.8-37 \citep{R:maptools} and \pkg{spdep}~0.5-88 \citep{R:spdep}.

\proglang{R} itself, the \pkg{surveillance} package, and all other
aforementioned packages are available from
the Comprehensive \proglang{R} Archive Network (CRAN) at \url{https://CRAN.R-project.org/}.
The development of \pkg{surveillance} is hosted at
\url{http://surveillance.r-forge.r-project.org/}.


\bibliography{references,R}

\begin{thebibliography}{90}
\newcommand{\enquote}[1]{``#1''}
\providecommand{\natexlab}[1]{#1}
\providecommand{\url}[1]{\texttt{#1}}
\providecommand{\urlprefix}{URL }
\expandafter\ifx\csname urlstyle\endcsname\relax
  \providecommand{\doi}[1]{doi:\discretionary{}{}{}#1}\else
  \providecommand{\doi}{doi:\discretionary{}{}{}\begingroup
  \urlstyle{rm}\Url}\fi
\providecommand{\eprint}[2][]{\url{#2}}

\bibitem[{Adelfio and Chiodi(2015)}]{adelfio.chiodi2015}
Adelfio G, Chiodi M (2015).
\newblock \enquote{{FLP} {E}stimation of {S}emi-parametric {M}odels for
  {S}pace–time {P}oint {P}rocesses and {D}iagnostic {T}ools.}
\newblock \emph{Spatial Statistics}.
\newblock ISSN 2211-6753.
\newblock \doi{10.1016/j.spasta.2015.06.004}.
\newblock In press.

\bibitem[{Anderson and May(1991)}]{Anderson.May1991}
Anderson RM, May RM (1991).
\newblock \emph{{I}nfectious {D}iseases of {H}umans: {D}ynamics and {C}ontrol}.
\newblock Oxford University Press.

\bibitem[{Auguie(2015)}]{R:gridExtra}
Auguie B (2015).
\newblock \emph{{\pkg{gridExtra}}: Miscellaneous Functions for "Grid"
  Graphics}.
\newblock \proglang{R}~package version~2.0.0,
  \urlprefix\url{http://CRAN.R-project.org/package=gridExtra}.

\bibitem[{Baddeley \emph{et~al.}(2015)Baddeley, Rubak, and Turner}]{R:spatstat}
Baddeley A, Rubak E, Turner R (2015).
\newblock \emph{Spatial Point Patterns: Methodology and Applications with
  \proglang{R}}.
\newblock Chapman and Hall/CRC Press, London.
\newblock In press,
  \urlprefix\url{http://www.crcpress.com/Spatial-Point-Patterns-Methodology-and-Applications-with-R/Baddeley-Rubak-Turner/9781482210200/}.

\bibitem[{Balderama \emph{et~al.}(2012)Balderama, Schoenberg, Murray, and
  Rundel}]{balderama.etal2012}
Balderama E, Schoenberg FP, Murray E, Rundel PW (2012).
\newblock \enquote{{A}pplication of {B}ranching {M}odels in the {S}tudy of
  {I}nvasive {S}pecies.}
\newblock \emph{Journal of the American Statistical Association},
  \textbf{107}(498), 467--476.
\newblock \doi{10.1080/01621459.2011.641402}.

\bibitem[{Bates and Maechler(2015)}]{R:Matrix}
Bates D, Maechler M (2015).
\newblock \emph{{\pkg{Matrix}}: Sparse and Dense Matrix Classes and Methods}.
\newblock \proglang{R}~package version~1.2-2,
  \urlprefix\url{http://CRAN.R-project.org/package=Matrix}.

\bibitem[{Bernard \emph{et~al.}(2014)Bernard, Werber, and
  Höhle}]{bernard.etal2014}
Bernard H, Werber D, Höhle M (2014).
\newblock \enquote{{E}stimating the {U}nder-{R}eporting of {N}orovirus
  {I}llness in {G}ermany {U}tilizing {E}nhanced {A}wareness of {D}iarrhoea
  {D}uring a {L}arge {O}utbreak of {S}higa {T}oxin-{P}roducing {E}. {C}oli
  {O}104:{H}4 in 2011 -- {A} {T}ime {S}eries {A}nalysis.}
\newblock \emph{BMC Infectious Diseases}, \textbf{14}(1), 116.
\newblock \doi{10.1186/1471-2334-14-116}.

\bibitem[{Besag \emph{et~al.}(1991)Besag, York, and
  Mollié}]{BesagYorkMollie1991}
Besag J, York J, Mollié A (1991).
\newblock \enquote{{B}ayesian {I}mage-{R}estoration, with {T}wo {A}pplications
  in {S}patial {S}tatistics.}
\newblock \emph{The Annals of the Institute of Statistical Mathematics},
  \textbf{43}(1), 1--20.

\bibitem[{Bivand \emph{et~al.}(2015)Bivand, Keitt, and Rowlingson}]{R:rgdal}
Bivand R, Keitt T, Rowlingson B (2015).
\newblock \emph{{\pkg{rgdal}}: Bindings for the Geospatial Data Abstraction
  Library}.
\newblock \proglang{R}~package version~1.1-1,
  \urlprefix\url{http://CRAN.R-project.org/package=rgdal}.

\bibitem[{Bivand and Lewin-Koh(2015)}]{R:maptools}
Bivand R, Lewin-Koh N (2015).
\newblock \emph{{\pkg{maptools}}: Tools for Reading and Handling Spatial
  Objects}.
\newblock \proglang{R}~package version~0.8-37,
  \urlprefix\url{http://CRAN.R-project.org/package=maptools}.

\bibitem[{Bivand and Piras(2015)}]{R:spdep}
Bivand R, Piras G (2015).
\newblock \enquote{Comparing Implementations of Estimation Methods for Spatial
  Econometrics.}
\newblock \emph{Journal of Statistical Software}, \textbf{63}(18), 1--36.
\newblock \urlprefix\url{http://www.jstatsoft.org/v63/i18/}.

\bibitem[{Bivand \emph{et~al.}(2013)Bivand, Pebesma, and
  Gómez-Rubio}]{Bivand.etal2013}
Bivand RS, Pebesma E, Gómez-Rubio V (2013).
\newblock \emph{{A}pplied {S}patial {D}ata {A}nalysis with \proglang{R}},
  volume~10 of \emph{Use R!}
\newblock 2nd edition. Springer-Verlag, New York.
\newblock ISBN 1-4614-7617-8.
\newblock \urlprefix\url{http://www.asdar-book.org}.

\bibitem[{Brown(2015)}]{brown2015}
Brown PE (2015).
\newblock \enquote{{M}odel-{B}ased {G}eostatistics the {E}asy {W}ay.}
\newblock \emph{Journal of Statistical Software}, \textbf{63}(12), 1--24.
\newblock ISSN 1548-7660.
\newblock \urlprefix\url{http://www.jstatsoft.org/v63/i12}.

\bibitem[{Cori \emph{et~al.}(2013)Cori, Ferguson, Fraser, and
  Cauchemez}]{cori.etal2013}
Cori A, Ferguson NM, Fraser C, Cauchemez S (2013).
\newblock \enquote{{A} {N}ew {F}ramework and {S}oftware to {E}stimate
  {T}ime-{V}arying {R}eproduction {N}umbers {D}uring {E}pidemics.}
\newblock \emph{American Journal of Epidemiology}, \textbf{178}(9), 1505--1512.
\newblock \doi{10.1093/aje/kwt133}.

\bibitem[{Czado \emph{et~al.}(2009)Czado, Gneiting, and Held}]{czado-etal-2009}
Czado C, Gneiting T, Held L (2009).
\newblock \enquote{{P}redictive {M}odel {A}ssessment for {C}ount {D}ata.}
\newblock \emph{Biometrics}, \textbf{65}(4), 1254--1261.
\newblock \doi{10.1111/j.1541-0420.2009.01191.x}.

\bibitem[{Dahl(2015)}]{R:xtable}
Dahl DB (2015).
\newblock \emph{{\pkg{xtable}}: Export Tables to LaTeX or HTML}.
\newblock \proglang{R}~package version~1.8-0,
  \urlprefix\url{http://CRAN.R-project.org/package=xtable}.

\bibitem[{Daley and Gani(1999)}]{Daley.Gani1999}
Daley DJ, Gani J (1999).
\newblock \emph{{E}pidemic {M}odelling: {A}n {I}ntroduction}, volume~15 of
  \emph{Cambridge Studies in Mathematical Biology}.
\newblock Cambridge University Press.
\newblock ISBN 0-521-64079-2.
\newblock \doi{10.1017/CBO9780511608834}.

\bibitem[{Daley and Vere-Jones(2003)}]{Daley.Vere-Jones2003}
Daley DJ, Vere-Jones D (2003).
\newblock \emph{{A}n {I}ntroduction to the {T}heory of {P}oint {P}rocesses},
  volume I: Elementary Theory and Methods of \emph{Probability and its
  Applications}.
\newblock 2nd edition. Springer-Verlag, New York.
\newblock ISBN 0-387-95541-0.

\bibitem[{Diggle(2006)}]{diggle2006}
Diggle PJ (2006).
\newblock \enquote{{S}patio-{T}emporal {P}oint {P}rocesses, {P}artial
  {L}ikelihood, {F}oot and {M}outh {D}isease.}
\newblock \emph{Statistical Methods in Medical Research}, \textbf{15}(4),
  325--336.
\newblock \doi{10.1191/0962280206sm454oa}.

\bibitem[{Douglas and Peucker(1973)}]{douglas.peucker1973}
Douglas DH, Peucker TK (1973).
\newblock \enquote{{A}lgorithms for the {R}eduction of the {N}umber of {P}oints
  {R}equired to {R}epresent a {D}igitized {L}ine or its {C}aricature.}
\newblock \emph{Cartographica: The International Journal for Geographic
  Information and Geovisualization}, \textbf{10}(2), 112--122.
\newblock \doi{10.3138/FM57-6770-U75U-7727}.

\bibitem[{Fahrmeir \emph{et~al.}(2013)Fahrmeir, Kneib, Lang, and
  Marx}]{Fahrmeir.etal2013}
Fahrmeir L, Kneib T, Lang S, Marx B (2013).
\newblock \emph{{R}egression: {M}odels, {M}ethods and {A}pplications}.
\newblock Springer-Verlag.
\newblock ISBN 3-642-34332-5.
\newblock \doi{10.1007/978-3-642-34333-9}.

\bibitem[{Fine and Clarkson(1982)}]{fine.clarkson1982}
Fine PEM, Clarkson JA (1982).
\newblock \enquote{{M}easles in {E}ngland and {W}ales---{I}: {A}n {A}nalysis of
  {F}actors {U}nderlying {S}easonal {P}atterns.}
\newblock \emph{International Journal of Epidemiology}, \textbf{11}(1), 5--14.
\newblock \doi{10.1093/ije/11.1.5}.

\bibitem[{Finkenstädt \emph{et~al.}(2002)Finkenstädt, Bjørnstad, and
  Grenfell}]{finkenstaedt.etal2002}
Finkenstädt BF, Bjørnstad ON, Grenfell BT (2002).
\newblock \enquote{{A} {S}tochastic {M}odel for {E}xtinction and {R}ecurrence
  of {E}pidemics: {E}stimation and {I}nference for {M}easles {O}utbreaks.}
\newblock \emph{Biostatistics}, \textbf{3}(4), 493--510.
\newblock \doi{10.1093/biostatistics/3.4.493}.

\bibitem[{Finkenstädt and Grenfell(2000)}]{finkenstaedt.grenfell2000}
Finkenstädt BF, Grenfell BT (2000).
\newblock \enquote{{T}ime {S}eries {M}odelling of {C}hildhood {D}iseases: {A}
  {D}ynamical {S}ystems {A}pproach.}
\newblock \emph{Journal of the Royal Statistical Society C}, \textbf{49}(2),
  187--205.
\newblock \doi{10.1111/1467-9876.00187}.

\bibitem[{Geilhufe \emph{et~al.}(2014)Geilhufe, Held, Skrøvseth, Simonsen, and
  Godtliebsen}]{geilhufe.etal2012}
Geilhufe M, Held L, Skrøvseth SO, Simonsen GS, Godtliebsen F (2014).
\newblock \enquote{{P}ower {L}aw {A}pproximations of {M}ovement {N}etwork
  {D}ata for {M}odeling {I}nfectious {D}isease {S}pread.}
\newblock \emph{Biometrical Journal}, \textbf{56}(3), 363--382.
\newblock \doi{10.1002/bimj.201200262}.

\bibitem[{Gneiting and Katzfuss(2014)}]{gneiting.katzfuss2014}
Gneiting T, Katzfuss M (2014).
\newblock \enquote{{P}robabilistic {F}orecasting.}
\newblock \emph{Annual Review of Statistics and Its Application},
  \textbf{1}(1), 125--151.
\newblock \doi{10.1146/annurev-statistics-062713-085831}.

\bibitem[{Greven and Kneib(2010)}]{greven.kneib2010}
Greven S, Kneib T (2010).
\newblock \enquote{{O}n the {B}ehaviour of {M}arginal and {C}onditional {AIC}
  in {L}inear {M}ixed {M}odels.}
\newblock \emph{Biometrika}, \textbf{97}(4), 773--789.
\newblock \doi{10.1093/biomet/asq042}.

\bibitem[{Groendyke \emph{et~al.}(2012)Groendyke, Welch, and
  Hunter}]{groendyke.etal2012}
Groendyke C, Welch D, Hunter DR (2012).
\newblock \enquote{{A} {N}etwork-{B}ased {A}nalysis of the 1861 {H}agelloch
  {M}easles {D}ata.}
\newblock \emph{Biometrics}, \textbf{68}(3), 755--765.
\newblock \doi{10.1111/j.1541-0420.2012.01748.x}.

\bibitem[{Harrower and Bloch(2006)}]{harrower.bloch2006}
Harrower M, Bloch M (2006).
\newblock \enquote{{M}apshaper.org: {A} {M}ap {G}eneralization {W}eb
  {S}ervice.}
\newblock \emph{IEEE Computer Graphics and Applications}, \textbf{26}(4),
  22--27.
\newblock \doi{10.1109/MCG.2006.85}.

\bibitem[{Held \emph{et~al.}(2006)Held, Hofmann, Höhle, and
  Schmid}]{held.etal2006a}
Held L, Hofmann M, Höhle M, Schmid V (2006).
\newblock \enquote{{A} {T}wo-{C}omponent {M}odel for {C}ounts of {I}nfectious
  {D}iseases.}
\newblock \emph{Biostatistics}, \textbf{7}(3), 422--437.
\newblock \doi{10.1093/biostatistics/kxj016}.

\bibitem[{Held \emph{et~al.}(2005)Held, Höhle, and Hofmann}]{held.etal2005}
Held L, Höhle M, Hofmann M (2005).
\newblock \enquote{{A} {S}tatistical {F}ramework for the {A}nalysis of
  {M}ultivariate {I}nfectious {D}isease {S}urveillance {C}ounts.}
\newblock \emph{Statistical Modelling}, \textbf{5}(3), 187--199.
\newblock \doi{10.1191/1471082X05st098oa}.

\bibitem[{Held and Paul(2012)}]{held.paul2012}
Held L, Paul M (2012).
\newblock \enquote{{M}odeling {S}easonality in {S}pace-{T}ime {I}nfectious
  {D}isease {S}urveillance {D}ata.}
\newblock \emph{Biometrical Journal}, \textbf{54}(6), 824--843.
\newblock \doi{10.1002/bimj.201200037}.

\bibitem[{Herzog \emph{et~al.}(2011)Herzog, Paul, and Held}]{herzog.etal2011}
Herzog SA, Paul M, Held L (2011).
\newblock \enquote{{H}eterogeneity in {V}accination {C}overage {E}xplains the
  {S}ize and {O}ccurrence of {M}easles {E}pidemics in {G}erman {S}urveillance
  {D}ata.}
\newblock \emph{Epidemiology and Infection}, \textbf{139}(04), 505--515.
\newblock \doi{10.1017/S0950268810001664}.

\bibitem[{Hughes and King(2003)}]{hughes.king2003}
Hughes AW, King ML (2003).
\newblock \enquote{{M}odel {S}election {U}sing {AIC} in the {P}resence of
  {O}ne-{S}ided {I}nformation.}
\newblock \emph{Journal of Statistical Planning and Inference},
  \textbf{115}(2), 397--411.
\newblock \doi{10.1016/S0378-3758(02)00159-3}.

\bibitem[{Höhle(2007)}]{hoehle2007}
Höhle M (2007).
\newblock \enquote{\pkg{surveillance}: {A}n \proglang{R} package for the
  monitoring of infectious diseases.}
\newblock \emph{Computational Statistics}, \textbf{22}(4), 571--582.
\newblock \doi{10.1007/s00180-007-0074-8}.

\bibitem[{Höhle(2009)}]{hoehle2009}
Höhle M (2009).
\newblock \enquote{{A}dditive-{M}ultiplicative {R}egression {M}odels for
  {S}patio-{T}emporal {E}pidemics.}
\newblock \emph{Biometrical Journal}, \textbf{51}(6), 961--978.
\newblock \doi{10.1002/bimj.200900050}.

\bibitem[{Höhle(2016)}]{Hoehle2016}
Höhle M (2016).
\newblock \enquote{{I}nfectious {D}isease {M}odelling.}
\newblock In AB~Lawson, S~Banerjee, RP~Haining, MD~Ugarte (eds.),
  \emph{Handbook of Spatial Epidemiology}, Chapman \& Hall/CRC Handbooks of
  Modern Statistical Methods. Chapman and Hall/CRC.
\newblock ISBN 1-4822-5301-1.
\newblock Forthcoming,
  \eprint{http://www.math.su.se/~hoehle/pubs/Hoehle_SpaMethInfEpiModelling2015.pdf}.

\bibitem[{Höhle and Mazick(2010)}]{hoehle.mazick2010}
Höhle M, Mazick A (2010).
\newblock \enquote{{A}berration {D}etection in \proglang{R} {I}llustrated by
  {D}anish {M}ortality {M}onitoring.}
\newblock In TA~Kass-Hout, X~Zhang (eds.), \emph{Biosurveillance: Methods and
  Case Studies}, pp. 215--238. Chapman and Hall/CRC.

\bibitem[{Höhle \emph{et~al.}(2015)Höhle, Meyer, and Paul}]{R:surveillance}
Höhle M, Meyer S, Paul M (2015).
\newblock \emph{{\pkg{surveillance}}: Temporal and Spatio-Temporal Modeling and
  Monitoring of Epidemic Phenomena}.
\newblock \proglang{R}~package version~1.10-0,
  \urlprefix\url{http://surveillance.r-forge.r-project.org/}.

\bibitem[{Höhle \emph{et~al.}(2009)Höhle, Paul, and Held}]{hoehle.etal2009}
Höhle M, Paul M, Held L (2009).
\newblock \enquote{{S}tatistical {A}pproaches to the {M}onitoring and
  {S}urveillance of {I}nfectious {D}iseases for {V}eterinary {P}ublic
  {H}ealth.}
\newblock \emph{Preventive Veterinary Medicine}, \textbf{91}(1), 2--10.
\newblock \doi{10.1016/j.prevetmed.2009.05.017}.

\bibitem[{Ihaka \emph{et~al.}(2015)Ihaka, Murrell, Hornik, Fisher, and
  Zeileis}]{R:colorspace}
Ihaka R, Murrell P, Hornik K, Fisher JC, Zeileis A (2015).
\newblock \emph{{\pkg{colorspace}}: Color Space Manipulation}.
\newblock \proglang{R}~package version~1.2-6,
  \urlprefix\url{http://CRAN.R-project.org/package=colorspace}.

\bibitem[{Johnson(2015)}]{R:polyclip}
Johnson A (2015).
\newblock \emph{{\pkg{polyclip}}: Polygon Clipping}.
\newblock \proglang{R}~package version~1.3-2, ported to \proglang{R} by Adrian
  Baddeley and Brian Ripley,
  \urlprefix\url{http://CRAN.R-project.org/package=polyclip}.

\bibitem[{Johnson(2010)}]{johnson2010}
Johnson SD (2010).
\newblock \enquote{{A} {B}rief {H}istory of the {A}nalysis of {C}rime
  {C}oncentration.}
\newblock \emph{European Journal of Applied Mathematics}, \textbf{21}(Special
  Double Issue 4-5), 349--370.
\newblock ISSN 1469-4425.
\newblock \doi{10.1017/S0956792510000082}.

\bibitem[{Jombart \emph{et~al.}(2014)Jombart, Cori, Didelot, Cauchemez, Fraser,
  and Ferguson}]{jombart.etal2014a}
Jombart T, Cori A, Didelot X, Cauchemez S, Fraser C, Ferguson N (2014).
\newblock \enquote{{B}ayesian {R}econstruction of {D}isease {O}utbreaks by
  {C}ombining {E}pidemiologic and {G}enomic {D}ata.}
\newblock \emph{PLOS Computational Biology}, \textbf{10}(1), e1003457.
\newblock \doi{10.1371/journal.pcbi.1003457}.

\bibitem[{Keeling and Rohani(2008)}]{Keeling.Rohani2008}
Keeling MJ, Rohani P (2008).
\newblock \emph{{M}odeling {I}nfectious {D}iseases in {H}umans and {A}nimals}.
\newblock Princeton University Press.
\newblock ISBN 0-691-11617-2.
\newblock \urlprefix\url{http://www.modelinginfectiousdiseases.org/}.

\bibitem[{Kermack and McKendrick(1927)}]{kermack.mckendrick1927}
Kermack WO, McKendrick AG (1927).
\newblock \enquote{{A} {C}ontribution to the {M}athematical {T}heory of
  {E}pidemics.}
\newblock \emph{Proceedings of the Royal Society of London A},
  \textbf{115}(772), 700--721.
\newblock \doi{10.1098/rspa.1927.0118}.

\bibitem[{Lawson and Leimich(2000)}]{lawson.leimich2000}
Lawson AB, Leimich P (2000).
\newblock \enquote{{A}pproaches to the {S}pace-{T}ime {M}odelling of
  {I}nfectious {D}isease {B}ehaviour.}
\newblock \emph{IMA Journal of Mathematics Applied in Medicine and Biology},
  \textbf{17}(1), 1--13.

\bibitem[{Liboschik \emph{et~al.}(2015)Liboschik, Fokianos, and
  Fried}]{liboschik.etal2015}
Liboschik T, Fokianos K, Fried R (2015).
\newblock \enquote{\pkg{tscount}: {A}n \textsf{R} package for analysis of count
  time series following generalized linear models.}
\newblock \emph{SFB 823 Discussion Paper 6/2015}, TU Dortmund.
\newblock \urlprefix\url{http://CRAN.R-project.org/package=tscount}.

\bibitem[{Malesios \emph{et~al.}(2014)Malesios, Demiris, Kalogeropoulos, and
  Ntzoufras}]{malesios.etal2014}
Malesios C, Demiris N, Kalogeropoulos K, Ntzoufras I (2014).
\newblock \enquote{{B}ayesian {S}patio-{T}emporal {E}pidemic {M}odels with
  {A}pplications to {S}heep {P}ox.}
\newblock \eprint{http://arxiv.org/abs/1403.1783}.

\bibitem[{Martinussen and Scheike(2002)}]{martinussen_scheike2002}
Martinussen T, Scheike TH (2002).
\newblock \enquote{{A} {F}lexible {A}dditive {M}ultiplicative {H}azard
  {M}odel.}
\newblock \emph{Biometrika}, \textbf{89}(2), 283--298.
\newblock ISSN 0006-3444.
\newblock \doi{10.1093/biomet/89.2.283}.

\bibitem[{Merl \emph{et~al.}(2010)Merl, Johnson, Gramacy, and
  Mangel}]{merl.etal2010}
Merl D, Johnson LR, Gramacy RB, Mangel M (2010).
\newblock \enquote{\pkg{amei}: {A}n \textsf{R} {P}ackage for the {A}daptive
  {M}anagement of {E}pidemiological {I}nterventions.}
\newblock \emph{Journal of Statistical Software}, \textbf{36}(6), 1--32.
\newblock ISSN 1548-7660.
\newblock \urlprefix\url{http://www.jstatsoft.org/v36/i06}.

\bibitem[{Meyer(2015)}]{R:polyCub}
Meyer S (2015).
\newblock \emph{{\pkg{polyCub}}: Cubature over Polygonal Domains}.
\newblock \proglang{R}~package version~0.5-2,
  \urlprefix\url{http://CRAN.R-project.org/package=polyCub}.

\bibitem[{Meyer \emph{et~al.}(2012)Meyer, Elias, and Höhle}]{meyer.etal2011}
Meyer S, Elias J, Höhle M (2012).
\newblock \enquote{{A} {S}pace-{T}ime {C}onditional {I}ntensity {M}odel for
  {I}nvasive {M}eningococcal {D}isease {O}ccurrence.}
\newblock \emph{Biometrics}, \textbf{68}(2), 607--616.
\newblock \doi{10.1111/j.1541-0420.2011.01684.x}.
\newblock \eprint{http://arxiv.org/abs/1508.05740}.

\bibitem[{Meyer and Held(2014{\natexlab{a}})}]{meyer.held2013}
Meyer S, Held L (2014{\natexlab{a}}).
\newblock \enquote{{P}ower-{L}aw {M}odels for {I}nfectious {D}isease {S}pread.}
\newblock \emph{The Annals of Applied Statistics}, \textbf{8}(3), 1612--1639.
\newblock \doi{10.1214/14-AOAS743}.

\bibitem[{Meyer and Held(2014{\natexlab{b}})}]{meyer.held2013:suppB}
Meyer S, Held L (2014{\natexlab{b}}).
\newblock \enquote{{S}upplement {B} of `{P}ower-{L}aw {M}odels for {I}nfectious
  {D}isease {S}pread'.}
\newblock \doi{10.1214/14-AOAS743SUPPB}.
\newblock
  \eprint{http://www.biostat.uzh.ch/research/manuscripts/powerlaw.html}.

\bibitem[{Meyer \emph{et~al.}(2015)Meyer, Warnke, Rössler, and
  Held}]{meyer.etal2015}
Meyer S, Warnke I, Rössler W, Held L (2015).
\newblock \enquote{{M}odel-based testing for space-time interaction using point
  processes: {A}n application to psychiatric hospital admissions in an urban
  area.}
\newblock Submitted to \textit{Spatial and Spatio-temporal Epidemiology}.

\bibitem[{Mohler \emph{et~al.}(2011)Mohler, Short, Brantingham, Schoenberg, and
  Tita}]{mohler.etal2011}
Mohler GO, Short MB, Brantingham PJ, Schoenberg FP, Tita GE (2011).
\newblock \enquote{{S}elf-exciting {P}oint {P}rocess {M}odeling of {C}rime.}
\newblock \emph{Journal of the American Statistical Association},
  \textbf{106}(493), 100--108.
\newblock \doi{10.1198/jasa.2011.ap09546}.

\bibitem[{Mossong \emph{et~al.}(2008)Mossong, Hens, Jit, Beutels, Auranen,
  Mikolajczyk, Massari, Salmaso, Tomba, Wallinga, Heijne, Sadkowska-Todys,
  Rosinska, and Edmunds}]{mossong.etal2008}
Mossong J, Hens N, Jit M, Beutels P, Auranen K, Mikolajczyk R, Massari M,
  Salmaso S, Tomba GS, Wallinga J, Heijne J, Sadkowska-Todys M, Rosinska M,
  Edmunds WJ (2008).
\newblock \enquote{{S}ocial {C}ontacts and {M}ixing {P}atterns {R}elevant to
  the {S}pread of {I}nfectious {D}iseases.}
\newblock \emph{PLoS Medicine}, \textbf{5}(3), e74.
\newblock \doi{10.1371/journal.pmed.0050074}.

\bibitem[{Neal and Roberts(2004)}]{neal.roberts2004}
Neal PJ, Roberts GO (2004).
\newblock \enquote{{S}tatistical {I}nference and {M}odel {S}election for the
  1861 {H}agelloch {M}easles {E}pidemic.}
\newblock \emph{Biostatistics}, \textbf{5}(2), 249--261.
\newblock \doi{10.1093/biostatistics/5.2.249}.

\bibitem[{Obadia \emph{et~al.}(2012)Obadia, Haneef, and
  Boelle}]{obadia.etal2012}
Obadia T, Haneef R, Boelle PY (2012).
\newblock \enquote{{T}he \pkg{R0} {P}ackage: {A} {T}oolbox to {E}stimate
  {R}eproduction {N}umbers for {E}pidemic {O}utbreaks.}
\newblock \emph{BMC Medical Informatics and Decision Making}, \textbf{12}(147).
\newblock ISSN 1472-6947.
\newblock \doi{10.1186/1472-6947-12-147}.

\bibitem[{Ogata(1988)}]{ogata1988}
Ogata Y (1988).
\newblock \enquote{{S}tatistical {M}odels for {E}arthquake {O}ccurrences and
  {R}esidual {A}nalysis for {P}oint {P}rocesses.}
\newblock \emph{Journal of the American Statistical Association},
  \textbf{83}(401), 9--27.
\newblock \urlprefix\url{http://www.jstor.org/stable/2288914}.

\bibitem[{Ogata(1999)}]{ogata1999}
Ogata Y (1999).
\newblock \enquote{{S}eismicity {A}nalysis {T}hrough {P}oint-process
  {M}odeling: {A} {R}eview.}
\newblock \emph{Pure and Applied Geophysics}, \textbf{155}(2), 471--507.
\newblock ISSN 0033-4553.
\newblock \doi{10.1007/s000240050275}.

\bibitem[{Paul and Held(2011)}]{paul.held2011}
Paul M, Held L (2011).
\newblock \enquote{{P}redictive {A}ssessment of a {N}on-{L}inear {R}andom
  {E}ffects {M}odel for {M}ultivariate {T}ime {S}eries of {I}nfectious
  {D}isease {C}ounts.}
\newblock \emph{Statistics in Medicine}, \textbf{30}(10), 1118--1136.
\newblock \doi{10.1002/sim.4177}.

\bibitem[{Paul \emph{et~al.}(2008)Paul, Held, and Toschke}]{paul.etal2008}
Paul M, Held L, Toschke A (2008).
\newblock \enquote{{M}ultivariate {M}odelling of {I}nfectious {D}isease
  {S}urveillance {D}ata.}
\newblock \emph{Statistics in Medicine}, \textbf{27}(29), 6250--6267.
\newblock \doi{10.1002/sim.3440}.

\bibitem[{Pebesma(2012)}]{pebesma2012}
Pebesma E (2012).
\newblock \enquote{\pkg{spacetime}: {S}patio-{T}emporal {D}ata in
  \proglang{R}.}
\newblock \emph{Journal of Statistical Software}, \textbf{51}(7), 1--30.
\newblock \urlprefix\url{http://www.jstatsoft.org/v51/i07/}.

\bibitem[{Pebesma(2015)}]{CTV:SpatioTemporal}
Pebesma E (2015).
\newblock \enquote{CRAN Task View: Handling and Analyzing Spatio-Temporal
  Data.}
\newblock \url{http://CRAN.R-project.org/web/views/SpatioTemporal.html}.
\newblock Version 2015-09-01.

\bibitem[{Pebesma and Bivand(2005)}]{R:sp}
Pebesma EJ, Bivand RS (2005).
\newblock \enquote{Classes and methods for spatial data in \proglang{R}.}
\newblock \emph{R News}, \textbf{5}(2), 9--13.
\newblock \urlprefix\url{http://CRAN.R-project.org/doc/Rnews/}.

\bibitem[{{R Core Team}(2015)}]{R:base}
{R Core Team} (2015).
\newblock \emph{\proglang{R}: A Language and Environment for Statistical
  Computing}.
\newblock \proglang{R}~Foundation for Statistical Computing, Vienna, Austria.
\newblock \urlprefix\url{https://www.R-project.org/}.

\bibitem[{Rowlingson and Diggle(2015)}]{R:splancs}
Rowlingson B, Diggle P (2015).
\newblock \emph{{\pkg{splancs}}: Spatial and Space-Time Point Pattern
  Analysis}.
\newblock \proglang{R}~package version~2.01-38,
  \urlprefix\url{http://CRAN.R-project.org/package=splancs}.

\bibitem[{Ryan and Ulrich(2014)}]{R:xts}
Ryan JA, Ulrich JM (2014).
\newblock \emph{{\pkg{xts}}: eXtensible Time Series}.
\newblock \proglang{R}~package version~0.9-7,
  \urlprefix\url{http://CRAN.R-project.org/package=xts}.

\bibitem[{Salmon \emph{et~al.}(2015)Salmon, Schumacher, and
  H\"{o}hle}]{salmon.etal2014}
Salmon M, Schumacher D, H\"{o}hle M (2015).
\newblock \enquote{{M}onitoring {C}ount {T}ime {S}eries in \proglang{R}:
  {A}berration {D}etection in {P}ublic {H}ealth {S}urveillance.}
\newblock \emph{Journal of Statistical Software}.
\newblock In press, \eprint{http://arxiv.org/abs/1411.1292}.

\bibitem[{Sarkar(2008)}]{R:lattice}
Sarkar D (2008).
\newblock \emph{Lattice: Multivariate Data Visualization with R}.
\newblock Springer, New York.
\newblock ISBN 978-0-387-75968-5,
  \urlprefix\url{http://lmdvr.r-forge.r-project.org}.

\bibitem[{Scheike and Martinussen(2006)}]{R:timereg}
Scheike TH, Martinussen T (2006).
\newblock \emph{Dynamic Regression models for survival data}.
\newblock Springer, NY.

\bibitem[{Scheike and Zhang(2011)}]{scheike.zhang2011}
Scheike TH, Zhang MJ (2011).
\newblock \enquote{{A}nalyzing {C}ompeting {R}isk {D}ata {U}sing the
  \proglang{R} \pkg{timereg} package.}
\newblock \emph{Journal of Statistical Software}, \textbf{38}(2), 1--15.
\newblock \urlprefix\url{http://www.jstatsoft.org/v38/i02}.

\bibitem[{Schrödle \emph{et~al.}(2012)Schrödle, Held, and
  Rue}]{schroedle.etal2012}
Schrödle B, Held L, Rue H (2012).
\newblock \enquote{{A}ssessing the {I}mpact of a {M}ovement {N}etwork on the
  {S}patiotemporal {S}pread of {I}nfectious {D}iseases.}
\newblock \emph{Biometrics}, \textbf{68}(3), 736--744.
\newblock \doi{10.1111/j.1541-0420.2011.01717.x}.

\bibitem[{Silvapulle and Sen(2005)}]{Silvapulle.Sen2005}
Silvapulle MJ, Sen PK (2005).
\newblock \emph{{C}onstrained {S}tatistical {I}nference: {O}rder, {I}nequality,
  and {S}hape {C}onstraints}.
\newblock Wiley Series in Probability and Statistics. John Wiley \& Sons.
\newblock ISBN 0-471-20827-2.

\bibitem[{Sommariva and Vianello(2007)}]{sommariva.vianello2007}
Sommariva A, Vianello M (2007).
\newblock \enquote{{P}roduct {G}auss {C}ubature over {P}olygons based on
  {G}reen's {I}ntegration {F}ormula.}
\newblock \emph{Bit Numerical Mathematics}, \textbf{47}(2), 441--453.
\newblock ISSN 0006-3835.
\newblock \doi{10.1007/s10543-007-0131-2}.

\bibitem[{Stadler and Bonhoeffer(2013)}]{stadler.bonhoeffer2013}
Stadler T, Bonhoeffer S (2013).
\newblock \enquote{{U}ncovering {E}pidemiological {D}ynamics in {H}eterogeneous
  {H}ost {P}opulations {U}sing {P}hylogenetic {M}ethods.}
\newblock \emph{Philosophical Transactions of the Royal Society of London B:
  Biological Sciences}, \textbf{368}(1614), 20120198.
\newblock \doi{10.1098/rstb.2012.0198}.

\bibitem[{Therneau(2015)}]{R:survival}
Therneau TM (2015).
\newblock \emph{A Package for Survival Analysis in S}.
\newblock Version 2.38,
  \urlprefix\url{http://CRAN.R-project.org/package=survival}.

\bibitem[{Turlach(2013)}]{R:quadprog}
Turlach BA (2013).
\newblock \emph{{\pkg{quadprog}}: Functions to solve Quadratic Programming
  Problems.}
\newblock \proglang{R}~package version~1.5-5, ported to \proglang{R} by Andreas
  Weingessel, \urlprefix\url{http://CRAN.R-project.org/package=quadprog}.

\bibitem[{Utsu \emph{et~al.}(1995)Utsu, Ogata, and Matsu'ura}]{utsu.etal1995}
Utsu T, Ogata Y, Matsu'ura RS (1995).
\newblock \enquote{{T}he {C}entenary of the {O}mori {F}ormula for a {D}ecay
  {L}aw of {A}ftershock {A}ctivity.}
\newblock \emph{Journal of Physics of the Earth}, \textbf{43}(1), 1--33.
\newblock \doi{10.4294/jpe1952.43.1}.

\bibitem[{Venables and Ripley(2002)}]{R:MASS}
Venables WN, Ripley BD (2002).
\newblock \emph{Modern Applied Statistics with S}.
\newblock Fourth edition. Springer, New York.
\newblock ISBN 0-387-95457-0,
  \urlprefix\url{http://www.stats.ox.ac.uk/pub/MASS4}.

\bibitem[{Vrbik \emph{et~al.}(2012)Vrbik, Deardon, Feng, Gardner, and
  Braun}]{vrbik.etal2012}
Vrbik I, Deardon R, Feng Z, Gardner A, Braun J (2012).
\newblock \enquote{{U}sing {I}ndividual-level {M}odels for {I}nfectious
  {D}isease {S}pread to {M}odel {S}patio-temporal {C}ombustion {D}ynamics.}
\newblock \emph{Bayesian Analysis}, \textbf{7}(3), 615--638.
\newblock \doi{10.1214/12-BA721}.

\bibitem[{Waller and Gotway(2004)}]{Waller.Gotway2004}
Waller LA, Gotway CA (2004).
\newblock \emph{{A}pplied {S}patial {S}tatistics for {P}ublic {H}ealth {D}ata}.
\newblock Wiley Series in Probability and Statistics. John Wiley \& Sons.
\newblock ISBN 0-471-66267-4.
\newblock \doi{10.1002/0471662682}.

\bibitem[{Wei and Held(2014)}]{wei.held2013}
Wei W, Held L (2014).
\newblock \enquote{{C}alibration {T}ests for {C}ount {D}ata.}
\newblock \emph{TEST}, \textbf{23}(4), 787--805.
\newblock \doi{10.1007/s11749-014-0380-8}.

\bibitem[{Wickham(2014)}]{R:memoise}
Wickham H (2014).
\newblock \emph{{\pkg{memoise}}: Memoise functions}.
\newblock \proglang{R}~package version~0.2.1,
  \urlprefix\url{http://CRAN.R-project.org/package=memoise}.

\bibitem[{Wickham(2015)}]{R:scales}
Wickham H (2015).
\newblock \emph{{\pkg{scales}}: Scale Functions for Visualization}.
\newblock \proglang{R}~package version~0.3.0,
  \urlprefix\url{http://CRAN.R-project.org/package=scales}.

\bibitem[{Xia \emph{et~al.}(2004)Xia, Bjørnstad, and Grenfell}]{xia.etal2004}
Xia Y, Bjørnstad ON, Grenfell BT (2004).
\newblock \enquote{{M}easles {M}etapopulation {D}ynamics: {A} {G}ravity {M}odel
  for {E}pidemiological {C}oupling and {D}ynamics.}
\newblock \emph{The American Naturalist}, \textbf{164}(2), 267--281.
\newblock ISSN 0003-0147.
\newblock \urlprefix\url{http://www.jstor.org/stable/10.1086/422341}.

\bibitem[{Xie(2013)}]{R:animation}
Xie Y (2013).
\newblock \enquote{{\pkg{animation}}: An \proglang{R} Package for Creating
  Animations and Demonstrating Statistical Methods.}
\newblock \emph{Journal of Statistical Software}, \textbf{53}(1), 1--27.
\newblock \urlprefix\url{http://www.jstatsoft.org/v53/i01/}.

\bibitem[{Xie(2015)}]{Xie2015}
Xie Y (2015).
\newblock \emph{{D}ynamic {D}ocuments with \proglang{R} and \pkg{knitr}}.
\newblock The R Series, 2nd edition. Chapman and Hall/CRC, Boca Raton, Florida.
\newblock ISBN 1-4987-1696-2.
\newblock \urlprefix\url{http://yihui.name/knitr/}.

\end{thebibliography}

\end{document}